\documentclass{article}

    \PassOptionsToPackage{numbers, compress}{natbib}
 \usepackage[preprint]{neurips_2025}

\usepackage[utf8]{inputenc} 
\usepackage[T1]{fontenc}    
\usepackage{hyperref}       
\usepackage{url}            
\usepackage{booktabs}       
\usepackage{amsfonts}       
\usepackage{nicefrac}       
\usepackage{microtype}      
\usepackage[dvipsnames,table,xcdraw]{xcolor}
\usepackage[utf8]{inputenc} 
\usepackage[T1]{fontenc}    
\usepackage{hyperref}       
\usepackage{url}            
\usepackage{booktabs}       
\usepackage{amsfonts}       
\usepackage{nicefrac}       
\usepackage{microtype}      
\usepackage{graphicx}
\usepackage{xspace}
\usepackage{authblk}
\usepackage{multirow}
\usepackage{multicol}
\usepackage{subcaption}
\usepackage{caption} 
\usepackage{amsfonts}     
\usepackage{amsmath}  

\usepackage{enumitem} 

\newenvironment{smitemize}{
  \begin{itemize}[topsep=1pt, partopsep=0pt, itemsep=1pt, parsep=0pt, itemindent=1pt]
}{\end{itemize}}

\usepackage[color=green]{todonotes}

\newcommand{\system}{{LAMDA}\xspace}

\title{{\system: A Longitudinal Android Malware Benchmark for Concept Drift Analysis}}

\author{%
  Md Ahsanul Haque$^{1}$, Ismail Hossain$^{1}$, Md Mahmuduzzaman Kamol$^{1}$, Md Jahangir Alam$^{1}$, Suresh Kumar Amalapuram$^{2}$, Sajedul Talukder$^{1}$, Mohammad Saidur Rahman$^{1}$ \\
    $^1$Department of Computer Science, University of Texas at El Paso\\
    $^2$School of Informatics, University of Edinburgh\\
   \texttt{\{mhaque3,ihossain,mkamol,malam10\}@miners.utep.edu}\\ \texttt{samalapu@ed.ac.uk}, \texttt{\{stalukder,msrahman3\}@utep.edu}\\
}

\begin{document}

\maketitle

\begin{abstract}

Machine learning (ML)-based malware detection systems often fail to account for the dynamic nature of real-world training and test data distributions. In practice, these distributions evolve due to frequent changes in the Android ecosystem, adversarial development of new malware families, and the continuous emergence of both benign and malicious applications. Prior studies have shown that such concept drift—distributional shifts in benign and malicious samples, leads to significant degradation in detection performance over time. Despite the practical importance of this issue, existing datasets are often outdated and limited in temporal scope, diversity of malware families, and sample scale, making them insufficient for the systematic evaluation of concept drift in malware detection.

To address this gap, we present LAMDA, the largest and most temporally diverse Android malware benchmark to date, designed specifically for concept drift analysis. LAMDA spans 12 years (2013–2025, excluding 2015), includes over 1 million samples (approximately 37\% labeled as malware), and covers 1,380 malware families and 150,000 singleton samples, reflecting the natural distribution and evolution of real-world Android applications. We empirically demonstrate LAMDA's utility by quantifying the performance degradation of standard ML models over time and analyzing feature stability across years. As the most comprehensive Android malware dataset to date, LAMDA enables in-depth research into temporal drift, generalization, explainability, and evolving detection challenges. The dataset and code are available at: https://iqsec-lab.github.io/LAMDA/.
\end{abstract}

\section{Introduction}




Android malware poses a growing threat to user privacy and security, with over 33 million attacks blocked in 2024 alone~\cite{kaspersky2024, gdata, av-test, counterpoint2025}. Machine learning (ML)-based detectors, which analyze features extracted from Android application packages (APKs), have emerged as a promising defense mechanism~\cite{arp2014drebin, amos2013applying, sahs2012machine, alzaylaee2017emulator, malwareguard, aafer2013droidapiminer, mamadroid}. However, these detectors often suffer performance degradation over time due to {\em concept drift} --- gradual shifts in the feature distribution caused by the evolving nature of both malicious and benign software~\cite{chow2023drift, chen2023continuous, cade, transcendingtranscend, Transcend, tesseract}.

Concept drift can result from several factors, including changes in developer practices, updates to Android APIs, and, most significantly, the evolving and adaptive strategies of malware authors~\cite{soumnibot2024,wired2020mallockerb}. 
To evade detection, adversaries frequently obfuscate or modify their code by injecting alternative API calls, altering manifest components, or exploiting newly introduced services~\cite{wired2020mallockerb, soumnibot2024}. For example, the Android trojan {\em SoumniBot} obfuscates its manifest file to evade analysis and detection~\cite{soumnibot2024}. These tactics lead to observable shifts in static features over time, undermining the robustness of ML-based detection systems.
Prior studies have shown that malware families (i.e., clusters of samples exhibiting similar behavioral traits) play a central role in driving such drifts~\cite{chow2023drift, transcendingtranscend, arp2014drebin}.

Although concept drift plays a central role in the evolution of Android malware, most existing datasets are not designed to support its analysis. Datasets such as Drebin~\cite{arp2014drebin}, TESSERACT~\cite{tesseract}, and API Graph~\cite{api_graph_dataset} are limited in temporal coverage, family diversity, or structural organization for studying drift. Similarly, Windows-based datasets like EMBER~\cite{ember}, SOREL-20M~\cite{sorel20m}, and BODMAS~\cite{BODMAS} are constrained by short collection periods or focus on different ecosystems. While EMBERSim~\cite{corlatescu2023embersim}, MalNet~\cite{freitas2020malnet}, and AnoShift~\cite{civitarese2022anoshift} offer task-specific contributions, they do not provide longitudinal support for drift analysis in Android malware classification. To address these gaps, we introduce \system, a novel Android malware benchmark dataset curated for temporal drift analysis with family evolution. \system spans over 12 years (i.e., 2013–2025, excluding 2015 due to the unavailability of hashes in the AndroZoo repository~\cite{androzoo,androzooMetadata}), covering 1,008,381 APK samples across 1,380 unique malware families and over 150,000 Singleton samples (i..e, samples without {\em av class} labels) from AndroZoo repository~\cite{androzoo,androzooMetadata}. Each sample is labeled using VirusTotal’s \texttt{vt\_detection} count~\cite{virustotal} reported in AndroZoo database~\cite{androzoo,androzooMetadata}. The samples are decompiled to extract fine-grained static features based on the feature definitions of Drebin~\cite{arp2014drebin}.



We validate \system through a series of comprehensive evaluations, including longitudinal degradation analysis of the supervised binary classification under concept drift (AnoShift-style~\cite{anoshift}), temporally disjoint training (testing), and family-wise feature stability assessments. \system enables explanation-guided analysis of concept drift and combines long-term structural modeling with SHAP-based attributions~\cite{shap}, allowing researchers to trace how feature relevance shifts over time and better understand the underlying causes of model degradation.

In summary, the contributions of this paper are as follows:

\begin{smitemize}

    \item We present \system, a large-scale Android malware benchmark comprising over 1 million APKs across 1,380 unique families spanning for 12 years (2013 to 2025, excluding 2015), built on static features based on Drebin~\cite{arp2014drebin} features.


    \item We conduct longitudinal evaluations under structured temporal splits~\cite{anoshift}, analyze per-feature distribution shifts, and perform feature stability analysis of malware families~\cite{api_graph_dataset}.
    
    \item \system facilitates explanation-driven drift analysis through SHapley Additive exPlanations (SHAP)-based attributions~\cite{shap}, supporting investigations into how feature importance shifts as malware evolves.
    

\end{smitemize}

\paragraph{Contributions in the Appendices and Supplementary Materials.} We include additional analyses and supporting experiments in the appendices, including dataset statistics (Appendix~\ref{appendix:dataset-stats}), feature descriptions (Appendix~\ref{appendix:featdescription}), model architectures and evaluation results (Appendix~\ref{appendix:model-config}), practical considerations in dataset construction (Appendix~\ref{appendix:practical-issues}), handling of label noise (Appendix~\ref{appendix:label-noise}), temporal label drift (Appendix~\ref{appendix:label-drift}), dataset scalability (Appendix~\ref{appendix:scalability}), concept drift adaptation on LAMDA (Appendix~\ref{appendix:comparison-sota}), SHAP-based explanation drift for top 1000 features (Appendix~\ref{appendix:explanation-drift}), continual learning experiments (Appendix~\ref{appendix:cl-exps}), and computational resources used for LAMDA generation (Appendix~\ref{appendix:computation}).

\section{Related Work}

In this section, we discuss prior work and their limitations that motivate the creation of \system. Furthermore, we review prior work on continual learning (CL) to position \system as a timely benchmark for studying concept drift analysis and CL in malware analysis.

\textbf{Evolution of Malware Datasets and Benchmarks.} Early malware 
datasets such as Drebin~\cite{arp2014drebin} (Android) and EMBER~\cite{ember} (Windows) have played a pivotal role to study concept drift in malware analysis. More recent efforts—including SOREL-20M~\cite{sorel20m} and BODMAS~\cite{BODMAS} for Windows, and TESSERACT~\cite{tesseract}, API Graph~\cite{api_graph_dataset}, and AL-Chen~\cite{chen2023continuous} for Android attempt to address limitations in scale and recency. Nonetheless, these datasets suffer from one or more major limitations --- they are often outdated, contain either relatively few malware samples or families, or lack long-term temporal coverage necessary for studying the evolution of malware. For example, Drebin spans only 2010–2012 with 5,560 samples from 179 families; TESSERACT covers 2014–2016 with 12,735 samples; API Graph spans 2012–2018 with 32,089 samples from 1,120 families; and AL-Chen~\cite{chen2023continuous} includes 10,200 samples across 254 families from 2019–2021. Despite their temporal spread, these datasets are not explicitly structured to support longitudinal drift analysis or capture evolutionary patterns in malware behavior.

\textbf{Explainability and Semantic Features.} Explainability is critical for understanding how feature importance shifts under concept drift. While Drebin~\cite{arp2014drebin} and BODMAS~\cite{yang2021bodmas} introduced interpretable features and temporal structure, few studies have systematically used them to analyze drift. TRANSCENDENT~\cite{barbero2022transcendent} incorporates semantic reasoning for selective prediction, but longitudinal robustness of explanations remains underexplored due to limited dataset support. \system fills this gap by providing a temporally structured benchmark with interpretable features and SHapley Additive exPlanations (SHAP)-based explanations~\cite{shap}, enabling fine-grained, longitudinal analysis of model behavior and drift.

\textbf{Continual Learning for Malware Analysis.} Continual learning (CL) in malware analysis remains fairly underexplored. Rahman et al.~\cite{continual-learning-malware} is the first to explore CL to this domain, found that catastrophic forgetting (CF) occurs due to the diverse yet semantically limited nature of tabular malware features, with replay-based approaches showing better resilience. Chen et al.~\cite{chen2023continuous} used contrastive and active learning to detect drift but did not address CF. 
Recent work explores CL via diversity-aware and generative replay on modified EMBER and small-scale Android datasets~\cite{malcl, madar}.

\section{\system Creation}

In this section, we describe the construction process of the LAMDA. We have downloaded APKs from AndrooZoo repository~\cite{androzoo,androzooMetadata} and decompiled APKs to extract static Drebin~\cite{arp2014drebin} features and then transformed the features into binary vectors for downstream analysis.





\paragraph{Label Assignment and Collection Strategy.}

To construct a large-scale, temporally diverse dataset, we use metadata from AndroZoo~\cite{androzoo,androzooMetadata}, including APK hashes, VirusTotal (VT) results, and submission dates. For each year from 2013 to 2025 (excluding 2015, which lacks valid entries), we collect APKs and assign binary labels using the \texttt{vt\_detection} field. Following prior heuristics~\cite{arp2014drebin, tesseract}, we define: (i) \emph{Benign} for \texttt{vt\_detection} $=$ 0, (ii) \emph{Malware} for \texttt{vt\_detection} $\geq$ 4, and (iii) discard \emph{Uncertain} samples with scores in ${1, 3}$. The $\geq$ 4 threshold mitigates label noise by requiring stronger AV consensus~\cite{chen2016stormdroid}.

To reduce sampling bias in learning systems, we collected 50,000 malware and 50,000 benign samples per year, while preserving month-wise temporal distributions across both categories. Although prior work such as TESSERACT~\cite{tesseract, chen2023continuous} adopts a 90:10 benign-to-malware ratio, we attempt to maintain a balanced 50:50 ratio~\cite{ember}. This choice is motivated by the need to mitigate the risk of skewed learned representations (such as overfitting~\cite{shwartz2023simplifying}, disparity in learning~\cite{zhou2023combating}) that can arise from class imbalance. A balanced dataset helps ensure that the model learns meaningful distinctions between classes, captures a wider range of malware families, and is exposed to a broader spectrum of behaviors and evasive techniques. Such diversity not only enables longitudinal generalization studies but also increases the difficulty of the detection task, particularly for learning systems that must contend with rare, novel, or semantically similar malware families~\cite{ember}. Nonetheless, due to limited availability of malware samples in certain years such as 2017, 2023, 2024, and 2025, \system~still exhibits class imbalance, particularly in those periods.

Another practical challenge during data collection involved download and decompilation failures, requiring us to over-fetch APKs to meet target counts. To mitigate this, we included a 20\% overhead in the number of APK hashes per year. All APKs are retrieved via authenticated academic access to the AndroZoo repository~\footnote{\url{https://androzoo.uni.lu/access}} and stored in a consistent directory structure (\texttt{[year]/malware/}, \texttt{[year]/benign/}) to facilitate temporal slicing and cross-year analysis. Corrupted or undecompilable samples are excluded and logged for transparency. The final dataset comprises over one million APKs. A detailed year-wise breakdown is provided in Appendix~\ref{appendix:dataset-stats}.

\paragraph{Family Label Acquisition.}

\begin{figure}[!t]
\centering
\begin{minipage}{0.35\textwidth}
    \centering
    \includegraphics[width=0.8\linewidth]{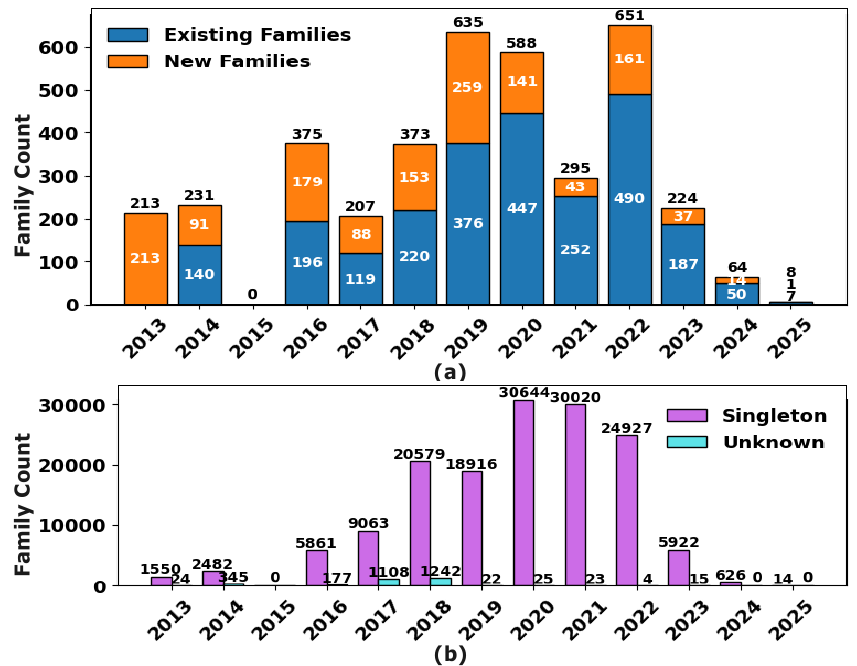}
    \caption{Temporal trends in malware family evolution.}
    \label{fig:malware_family}
\end{minipage}
\hfill 
\begin{minipage}{0.63\textwidth}
    \centering
    \begin{subfigure}{0.475\linewidth}
        \centering
        \includegraphics[width=\linewidth]{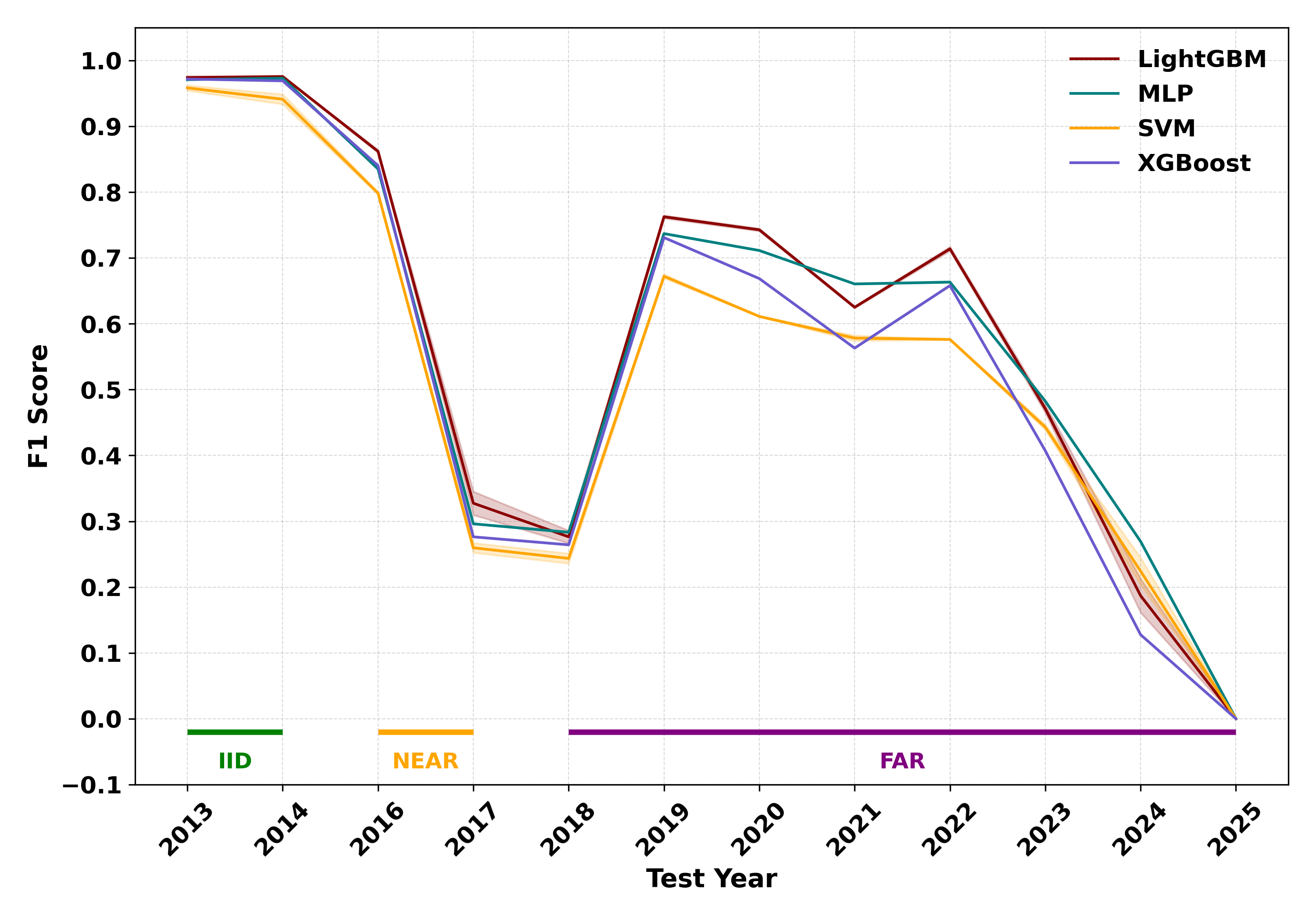}
        \subcaption{LAMDA.}
        \label{lamdaanoshift}
    \end{subfigure}
    \hfill
    \begin{subfigure}{0.475\linewidth}
        \centering
        \includegraphics[width=\linewidth]{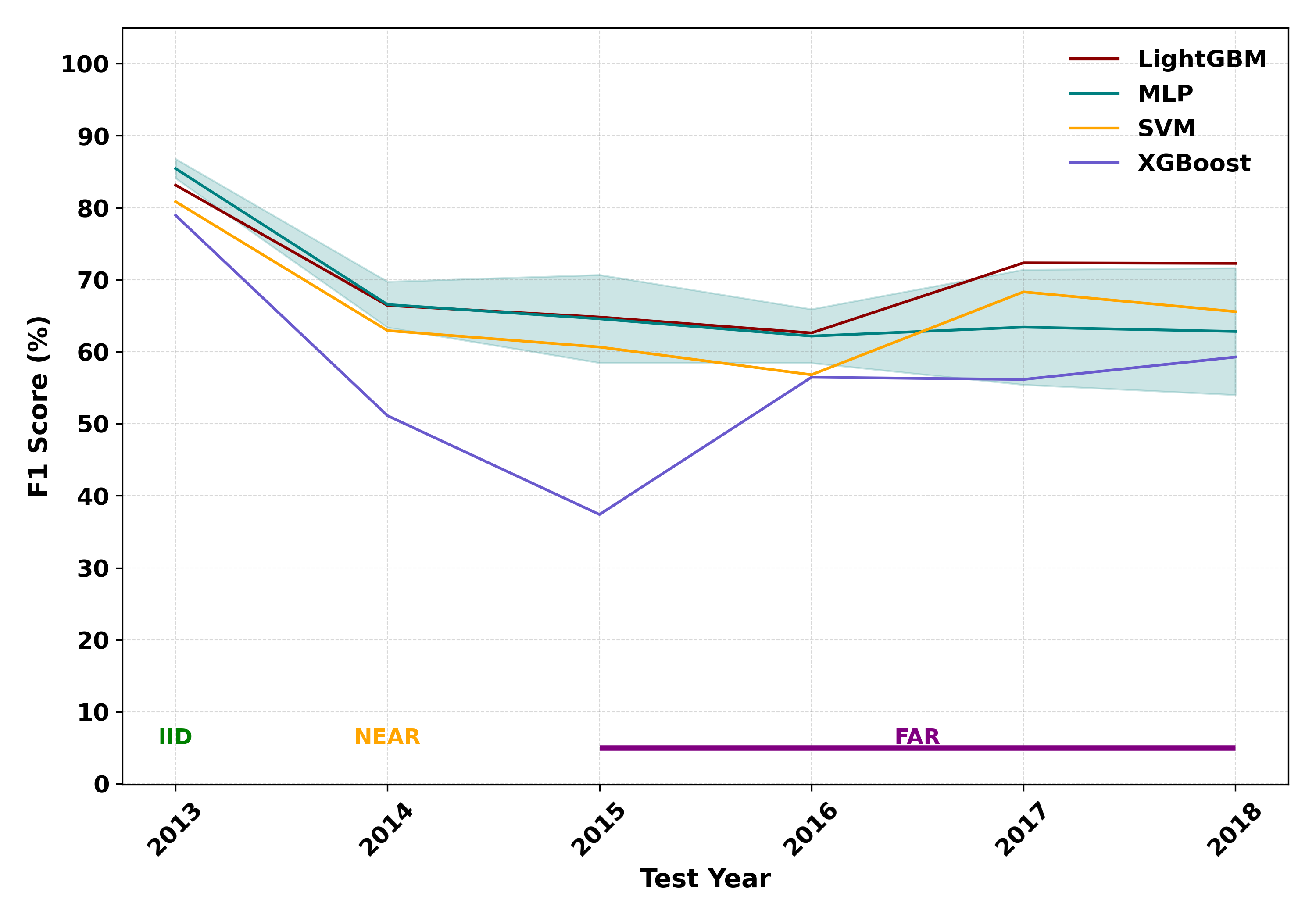}
        \subcaption{API Graph.}
        \label{apigraphanoshift}
    \end{subfigure}
    \caption{F1-score over time across different temporal splits.}
    \label{fig:anoshift_score}
\end{minipage}
\vspace{-0.3cm}
\end{figure}

To support more detailed analysis beyond binary classification, we assign family-level labels to all malware samples. These labels offer a finer-grained perspective on how malware behavior evolves over time, which is important for developing detection systems that can generalize to new threats. We use \texttt{AVClass2}~\cite{sebastian2016avclass}, a widely used tool that standardizes noisy and inconsistent labels from antivirus vendors into meaningful malware family names. The labeling process involves collecting VirusTotal~\cite{virustotal} reports for each sample, converting them to the required format, running AVClass2, and post-processing the output to retain links to individual files via SHA256 hashes. Figures~\ref{fig:malware_family}(a) and~\ref{fig:malware_family}(b) illustrate the yearly distribution of recurring versus newly observed families and the count of singleton families—those that appear only once—respectively. These trends highlight the inherent difficulty of LAMDA. A growing number of novel and singleton families suggests that many malware instances exhibit unique or rare behaviors, limiting the effectiveness of detection systems. Family labels will facilitate research into more complex tasks such as multi-class classification and the study of temporal trends across malware families. In LAMDA, the labels are provided as an optional extension, allowing researchers to conduct either standard binary classification or more advanced analyses, depending on their goals.

\paragraph{Decompilation and Static Feature Extraction.}

Each APK is statically decompiled using \texttt{apktool}~\cite{apktool}, producing a disassembled \texttt{smali} representation and the original \texttt{AndroidManifest.xml}. We parse these artifacts to extract a diverse set of static features commonly used in Android malware detection~\cite{arp2014drebin}. Specifically, the \texttt{AndroidManifest.xml} file is analyzed to obtain the list of requested permissions (e.g., \texttt{ACCESS\_FINE\_LOCATION}, \texttt{INTERNET}), declared activities and services, broadcast receivers, required hardware components, and intent filters~\cite{arp2014drebin}. Meanwhile, the disassembled \texttt{smali} code is scanned to identify invocations of restricted APIs (e.g., \texttt{NotificationManager.notify}), suspicious API usages (e.g., \texttt{getSystemService}), and embedded hardcoded IPs/URLs (e.g., \texttt{e.crashlytics.com}). The extracted Drebin~\cite{arp2014drebin} feature sets comprises several static categories derived from Android APKs~\cite{arp2014drebin, chen2016stormdroid}. A detailed list of features is provided in Appendix~\ref{appendix:featdescription}.

\paragraph{Vectorization and Temporal Feature Alignment.}

After decompiling each APK, we extract static features into a \texttt{.data} file (see Appendix~\ref{appendix:featdescription}). Each year’s data is split into 80\% training and 20\% testing sets using stratified sampling to preserve class balance. From the training set, we construct a global vocabulary by taking the union of unique tokens across all samples, yielding 9,690,482 ($\approx 9.69$ million) raw features~\cite{madar, cade}. Each APK is then represented as a high-dimensional binary vector using a bag-of-tokens model, where each token corresponds to a binary feature indicating its presence or absence in the sample~\cite{arp2014drebin, chen2016stormdroid}.

To reduce dimensionality and ensure computational feasibility, we apply \texttt{VarianceThreshold} from \texttt{scikit-learn} to eliminate low-variance features. For all experiments, we use the \texttt{Baseline} variant, which applies a threshold of 0.001~\cite{cade,continual-learning-malware,malcl,madar}, resulting in $4,561$ final features. This compact and consistent representation supports a range of downstream tasks, including supervised learning, drift analysis, and continual learning~\cite{madar, malcl}.



The dataset is initially created in a sparse matrix format, storing binary feature vectors and metadata as compressed \texttt{.npz} files to optimize for storage efficiency and computational performance. These NPZ files are organized by year and stratified into {\em training} and {\em test} splits. To enhance accessibility and integration with machine learning pipelines, particularly for the Hugging Face~\footnote{\url{https://huggingface.co/docs/hub/datasets-adding\#file-formats}} ecosystem, we have converted the data into tabular \texttt{.parquet} files.

The final dataset is also organized into year-based folders, each containing stratified {\em training} and {\em test} splits. For each split, we provide two \texttt{.parquet} files within the corresponding year-specific folder (e.g., \texttt{YYYY/X\_train.parquet} and \texttt{YYYY/X\_test.parquet}), where \texttt{YYYY} denotes the year. Each file is a tabular dataset in which the first five columns represent metadata fields: \texttt{hash}, \texttt{label}, \texttt{family}, \texttt{vt\_count}, and \texttt{year\_month}. The remaining columns contain binary features, with the exact number depending on the applied variance threshold. A detailed breakdown of feature dimensions under varying \texttt{VarianceThreshold} settings is provided in Appendix~\ref{appendix:featdescription}. For scalability of LAMDA, we also published global features, variance threshold objects and selected features after applying \texttt{VarianceThreshold}.

\section{Concept Drift Analysis}

We first examine performance degradation of supervised models across temporally distant splits~\cite{anoshift} (Section~\ref{sec:cdalearning}), followed by distributional shifts using Jeffreys divergence and t-SNE visualizations (Sections~\ref{sec:cdalearning} and~\ref{sec:visualcda}). We then assess feature stability in top malware families (Section~\ref{sec:featstability}), drift in common families (Section~\ref{sec:drifttest}), and conclude with SHAP-based explainability analysis (Section~\ref{sec:shap}). Each analysis includes a comparison with API Graph~\cite{api_graph_dataset}, which spans seven years, longer than prior datasets, and captures long-term API call evolution with greater malware family diversity.

\subsection{Concept Drift Analysis with Supervised Learning}
\label{sec:cdalearning}


\begin{table}[!t]
\vspace{-0.3cm}
\centering
\resizebox{\textwidth}{!}{%
\begin{tabular}{llccccc|ccccc}
\toprule
\multirow{2}{*}{\textbf{Split}} & \multirow{2}{*}{\textbf{Model}} 
& \multicolumn{5}{c|}{\textbf{LAMDA}} 
& \multicolumn{5}{c}{\textbf{API Graph}} \\
\cmidrule(lr){3-7} \cmidrule(lr){8-12}
 &  & F1 & ROC-AUC & PR-AUC & FNR & FPR 
    & F1 & ROC-AUC & PR-AUC & FNR & FPR \\
\midrule

\multirow{4}{*}{IID}
  & LightGBM & \textbf{97.49 $\pm$ 0.17} & \textbf{99.55 $\pm$ 0.03} & \textbf{99.50 $\pm$ 0.11} & \textbf{1.74 $\pm$ 0.34} & \textbf{2.69 $\pm$ 0.48}
              & \textbf{85.95 $\pm$ 0.00} & \textbf{98.91 $\pm$ 0.00} & \textbf{95.20 $\pm$ 0.00} & 22.39 $\pm$ 0.00 & \textbf{0.33 $\pm$ 0.00} \\
  & MLP      & 97.21 $\pm$ 0.12 & 99.48 $\pm$ 0.04 & 99.38 $\pm$ 0.20 & 2.50 $\pm$ 0.06 & 2.58 $\pm$ 0.85 
              & 85.79 $\pm$ 0.00 & 96.37 $\pm$ 0.00 & 88.49 $\pm$ 0.00 & 20.31 $\pm$ 0.00 & 0.67 $\pm$ 0.00 \\
  & SVM      & 94.98 $\pm$ 1.07 & 98.89 $\pm$ 0.28 & 98.75 $\pm$ 0.46 & 4.82 $\pm$ 0.76 & 4.09 $\pm$ 0.55 
              & 82.00 $\pm$ 0.00 & 97.33 $\pm$ 0.00 & 90.94 $\pm$ 0.00 & 26.74 $\pm$ 0.00 & 0.60 $\pm$ 0.00 \\
  & XGBoost  & 97.05 $\pm$ 0.14 & 99.15 $\pm$ 0.16 & 97.68 $\pm$ 1.16 & 2.20 $\pm$ 0.43 & 2.96 $\pm$ 0.01 
              & 80.33 $\pm$ 0.00 & 96.05 $\pm$ 0.00 & 89.74 $\pm$ 0.00 & 28.35 $\pm$ 0.00 & 0.75 $\pm$ 0.00 \\
\midrule

\multirow{4}{*}{NEAR}
  & LightGBM & \textbf{59.48 $\pm$ 28.20} & 74.05 $\pm$ 23.76 & \textbf{70.18 $\pm$ 27.10} & \textbf{50.51 $\pm$ 30.82} & \textbf{1.85 $\pm$ 0.95} 
              & 66.77 $\pm$ 0.00 & \textbf{95.94 $\pm$ 0.00} & \textbf{83.01 $\pm$ 0.00} & 47.68 $\pm$ 0.00 & \textbf{0.48 $\pm$ 0.00} \\
  & MLP      & 56.57 $\pm$ 28.41 & 82.71 $\pm$ 11.19 & 67.94 $\pm$ 24.59 & 51.95 $\pm$ 30.42 & 3.98 $\pm$ 1.19 
              & \textbf{68.72 $\pm$ 0.00} & 86.79 $\pm$ 0.00 & 68.31 $\pm$ 0.00 & 38.80 $\pm$ 0.00 & 1.87 $\pm$ 0.00 \\
  & SVM      & 52.91 $\pm$ 28.40 & 75.18 $\pm$ 17.98 & 62.53 $\pm$ 29.82 & 55.62 $\pm$ 28.88 & 4.71 $\pm$ 0.97 
              & 63.06 $\pm$ 0.00 & 89.25 $\pm$ 0.00 & 70.15 $\pm$ 0.00 & 45.48 $\pm$ 0.00 & 2.03 $\pm$ 0.00 \\
  & XGBoost  & 55.84 $\pm$ 29.73 & 77.75 $\pm$ 16.85 & 68.14 $\pm$ 26.55 & 53.84 $\pm$ 30.94 & 2.14 $\pm$ 0.86 
              & 51.47 $\pm$ 0.00 & 86.90 $\pm$ 0.00 & 65.69 $\pm$ 0.00 & 60.41 $\pm$ 0.00 & 1.57 $\pm$ 0.00 \\
\midrule

\multirow{4}{*}{FAR}
  & LightGBM & 47.24 $\pm$ 27.33 & 78.04 $\pm$ 20.83 & 63.45 $\pm$ 35.80 & 64.10 $\pm$ 22.97 & 1.30 $\pm$ 0.95 
              & \textbf{68.20 $\pm$ 4.63} & \textbf{95.68 $\pm$ 0.81} & \textbf{81.69 $\pm$ 1.59} & 45.04 $\pm$ 5.72 & \textbf{0.61 $\pm$ 0.07} \\
  & MLP      & \textbf{47.59 $\pm$ 25.30} & \textbf{84.04 $\pm$ 11.23} & \textbf{66.16 $\pm$ 34.34} & \textbf{64.40 $\pm$ 20.63} & \textbf{1.14 $\pm$ 0.71} 
              & 63.92 $\pm$ 5.39 & 87.10 $\pm$ 1.95 & 64.22 $\pm$ 4.19 & 46.45 $\pm$ 6.10 & 1.40 $\pm$ 0.15 \\
  & SVM      & 41.86 $\pm$ 22.55 & 79.07 $\pm$ 15.06 & 62.27 $\pm$ 34.09 & 69.93 $\pm$ 16.85 & 1.27 $\pm$ 0.76 
              & 66.18 $\pm$ 6.21 & 93.44 $\pm$ 0.52 & 75.46 $\pm$ 2.58 & 44.26 $\pm$ 8.35 & 1.23 $\pm$ 0.14 \\
  & XGBoost  & 42.75 $\pm$ 25.86 & 76.85 $\pm$ 16.49 & 60.33 $\pm$ 35.88 & 68.11 $\pm$ 20.43 & 1.69 $\pm$ 0.57 
              & 54.88 $\pm$ 9.26 & 83.44 $\pm$ 4.79 & 65.03 $\pm$ 6.67 & 56.68 $\pm$ 8.92 & 1.41 $\pm$ 0.35 \\
\bottomrule
\end{tabular}
}
\caption{Comparison of performances on LAMDA and API Graph across three temporal splits.
}
\label{tab:comparison_lamda_apigraph}
\vspace{-0.3cm}
\end{table}

\begin{figure}[!t]
\centering
\begin{minipage}{0.49\textwidth}
    \centering
    \begin{subfigure}{0.46\linewidth}
        \centering
        \includegraphics[width=\linewidth]{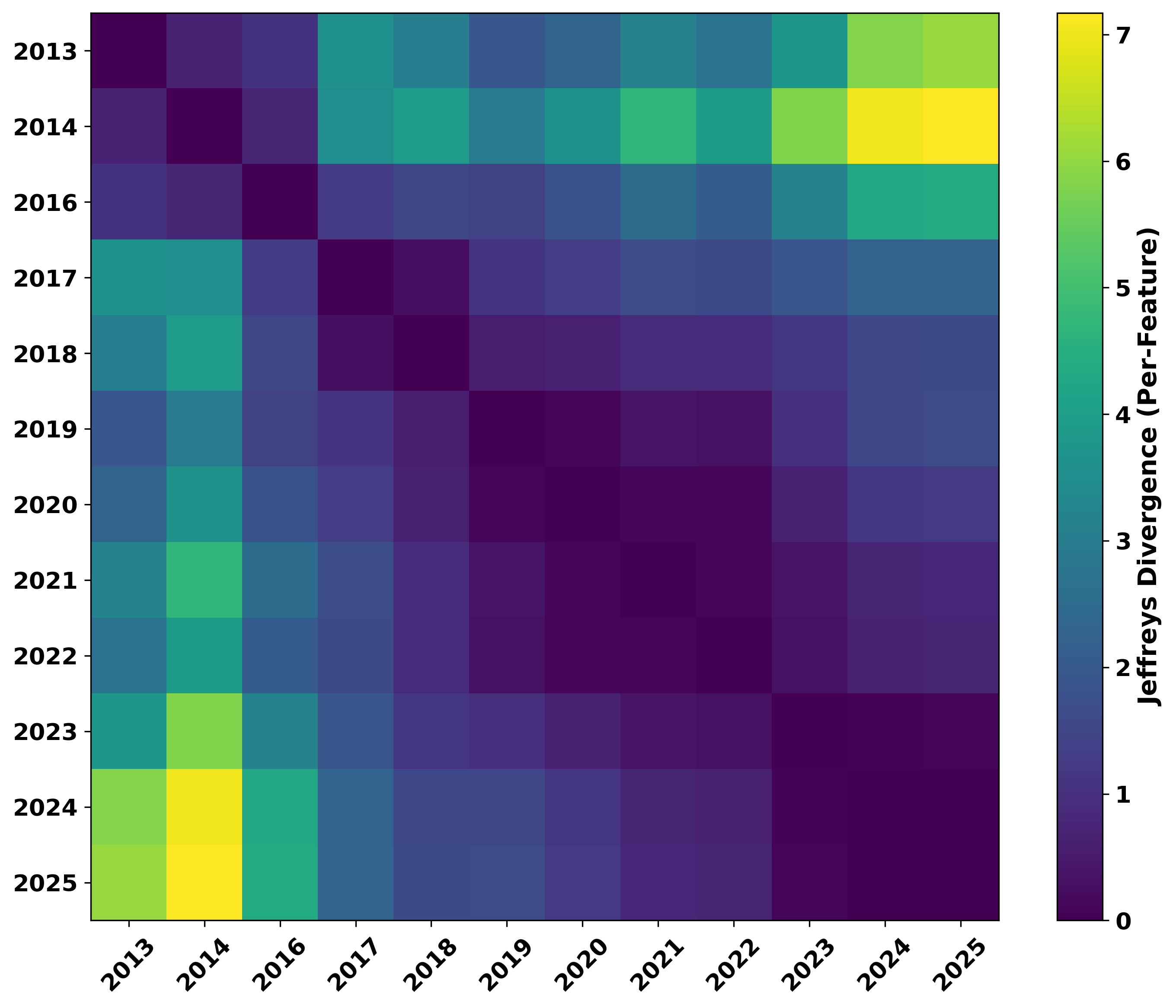}
        \subcaption{LAMDA}
    \end{subfigure}
    \hfill
    \begin{subfigure}{0.48\linewidth}
        \centering
        \includegraphics[width=\linewidth]{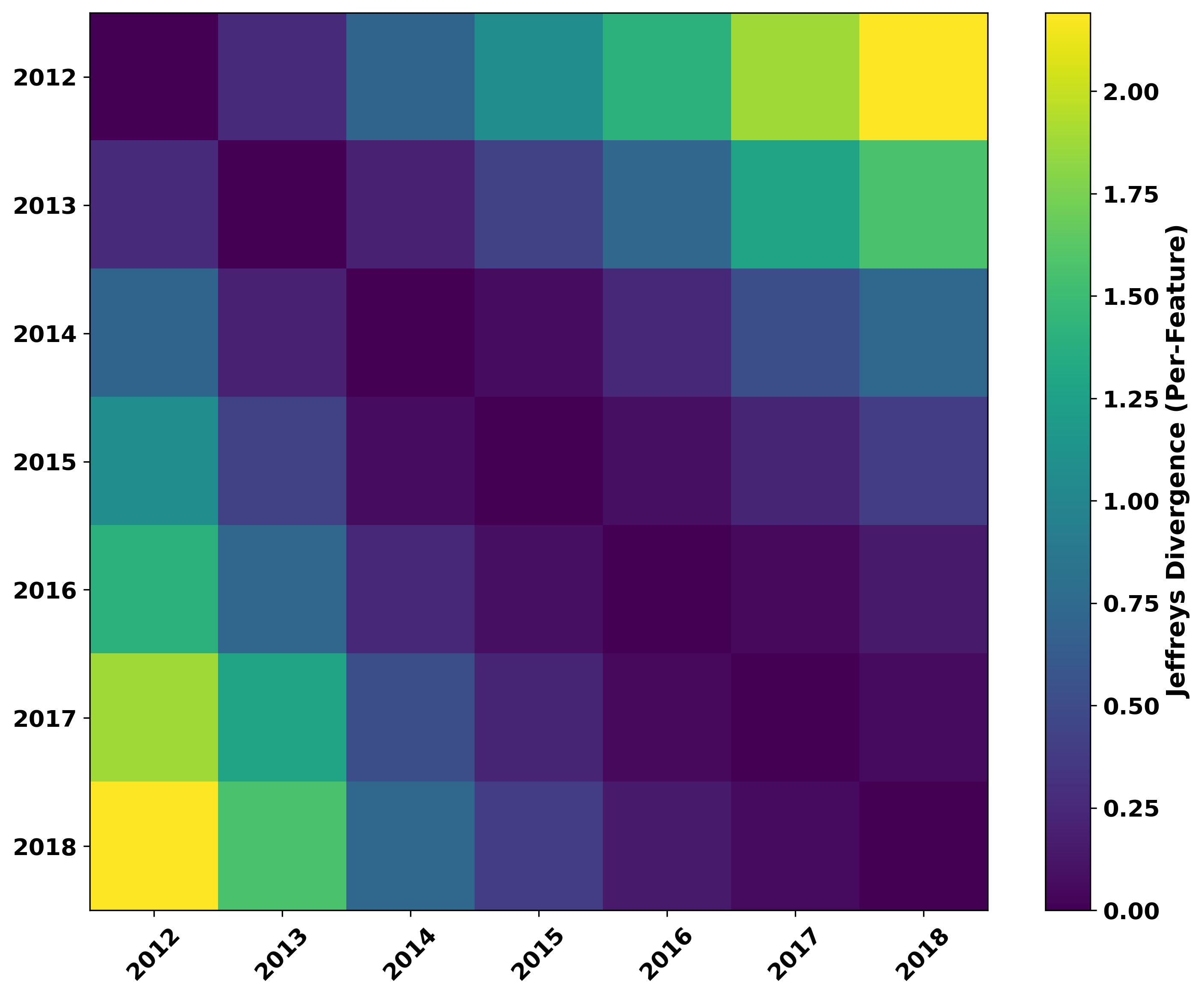}
        \subcaption{API Graph}
    \end{subfigure}
    \caption{Jeffreys divergence heatmaps across years for LAMDA and API Graph datasets.}
    \label{fig:jeffreys_figures}
\end{minipage}
\hfill
\begin{minipage}{0.49\textwidth}
    \centering
    \includegraphics[width=0.9\linewidth]{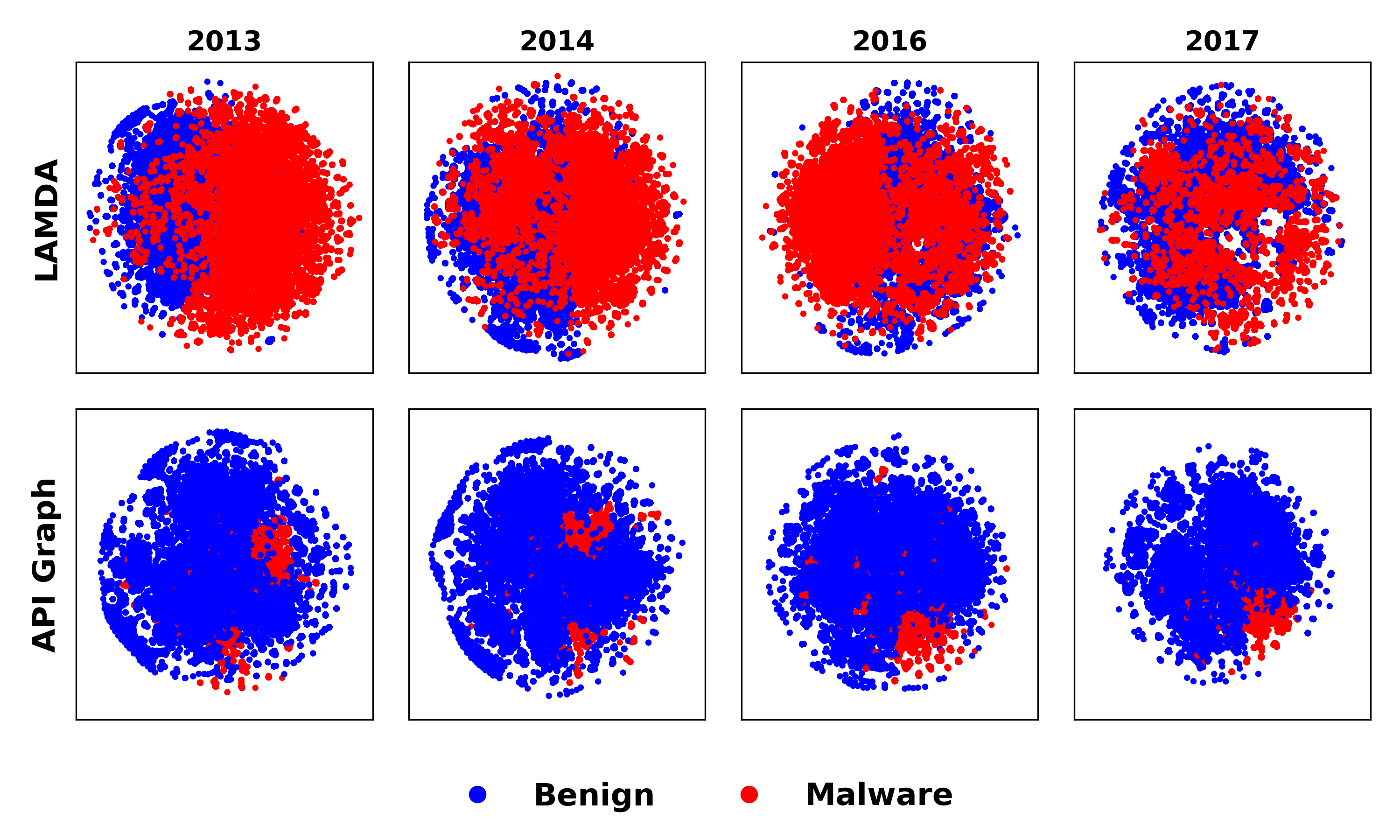}
    \vspace{-0.2cm}
    \caption{t-SNE projections showing feature space evolution for LAMDA and API Graph.}
    \label{fig:tsne_projection}
\end{minipage}
\vspace{-0.4cm}
\end{figure}

\paragraph{Experimental Setting.} To evaluate the robustness of malware detectors under temporal distribution shifts, we perform supervised learning experiments using four widely adapted detector models from the malware research --- Linear SVM, LightGBM, MLP, and XGBoost~\cite{arp2014drebin, ember, BODMAS, chen2023continuous}. 
Detailed model configurations are provided in Appendix~\ref{appendix:model-config}.

Inspired by the AnoShift benchmark~\cite{anoshift}, we divide LAMDA into three temporally separated regions: TRAIN (i.e., initial training set) with Independent and Identically Distributed (IID), NEAR, and FAR. We construct the TRAIN+IID set using samples from 2013 and 2014. Models are trained on samples from all months of these two years, excluding the final month of each, which is reserved for IID evaluation. This held-out portion serves as an in-distribution test set, allowing us to measure baseline performance on temporally adjacent, yet unseen data. To examine generalization under increasing drift, we define two additional test regions: NEAR (2016–2017) and FAR (2018–2025), both strictly used for evaluation. These splits enable a principled analysis of how detection performance degrades as the temporal gap from the training data widens, thereby reflecting progressively stronger distributional shifts.

For comparison, we evaluate the same malware detectors on the API Graph dataset~\cite{api_graph_dataset} using a similar AnoShift-style~\cite{anoshift} split: training on 2012, IID on 2013, NEAR on 2014, and FAR on 2015--2018. 
All experiments are repeated five times with different random seeds. For each split: IID (2013--2014), NEAR (2016--2017), and FAR (2018--2025). We report the results on different evaluation metrics in Table~\ref{tab:comparison_lamda_apigraph} as \textit{mean$\pm$std}, averaged over all runs and all years within each split. In Figure~\ref{fig:anoshift_score} results are averaged across all runs, but shown separately for each year.



\paragraph{Results.} 
Table~\ref{tab:comparison_lamda_apigraph} summarizes the performance of malware detectors on both LAMDA and API Graph under the IID, NEAR, and FAR evaluation splits. All detectors perform strongly under IID conditions, but their effectiveness declines sharply as the temporal gap from training increases. For instance, LightGBM's F1-score on LAMDA drops from 97.49\% (IID) to 59.48\% (NEAR) and 47.24\% (FAR), alongside a significant rise in the false negative rate, from 1.47\% to 50.51\% and 64.10\%, respectively,—demonstrating increased difficulty. In contrast, the false positive rate (FPR) remains low and stable, likely due to the more consistent behavior of benign apps over time. Figure~\ref{fig:anoshift_score}(a) further visualizes this trend, showing how F1-scores decline over time. Notably, we observe a sharp drop in performance between 2016 and 2017, indicating a significant distributional shift. A similar decline is evident from 2023 to 2024. In contrast, F1-scores increase from 2018 to the 2019--2022 period, suggesting that these intermediate years exhibit less drift relative to 2017 and 2018. 

In API Graph, LightGBM’s F1-score drops from 85.95\% (IID) to 66.77\% (NEAR), but stabilizes at 68.20\% on FAR. The F1-scores over the years in Figure~\ref{fig:anoshift_score}(b) indicate a smaller degree of temporal drift, with only modest changes in performance between years. Compared to the API graph, LAMDA shows a higher standard deviation in both NEAR and FAR, suggesting more pronounced and variable distributional shifts. This supports our claim that LAMDA introduces stronger concept drift, making it a more challenging and realistic benchmark for evaluating long-term malware detection.

\subsection{Visual Analysis of Concept Drift}
\label{sec:visualcda}

\paragraph{Visualization Setting.}

To better understand how malware and benign class distributions evolve over time, we employ two complementary visualization techniques: Jeffreys divergence heatmaps and t-SNE projections. Jeffreys divergence~\cite{jeffreys1946invariant, anoshift} is a symmetric information-theoretic measure that quantifies how the distribution of individual static features shifts across years. We compute this metric pairwise between all yearly combinations in both LAMDA (2013--2025, excluding 2015) and API Graph (2012--2018), producing yearly heatmaps that capture the extent and direction of temporal drift~\cite{anoshift}. 
Additionally, we use t-SNE~\cite{van2008visualizing}, a non-linear dimensionality reduction technique, to project high-dimensional feature vectors into 2D space. To ensure a fair comparison, we selected four common years available in both datasets -- 2013, 2014, 2016, and 2017. This setting has been widely adopted in prior work on malware drift and structure visualization~\cite{tesseract, xu2019droidevolver}. Full year-wise t-SNE projections for both datasets are provided in Appendix~\ref{appendix:featdescription}.

\paragraph{Analysis.}

Figure~\ref{fig:jeffreys_figures} presents the Jeffreys divergence heatmaps for both datasets. In both LAMDA and API Graph, we observe increasing divergence values as the gap between years widens, confirming the presence of non-trivial concept drift. However, LAMDA exhibits a broader range of divergence, particularly from 2022 to 2025, indicating substantial changes in the distribution of static features. These changes likely arise from evolving APIs, development practices, and new malware behaviors. In contrast, API Graph shows relatively stable patterns, with limited divergence in its final years. Figure~\ref{fig:tsne_projection} provides t-SNE projections for the selected years. While t-SNE visualizations suggest that malware samples in LAMDA appear more scattered in later years (2016--2017), this may reflect increasing structural diversity or sparsity in feature space. However, as t-SNE distorts global distances, we corroborate these patterns with quantitative measures like Jeffreys divergence to assess real distributional shifts. On the other hand, API Graph maintains tightly clustered and relatively static distributions throughout, indicating limited structural evolution. These visual trends reinforce LAMDA's value as a temporally rich benchmark for studying real-world concept drift.

\begin{figure}[!t]
    \centering
    \includegraphics[width=0.9\textwidth]{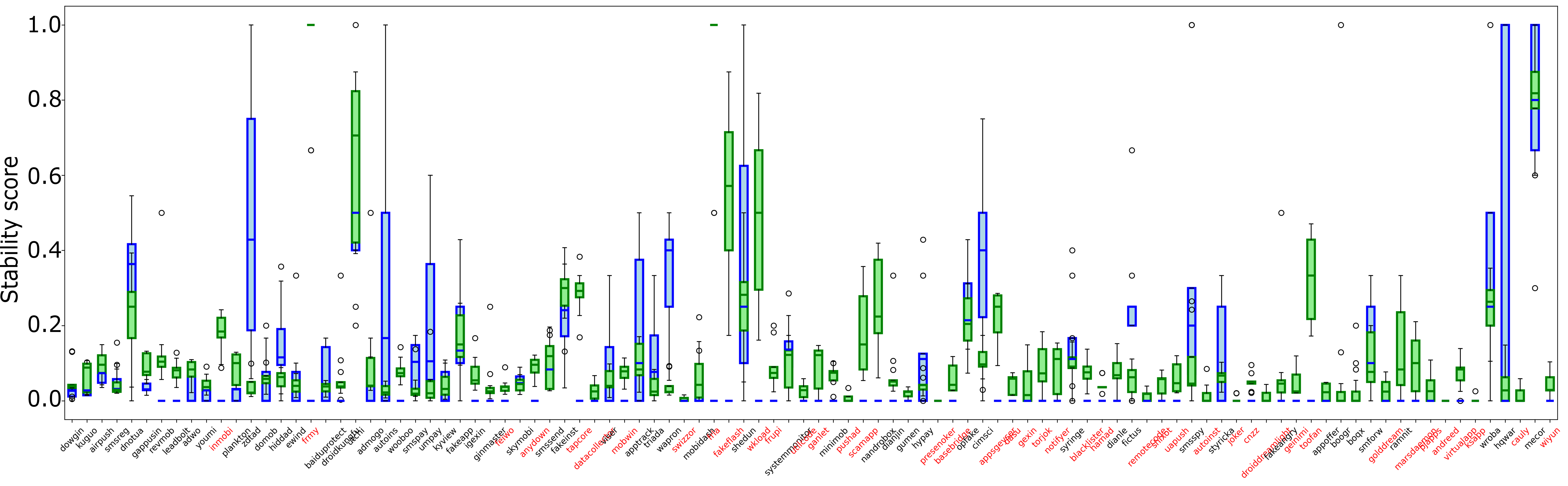}
    \vspace{-0.3cm}
    \caption{The distribution of feature stability scores for top 100 malware families. 58 families are common in both LAMDA (\textcolor{ForestGreen}{\bf green}) and API Graph (\textcolor{blue}{\bf blue}) datasets, and families marked as \textcolor{red}{red labels} along $x$-axis available in LAMDA with minimum family size criteria.}
    \label{fig:stability-scores-lamda-apigraph}
\vspace{-0.3cm}
\end{figure}

\begin{figure}[!t]
\centering
\includegraphics[width=\textwidth]{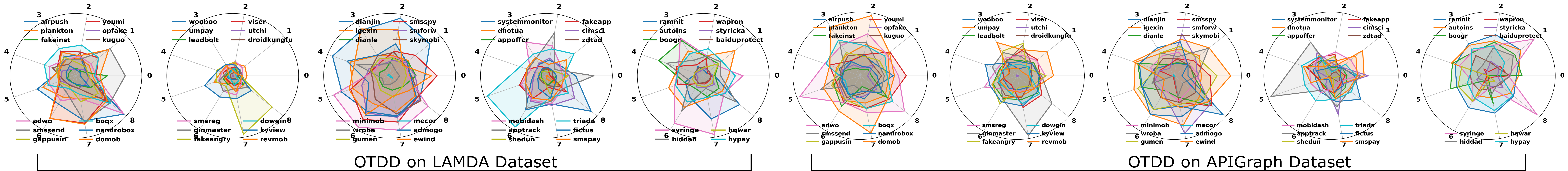}
\caption{Optimal Transport Distance of 60 common families. Each of the plots shows the area of nine OTDD scores of 10 groups of 10 families in LAMDA (left) and API Graph (right).}
\label{fig:otd-analysis}
\vspace{-0.3cm}
\end{figure}

\begin{figure}[!t]
\centering
\begin{minipage}[t]{0.48\textwidth}
    \centering
    \begin{subfigure}[t]{0.48\textwidth}
        \includegraphics[width=\linewidth]{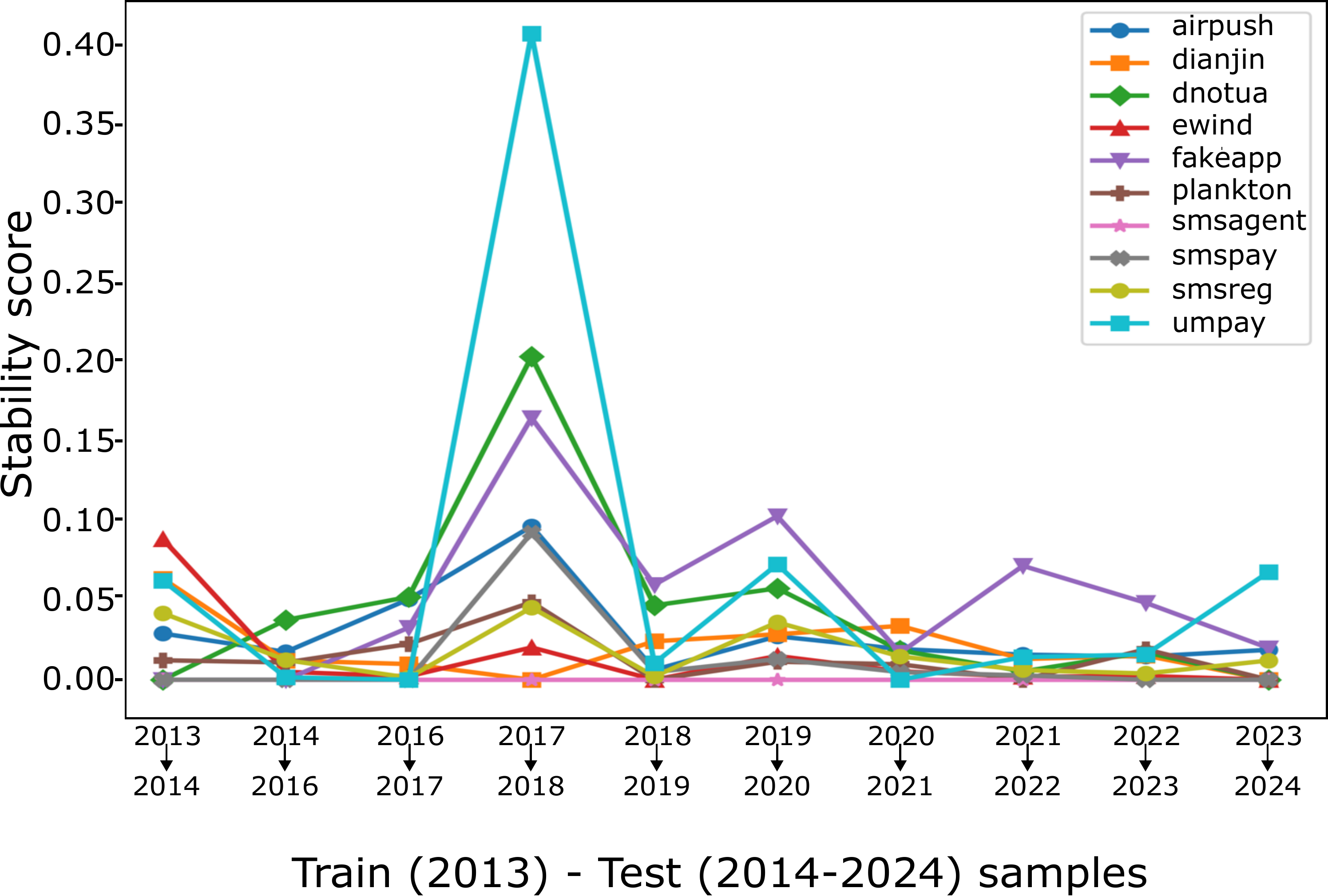}
        \caption{Stability scores.}
        \label{fig:stability-scores}
    \end{subfigure}
    \hfill
    \begin{subfigure}[t]{0.48\textwidth}
        \includegraphics[width=\linewidth]{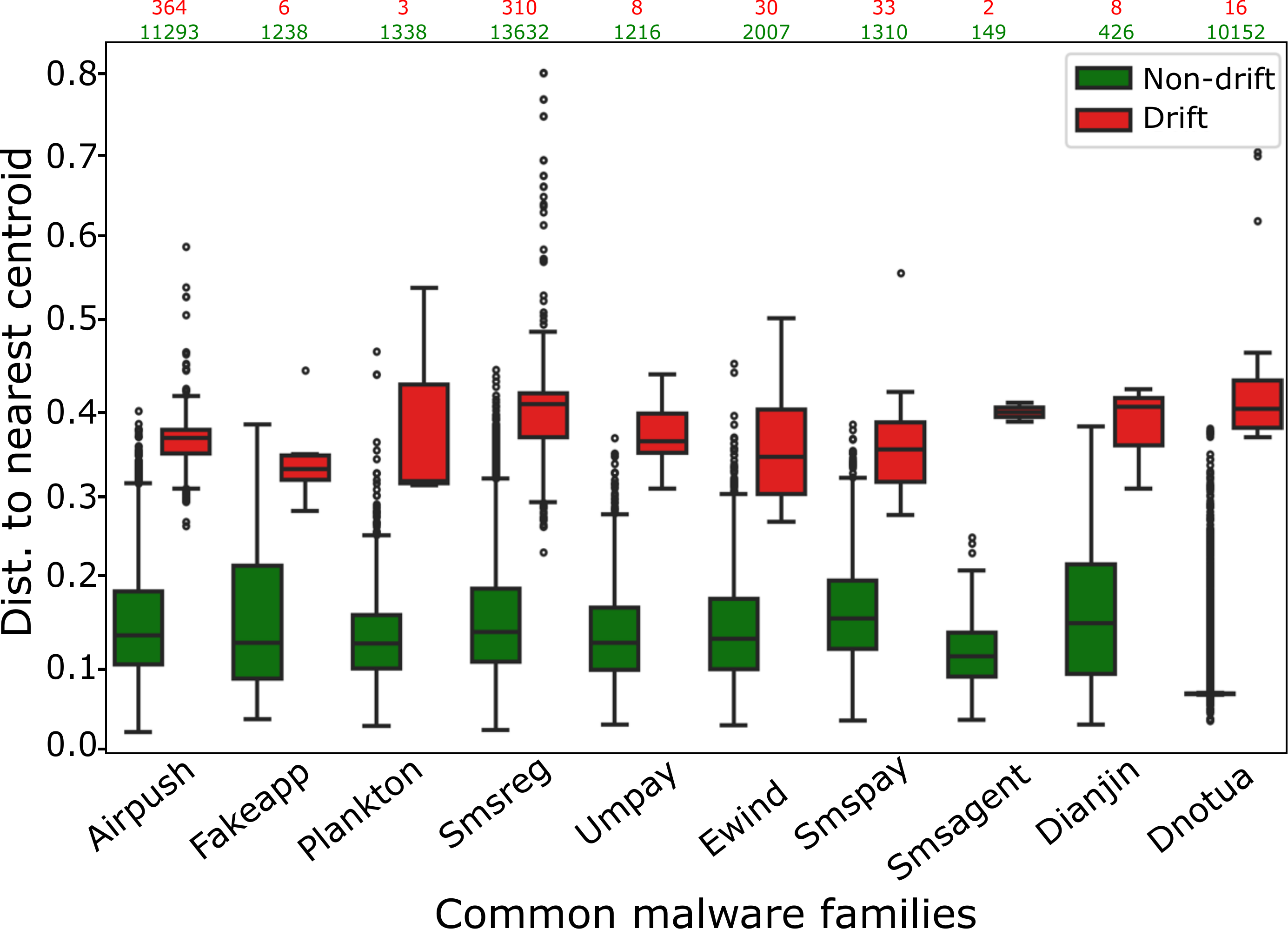}
        \caption{Distribution distances.}
        \label{fig:cade-distances}
    \end{subfigure}
    \caption{Stability and distribution analysis on malware families.}
    \label{fig:pair-stability}
\end{minipage}
\hfill
\begin{minipage}[t]{0.48\textwidth}
    \centering
    \begin{subfigure}[t]{0.48\textwidth}
        \includegraphics[width=\linewidth]{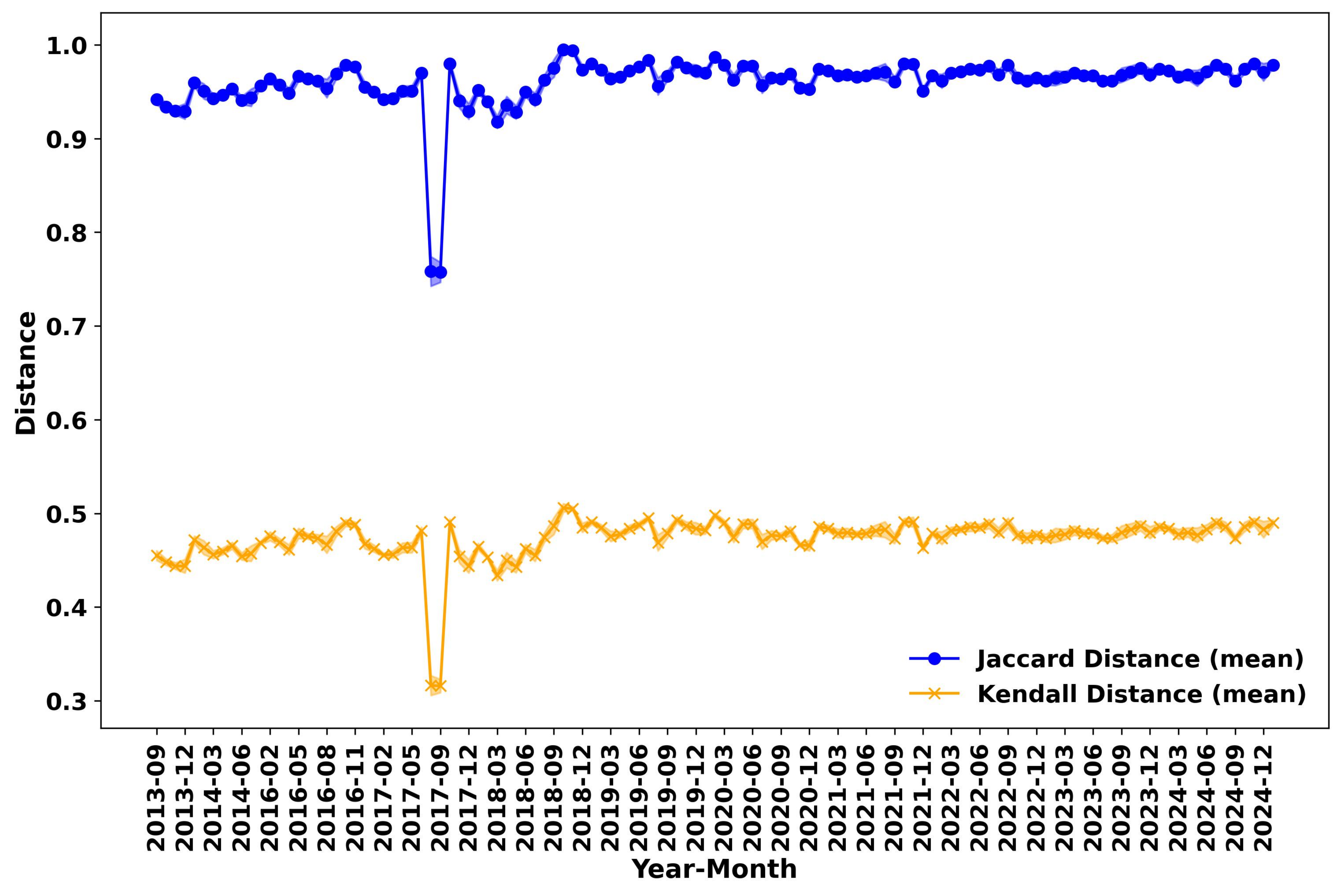}
        \caption{LAMDA.}
        \label{fig:lamda}
    \end{subfigure}
    \hfill
    \begin{subfigure}[t]{0.48\textwidth}
        \includegraphics[width=\linewidth]{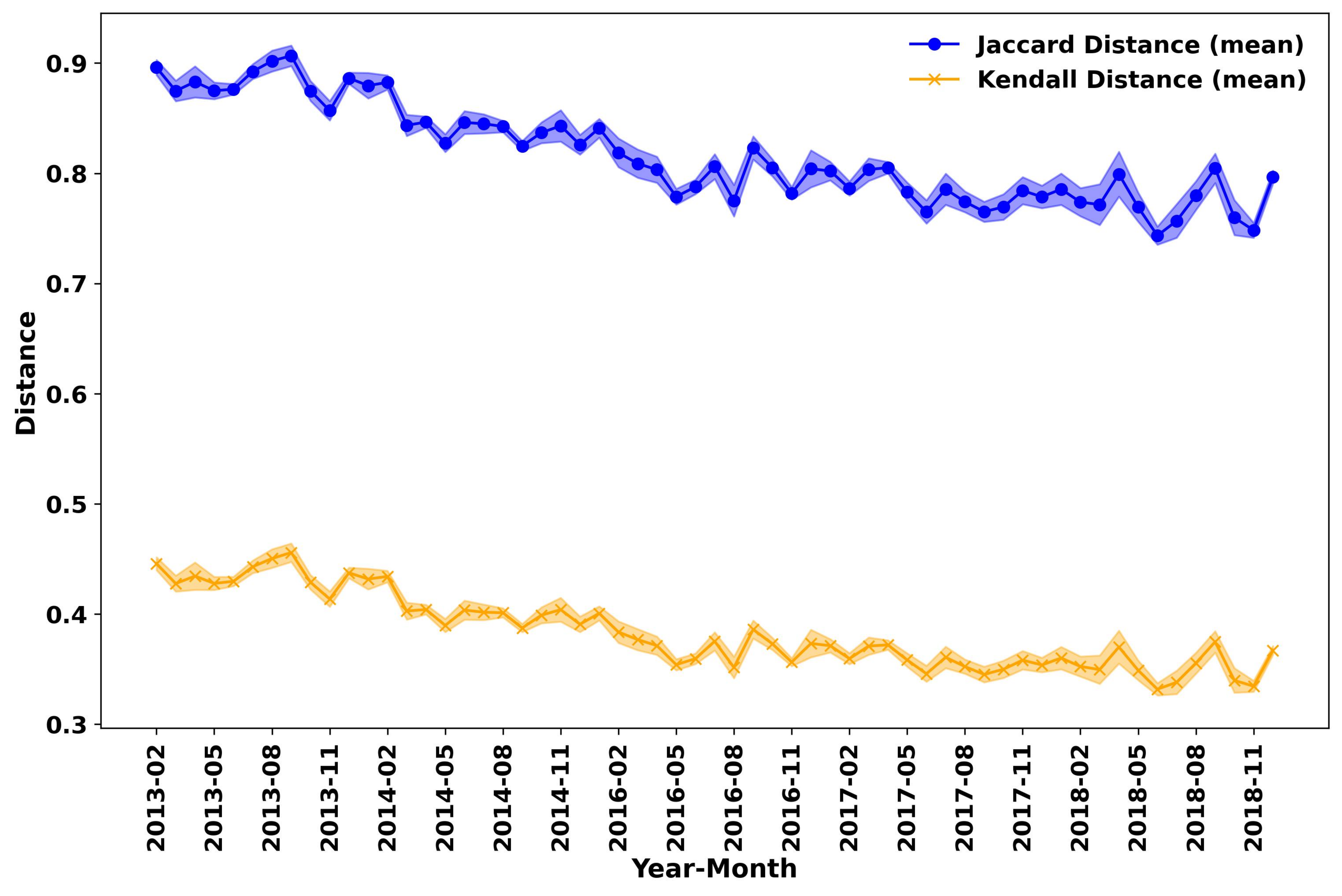}
        \caption{API Graph.}
        \label{fig:apigraph}
    \end{subfigure}
    \caption{SHAP-based explanation drift on LAMDA and API Graph datasets.}
    \label{fig:pair-drift}
\end{minipage}
\vspace{-0.4cm}
\end{figure}

\subsection{Feature Space Stability Analysis on Top Malware Families}
\label{sec:featstability}

\paragraph{Analysis Setting.}

Understanding the temporal consistency of malware families is essential for developing robust detection systems. We evaluate this using two complementary metrics --- stability scores and Optimal Transport Dataset Distance (OTDD)~\cite{alvarez2020geometric}, following prior work~\cite{api_graph_dataset, civitarese2022anoshift}. The analysis is performed in the original feature space, focusing on the top 100 malware families with the highest number of samples. Samples within each family are temporally ordered from 2013 to 2025; however, not all families have data for every year. Inspired by~\cite{api_graph_dataset}, we partition each family’s samples into ten equal subsets, each representing 10\% of the total. For API Graph, we identified 58 families with at least 10 samples within the year range 2013 and 2018, meeting the requirements for this subdivision. We then compute the stability score using the Jaccard similarity metric, as in~\cite{api_graph_dataset}, across the ten subsets for both LAMDA and API Graph.


\paragraph{Stability Scores Analysis.} 
Figure~\ref{fig:stability-scores-lamda-apigraph} shows the distribution of consecutive pairwise stability scores across ten groups for each of the top 100 malware families. The number of samples per family varies considerably, ranging from 186 to 32,475, with a mean of 1,984 and a median of 535. The green box plots correspond to LAMDA, while the blue box plots represent API Graph. Both datasets capture the temporal evolution of malware families, as reflected in the spread and median of stability scores. Broader spreads and lower medians in both datasets indicate greater behavioral variability over time. Notably, LAMDA includes more families and reflects broader evolutionary patterns than API Graph. These differences suggest that detection models trained on LAMDA may offer improved insight into concept drift, benefiting from greater sample diversity and family coverage.


\paragraph{Optimal Transport Dataset Distance (OTDD) Analysis.}

Figure~\ref{fig:otd-analysis} illustrates temporal distributional shifts using Optimal Transport Dataset Distance (OTDD)~\cite{alvarez2020geometric}, a geometric method for quantifying differences between probability distributions. To assess intra-family drift, we partition each malware family in the LAMDA and API Graph datasets into ten chronological subsets and compute OTDD between consecutive pairs. The results are visualized via radar plots, where each axis represents a subset transition. Compact, regular shapes indicate temporal stability, while larger or irregular shapes signal drift. Comparing the two datasets, the LAMDA radar plots show both regular and irregular patterns indicating temporal shifts of malware families that causes concept drift. Similar behavior is also observed in the API Graph dataset for the same families.

\subsection{Temporal Drift Analysis on Common Malware Families}
\label{sec:drifttest}
\paragraph{Analysis Setting.} We assess the drifting behavior over the years for the common families present from 2013 to 2025. We observe that {\em only 10} families appear consistently each year, except for 2025. Subsequently, we compute the year-wise stability score for the original feature set within each of these 10 common families. Additionally, we measure the distribution distances based on the CADE~\cite{yang2021cade} latent features in the test set. This experiment uses 2013 dataset for training and 2014 to 2024 samples serve as test sets.

\paragraph{Feature-Based Stability Evaluation.} In Figure~\ref{fig:stability-scores}, we present the stability scores ({\em jaccard similarity}) across consecutive year-wise malware sample sets for 10 common malware families. A flatter curve across the years indicates stronger temporal consistency within a family, whereas sharp variations reflect instability or feature drift over time. Most families, such as \texttt{airpush}, \texttt{dianjin}, \texttt{plankton}, \texttt{smsagent}, \texttt{smspay}, and \texttt{smsreg}, demonstrate relatively flat trends, suggesting stable feature distributions across years. In contrast, families like \texttt{umpay}, \texttt{fakeapp}, and \texttt{dnotua} exhibit significant fluctuations, notably a major spike around the fourth group (2017-2018), indicating periods of high instability. The especially large peak observed for \texttt{umpay} indicates a considerable temporal drift, which may be attributed to evolving malware behaviors during that time. Overall, the results show that while several malware families maintain stable characteristics over time, certain families undergo notable shifts, highlighting the need for dynamic adaptation in detection models.


\paragraph{Latent Space Drift Detection via Distance Metrics.} Figure~\ref{fig:cade-distances} presents the distribution of distances to the nearest class centroids for testing samples (2014-2024) across 10 common families, computed based on the contrastive latent space representations~\cite{cade}. We encode each test samples with the trained contrastive autoencoder and compute the Euclidean distance to each class centroid~\cite{cade}. Afterwards, we utilize the {\em Median Absolute Deviation }(MAD) to normalize these distances within each class. In particular, a test sample is classified as a \textit{drifted sample} if its normalized MAD score, $A^{(k)}$, exceed an empirically set threshold $T_{\text{MAD}} = 3.5$; otherwise, it is classified as a \textit{non-drifted} sample. This decision rule ensures that samples deviating significantly from the learned class distributions are detected as potential concept drift instances. The resulting boxplots show a clear separation between non-drifted (green) and drifted (red) samples across families. Drifted samples consistently exhibit higher distance values relative to non-drifted samples, with especially pronounced separation observed for families such as \texttt{plankton}, \texttt{umpay}, dianjin, and \texttt{dnotua}. Non-drifted samples demonstrate tight clustering around the respective centroids, indicating stability within the known malware families.

\subsection{Temporal Analysis of SHAP-based Explanation Drift}
\label{sec:shap}

\paragraph{Analysis Setting.} Explanation drift occurs when the features a malware detector (model) relies on for its predictions change over time, even if accuracy remains stable. To assess explanation drift over time in the LAMDA and API Graph~\cite{api_graph_dataset} datasets, we compute Jaccard and Kendall distances~\cite{burger2025improving,kosub2019note,kendall1938new} over SHapley Additive exPlanations (SHAP)~\cite{shap} feature attributions. Jaccard distance measures changes in the set of important features to find the feature set overlap, while Kendall distance captures shifts in their ranking to find feature ranking consistency. Low distances suggest consistent model reasoning, while high Jaccard or low Kendall values indicate significant drift in explanations. This can result from retraining, data shifts, or adversarial influence, and may signal the need for closer model monitoring. We generate SHAP values using \texttt{KernelExplainer} proposed in \cite{shap} with 100 background and 100 test samples per month. Kernel SHAP algorithm is a model agnostic optimization algorithm. Results using the top 1,000 features are reported in Appendix~\ref{appendix:explanation-drift}. For readability, the $x$-axis in Figure~\ref{fig:lamda} is labeled every third month, spanning June 2013 to January 2025 (labels shown from September 2013 to December 2024). For API Graph~\cite{api_graph_dataset}, the range is January 2013 to December 2018, with similar labeling (see Figure~\ref{fig:apigraph}).

\paragraph{Jaccard and Kendall Distance Analysis.}
To evaluate the temporal consistency of model explanations, we compute Jaccard and Kendall distances between consecutive months based on the top-100 SHAP feature indices. As shown in Figure~\ref{fig:lamda}, our proposed LAMDA dataset displays consistently high Jaccard distances (close to 0.9), indicating significant variability in the feature sets used by the model for prediction across time. A sharp drop around September 2017 suggests a rare period of stability or possibly an anomaly in model behavior. The corresponding Kendall distances show a moderate but steady pattern, further reinforcing that both the set and order of important features fluctuate over time. In contrast, the API Graph dataset, depicted in Figure~\ref{fig:apigraph}, exhibits a gradual downward trend in both Jaccard and Kendall distances. This suggests that the features influencing model predictions in API Graph remain relatively stable across time. SHAP-based explanation drift reveals that the LAMDA dataset induces significantly more volatile model behavior, as shown by higher and more variable Jaccard and Kendall distances (Figure~\ref{fig:lamda}), indicating greater temporal variation in feature importance. In contrast, the API Graph dataset (Figure~\ref{fig:apigraph}) exhibits more stable patterns over time. These results highlight LAMDA's suitability for evaluating concept drift, continual learning, and model robustness in dynamic malware detection settings~\cite{api_graph_dataset}.

\section{Discussion}


We introduce LAMDA, the most extensive and temporally diverse Android malware dataset to date, spanning 2013–2025 (excluding 2015). Unlike prior datasets with limited temporal scope or family diversity~\cite{arp2014drebin, api_graph_dataset}, LAMDA supports long-term, realistic evaluations of malware detectors under evolving threat landscapes. Through a series of systematic evaluations including supervised detection, per-feature divergence, and feature stability analysis across top and common malware families, SHAP based drift explanation analysis, we show that LAMDA is a strong benchmark for real-world concept drift analysis in malware detection. While supervised learners perform well on in-distribution data (IID), their effectiveness declines on temporally distant samples evident in sharp F1-score drops between 2016–2017 (NEAR) and 2023–2024 (FAR). This degradation aligns with Jeffreys divergence values up to 7.0 between 2014 and 2025, indicating significant shifts in static features like API calls and permissions. t-SNE visualizations further reveal increasingly fragmented malware clusters over time, underscoring growing behavioral diversity and structural sparsity.

Our study on feature stability reveal that the importance of certain features, such as permissions and intent filters, change over time. 
Furthermore, SHAP-based~\cite{shap} explanation drift analysis exhibits persistent shifts in feature sets and rankings especially around 2014, 2017, and 2021 indicating evolving model reasoning despite stable classification performance. In contrast, API Graph shows lower, more stable distances, reflecting limited temporal variation. 


\paragraph{Broader Adoptions.}In addition to concept drift analysis, supervised detection, and feature stability assessments, LAMDA can support a range of adjacent research directions in cybersecurity and machine learning. Its temporal structure enables evaluation of generalization under distributional shift, while its family diversity allows for studying malware evolution and model adaptation with limited data, including transfer and few-shot learning. As one of the largest Android malware datasets, LAMDA can also be used to train feature extractors for related tasks such as domain adaptation. Moreover, its global feature vocabulary allows researchers to align newly collected samples with LAMDA, supporting scalable benchmarking. The provided variance-threshold objects further enable transformation of future test features, facilitating continuous dataset expansion.



\paragraph{Limitations.} While \textsc{LAMDA} provides a strong foundation for studying concept drift in malware analysis, we acknowledge a few limitations. It relies exclusively on static features, omitting dynamic behaviors observable only at runtime. While prior work suggests a 10:90 malware-to-benign ratio, \textsc{LAMDA} attempts to maintain a 50:50 ratio, which may be viewed as downplaying the role of benign software distributions. However, we argue that \textsc{LAMDA} is constructed as a challenging dataset with greater family diversity and balanced class distribution. As such, LAMDA will facilitate the investigation of detectors that are more resilient to distributional shifts and capable of generalizing across a broad spectrum of evolving malware behaviors.

\section{Conclusion}
In this paper, we introduce \system, a large-scale and temporally structured Android malware dataset designed to support long-term evaluation of detection systems as threats evolve. Spanning over a decade, the dataset enables detailed analysis of how model performance changes over time due to shifts in malware behavior and feature distributions. Evaluations in supervised learning, feature stability, and explanation analysis highlight the impact of these shifts on detection performance. With extensive temporal coverage, diverse malware families, and static features, \system provides a practical and reproducible foundation for research in cybersecurity and machine learning. We envision that \system~benchmark would serve as a resource for advancing more resilient and adaptive malware detection systems against evolving threat landscape.

\bibliographystyle{plainnat}
\bibliography{main}

\newpage

\appendix

\section*{Overview of Appendix}

\paragraph{Supplementary Material.} The following appendices provide further information:

\begin{enumerate}
    \item \textbf{\ref{appendix:dataset-stats} Dataset Statistics}: Year-wise malware/benign counts and family distributions.
    \item \textbf{\ref{appendix:featdescription} Feature Description}: Overview of all static features extracted from APKs.
    \item \textbf{\ref{appendix:model-config} Model Architectures and Detail Results}: Architectures and extended evaluation on LAMDA variants.
    \item \textbf{\ref{appendix:practical-issues} Behind the Scenes: Practical Challenges}: Technical and operational challenges during dataset construction.
    \item \textbf{\ref{appendix:label-noise} Effect of Label Noise in Training Data}: Impact of different VirusTotal thresholds on labeling.
    \item \textbf{\ref{appendix:label-drift} Label Drift Across Years Based on VirusTotal Label Changes}: Year-wise analysis of evolving VirusTotal labels.
    \item \textbf{\ref{appendix:scalability} Scalability of LAMDA}: Instructions for extending LAMDA with new samples using our codebase.
    \item \textbf{\ref{appendix:comparison-sota} Concept Drift Adaptation on LAMDA}: Results of prior adaptation methods (e.g., CADE, Chen) on LAMDA.
    \item \textbf{\ref{appendix:explanation-drift} SHAP-Based Explanation Drift}: Temporal trends in top 1000 feature attributions.
    \item \textbf{\ref{appendix:cl-exps} Continual Learning on LAMDA}: Class- and domain-incremental learning benchmarks.
    \item \textbf{\ref{appendix:computation} Computational Resources}: Hardware and runtime configuration for dataset generation.
    \item \textbf{\ref{appendix:datasetdocument} Dataset Documentation}: Details about dataset documentation. 
\end{enumerate}

\section{Dataset Statistics}
\label{appendix:dataset-stats}

\textsc{LAMDA} benchmark is constructed from a total of 1,008,381 Android APKs, comprising 369,906 malware samples and 638,475 benign samples. Table~\ref{tab:yearly_distribution} summarizes the yearly distribution of both malware and benign APKs. To mitigate class imbalance during training, our initial goal was to collect approximately 50,000 malware and 50,000 benign samples per year; this target could not be met in certain years. Specifically, we were unable to collect sufficient samples for the years 2017, 2021, 2023, 2024, and 2025 due to our labeling criterion—requiring a VirusTotal detection count of 4 or more for malware—and the limited availability of up-to-date samples in the AndroZoo repository~\cite{androzoo,androzooMetadata}. Additional constraints, such as corrupted downloads and decompilation failures, further reduced the effective sample count in those years. Despite these limitations, \textsc{LAMDA} remains the largest Android malware dataset to date in terms of both total sample count and temporal coverage.




\begin{table}[!ht]
\centering
\caption{Year-wise distribution of total, malware, and benign samples.}
\label{tab:yearly_distribution}
\begin{tabular}{lrrr}
\toprule
\textbf{Year} & \textbf{Total Samples} & \textbf{Malware Samples} & \textbf{Benign Samples} \\
\midrule
2013 & 86,431  & 44,383  & 42,048  \\
2014 & 101,183 & 45,756  & 55,427  \\
2016 & 109,193 & 45,134  & 64,059  \\
2017 & 99,144  & 21,359  & 77,785  \\
2018 & 104,292 & 39,350  & 64,942  \\
2019 & 91,050  & 41,585  & 49,465  \\
2020 & 102,073 & 46,355  & 55,718  \\
2021 & 81,155  & 35,627  & 45,528  \\
2022 & 86,416  & 41,648  & 44,768  \\
2023 & 54,354  & 7,892   & 46,462  \\
2024 & 48,427  & 794     & 47,633  \\
2025 & 44,663  & 23      & 44,640  \\
\midrule
\textbf{Total} & \textbf{1,008,381} & \textbf{369,906} & \textbf{638,475} \\
\bottomrule
\end{tabular}
\end{table}

\begin{table}[!ht]
\centering
\caption{Year-wise breakdown of malware family distributions in LAMDA.}
\label{tab:lamda_metadata}
\begin{tabular}{cccccc}
\toprule
\textbf{Year} & \textbf{New} & \textbf{Existing} & \textbf{Valid Family} & \textbf{\#of Singleton} & \textbf{\#of Unknown} \\
\midrule
2013 & 213 & 0   & 213  & 1550  & 24 \\
2014 & 91  & 140 & 231  & 2482  & 345 \\
2016 & 179 & 196 & 375  & 5861  & 177 \\
2017 & 88  & 119 & 207  & 9063  & 1108 \\
2018 & 153 & 220 & 373  & 20579 & 1242 \\
2019 & 259 & 376 & 635  & 18916 & 22 \\
2020 & 141 & 447 & 588  & 30644 & 25 \\
2021 & 43  & 252 & 295  & 30020 & 23 \\
2022 & 161 & 490 & 651  & 24927 & 4 \\
2023 & 37  & 187 & 224  & 5922  & 15 \\
2024 & 14  & 50  & 64   & 626   & 0 \\
2025 & 1   & 7   & 8    & 14    & 0 \\
\midrule
\textbf{Total} & \textbf{1,380} & & & \textbf{150,604} & \textbf{2,985} \\
\bottomrule
\end{tabular}
\end{table}

\begin{table}[!ht]
\centering
\scriptsize
\caption{Distribution of unknown malware samples by VirusTotal detection count.}
\label{tab:vt_detection_unknown}
\begin{tabular}{lccccccccccccccc}
\toprule
\textbf{VT Detection} & 4 & 5 & 6 & 7 & 8 & 9 & 10 & 11 & 12 & 13 & 14 & 15 & 18 & 19 & \textbf{Total} \\
\midrule
\textbf{\# of Unknown Sample} & 1643 & 664 & 226 & 153 & 133 & 65 & 68 & 15 & 3 & 4 & 5 & 4 & 1 & 1 & \textbf{2,985} \\
\bottomrule
\end{tabular}
\end{table}

Beyond binary labels, \textsc{LAMDA} also includes family-level annotations for malware samples. As shown in Table~\ref{tab:lamda_metadata}, the dataset spans 1,380 distinct malware families, offering rich diversity for future analysis. Additionally, 150,604 samples are singletons, belonging to families that appear only once in the dataset, representing rare or unique variants. Moreover, 2,985 samples are marked as ``unknown", where AVClass2 is unable to confidently assign a family label. Table~\ref{tab:vt_detection_unknown} reports the VirusTotal~\cite{virustotal} detection counts for these unknown-labeled samples, offering insight into their potential threat level even in the absence of a family tag.

This comprehensive summary, encompassing both class labels and family-level information, supports a wide range of research directions, including supervised detection, rare variant modeling, family classification, and concept drift analysis across diverse malware behaviors.

\section{Feature Description}
\label{appendix:featdescription}

Built upon static analysis of Android APKs, LAMDA incorporates a broad spectrum of execution-free features based on the features of Drebin~\cite{arp2014drebin}. Table~\ref{tab:staticfeat} summarizes the key categories of static features used in \system~\cite{arp2014drebin}. These include declared components (e.g., services, activities), permissions (requested and used), intent filters, restricted or suspicious API calls, and embedded network indicators such as hardcoded IPs and URLs.

Each APK is converted into a binary feature vector using a bag-of-tokens representation. Tokens are derived from the presence or absence of the static properties listed in Table~\ref{tab:staticfeat}. Since each application typically uses only a small fraction of the global feature space, the resulting vectors are sparse and high-dimensional. To address this, we apply different \texttt{VarianceThreshold} feature selection~\cite{scikit-learn}, resulting in three dataset variants with different dimensionalities and sizes. Table~\ref{tab:lamda_variants} summarizes these variants. The \texttt{Baseline} variant uses a threshold of 0.001~\cite{madar} and yields 4,561 binary features. Increasing the threshold to 0.01 results in a smaller, more compressed feature space with 925 features, while lowering it to 0.0001 expands the feature space to over 25,000 features.

\begin{figure}[!t]
    \centering
    \begin{subfigure}[t]{0.48\linewidth}
        \centering
        \includegraphics[width=\linewidth]{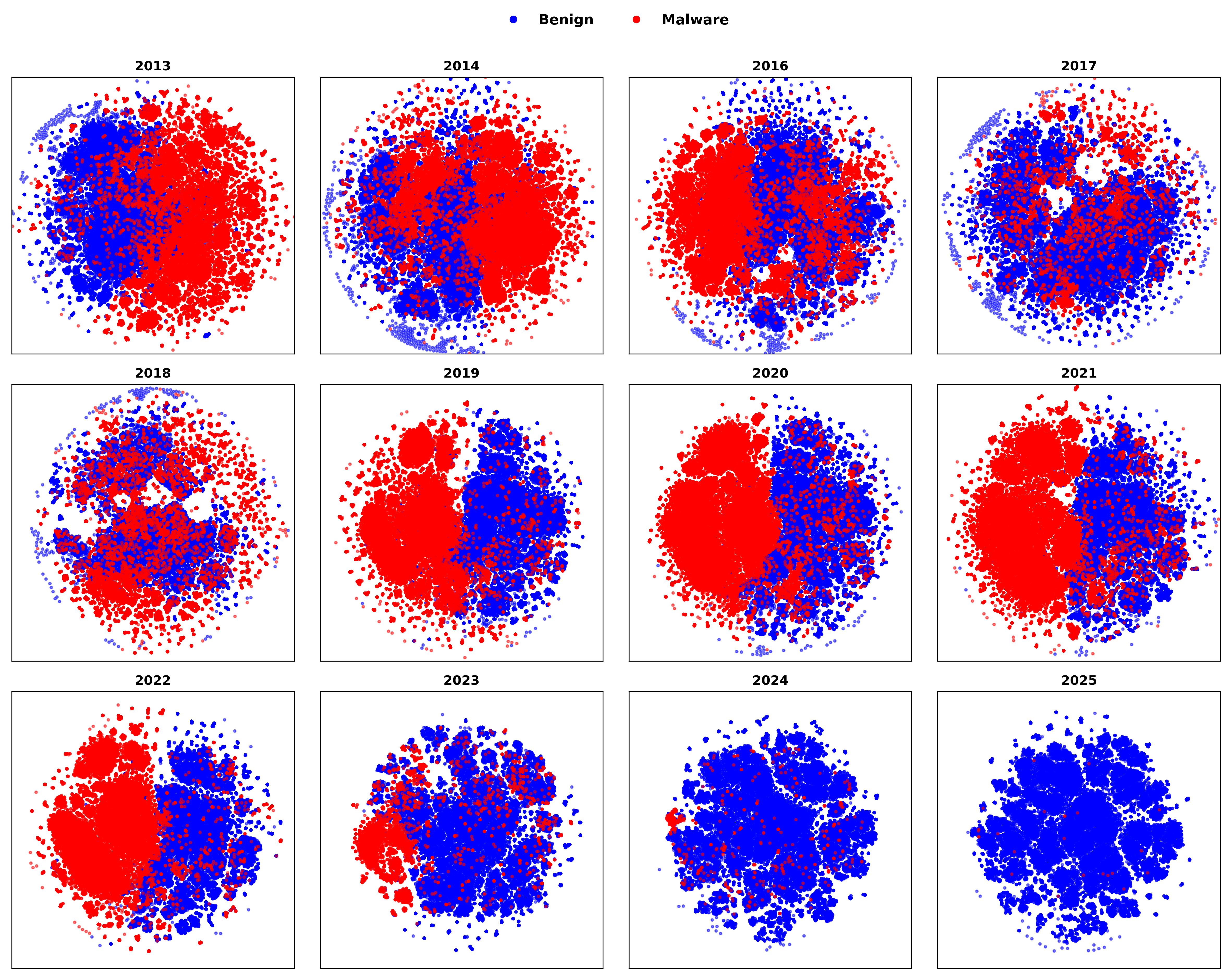}
        \caption{LAMDA.}
        \label{fig:figure1}
    \end{subfigure}
    \hfill
    \begin{subfigure}[t]{0.48\linewidth}
        \centering
        \includegraphics[width=\dimexpr0.75\linewidth,height=\dimexpr0.75\linewidth]{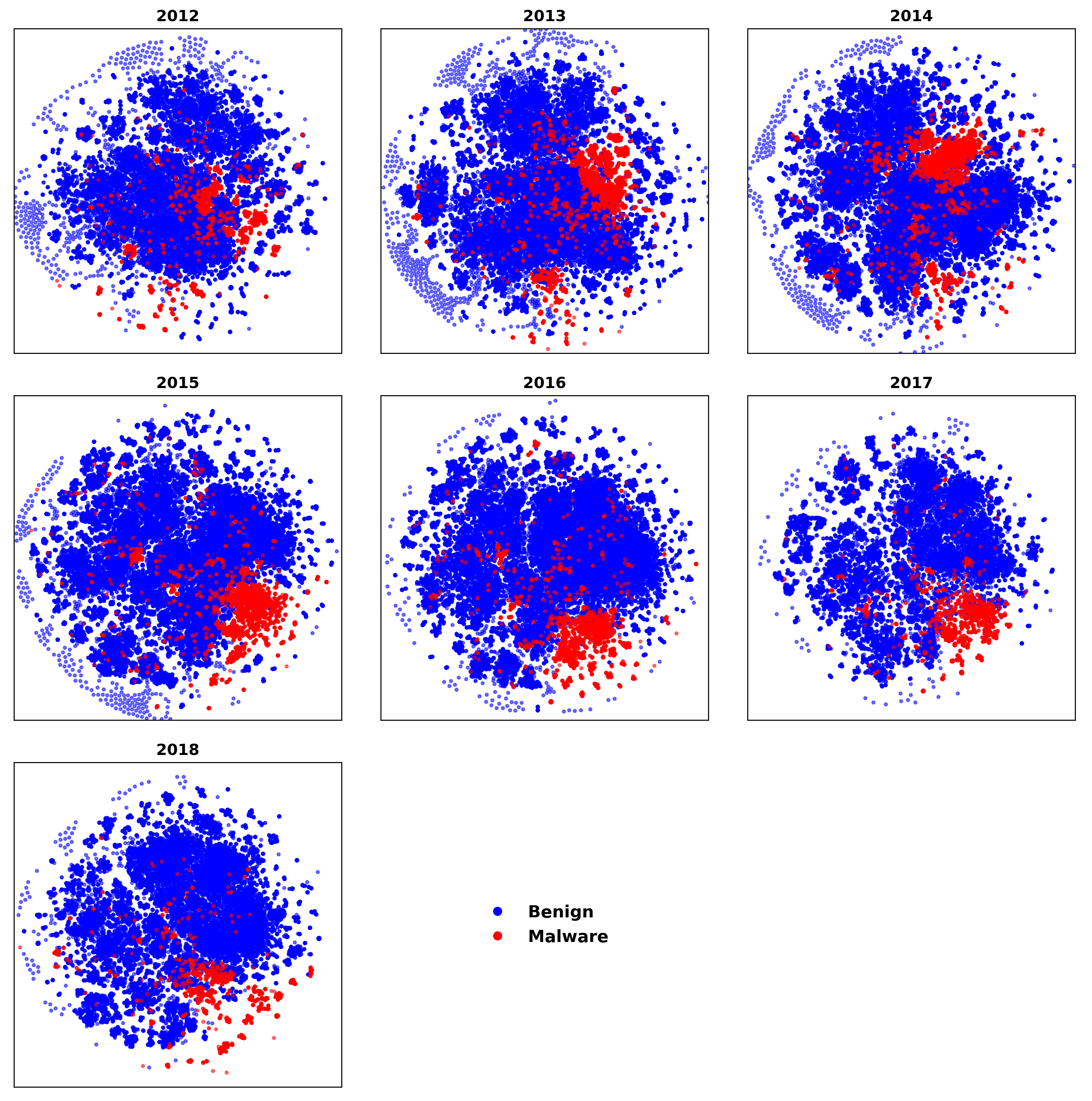}
        \caption{API Graph.}
        \label{fig:figure2}
    \end{subfigure}
    \caption{t-SNE projection of LAMDA and API Graph dataset, (a) t-SNE project of LAMDA from 2013 to 2025 (excluding 2015) and (b) t-SNE projection of API Graph from 2012 to 2018.}
    \label{fig:combined_figure1}
\end{figure}

To visualize the structural differences these features capture, we present t-SNE projections comparing LAMDA and API Graph~\cite{anoshift} in Figure~\ref{fig:combined_figure1}. LAMDA shows more scattered and diverse malware clusters over time, suggesting richer feature representations and stronger concept drift compared to the relatively compact structure in API Graph. This diversity, driven by the dynamic use of static tokens such as APIs and permissions, highlights the importance of broad and representative feature sets for modeling evolving malware behavior. Figure~\ref{fig:vt_detection_tsne_plot} further validates this hypothesis with varying number of virus total engine detection count.

\begin{table}[!t]
\centering
\caption{Static Features and Their Descriptions.}
\begin{tabular}{p{0.275\textwidth} p{0.675\textwidth}}
\toprule
\textbf{Feature} & \textbf{Description} \\
\midrule
\textbf{Requested permissions} & Permissions declared in the manifest (e.g., \texttt{CAMERA}, \texttt{BLUETOOTH}) indicating intended access to sensitive resources. \\ \hline
\textbf{Declared activities and services} & Registered components of the application, providing insight into its structural and behavioral composition. \\ \hline
\textbf{Broadcast receivers} & Components that handle specific system or custom intents (e.g., \texttt{BOOT\_COMPLETED}), often linked to persistence or event-driven behavior. \\ \hline
\textbf{Hardware components} & Device capabilities required by the app (e.g., camera, Bluetooth), implying functional intent. \\ \hline
\textbf{Intent filters} & Define the types of intents components can respond to; critical for modeling potential entry points. \\ \hline
\textbf{Used permissions} & Permissions referenced in the smali code, reflecting actual permission usage. \\ \hline
\textbf{Restricted API calls} & APIs that are protected by system permissions or grant access to sensitive resources. \\ \hline
\textbf{Suspicious API calls} & APIs heuristically associated with malicious or abnormal behavior. \\ \hline
\textbf{Embedded IP addresses and URL domains} & Hardcoded network endpoints that may indicate command-and-control (C\&C) servers or tracking mechanisms. \\
\bottomrule
\end{tabular}
\label{tab:staticfeat}
\end{table}

\subsection{LAMDA Variants}
\label{appendix:lamdavariants}

For the LAMDA dataset variants, we apply different thresholds using the \texttt{VarianceThreshold} (varTh) feature selector. In the baseline configuration (\texttt{varTh = 0.001}), we retain 4,561 features with a total in-memory size of 222\,MB. For a more relaxed threshold (\texttt{varTh = 0.0001}), we preserve 25,460 features, resulting in a memory size of 554\,MB. Conversely, applying a stricter threshold (\texttt{varTh = 0.01}) yields 915 features with a reduced storage size of 138\,MB. These information are summarized in Table~\ref{tab:lamda_variants}.

\begin{table}[h]
\centering
\caption{Summary of Dataset Variants by Variance Threshold.}
\label{tab:lamda_variants}
\footnotesize
\begin{tabular}{lcccc}
\toprule
\textbf{Variant} & \textbf{Threshold} & \textbf{\# Metadata} & \textbf{\# Binary Features} & \textbf{Size} \\
\midrule
{Baseline}             & 0.001   & 5 & 4561   & 222MB \\
{var\_thresh\_0.0001}  & 0.0001  & 5 & 25460 & 554MB \\
{var\_thresh\_0.01}    & 0.01    & 5 & 925 & 138MB \\
\bottomrule
\end{tabular}
\end{table}

\section{Additional Experimental Details}
\label{appendix:model-config}

In this section, we summarize the details of the model architectures and the training setup for each method used in our experiments. In addition, we present supplementary results.

\subsection{Details of the baseline methods}

\paragraph{Multi-Layer Perceptron (MLP).} The MLP model used for the experiments is adapted from prior work~\cite{continual-learning-malware,madar} and is composed of four fully connected layers with the following sizes: 1024, 512, 256, and 128. Each hidden layer is followed by batch normalization, ReLU activation, and a dropout layer with a dropout rate of 0.5. The final output layer uses a \texttt{sigmoid} activation function for binary classification. The model is trained 
using Adam optimizer with a learning rate of 0.001, and batch size 512, as it stabilizes by this point, avoiding unnecessary GPU time.




\paragraph{LightGBM.} 
In addition to MLP, we also use LightGBM~\cite{ke2017lightgbm}, gradient-boosted decision tree ensemble, for binary classification.
LightGBM is trained with up to 5000 estimators and a learning rate of 0.02, with early stopping based on Area Under the Curve (AUC) metric if no improvement is observed for 100 rounds. Each tree is allowed up to 256 leaves to provide high capacity for learning complex patterns. We apply 80\% subsampling of both rows and features to mitigate overfitting. We also include $L_1$ and $L_2$ regularization to further penalize methods complexity to prevent overfitting. These hyperparameters are selected based on practices in malware detection benchmarks such as EMBER~\cite{anderson2018ember} and TESSERACT~\cite{tesseract}. 



\paragraph{XGBoost.} 

The adapted XGBoost is configured with a tree depth of 12 and a learning rate of 0.05. We use log loss objective for binary classification~\cite{chen2016xgboost, ember}. The method is trained for up to 3000 boosting rounds and uses the \texttt{gpu\_hist} tree construction method to accelerate training. The input data is loaded in XGBoost's \texttt{DMatrix} format, which is optimized for memory efficiency and fast training. We train the method on the full training data without applying early stopping and evaluate using log loss.




\paragraph{Support Vector Machine (SVM).} A linear SVM model is implemented using \texttt{LinearSVC} and calibrated using \texttt{CalibratedClassifierCV} to enable probability outputs. This is essential for downstream evaluation where probabilistic thresholds or ranking-based metrics are used. Following prior work, the method is trained on the full dataset with a maximum of 10,000 iterations~\cite{chen2023continuous}. Post-training, model memory usage is reported using \texttt{psutil} to assess resource footprint. 




All models are trained on three different LAMDA variants with \texttt{VarianceThreshold} (VarTh) $\in \{0.01, 0.001, 0.0001\}$ where VarTh $= 0.001$ is the baseline. No task-specific tuning or dataset-specific hyperparameter adjustments are performed to ensure fair comparisons across splits and datasets.

\subsection{Baseline Performance}

We compare LAMDA baseline with API Graph~\cite{api_graph_dataset} dataset and provide a comprehensive results on four methods discussed above using AnoShift-style~\cite{anoshift} splits. A subset of Table~\ref{tab:model_performance} and Table~\ref{tab:api_graph_performance} are explained in the main body of the paper. We present the results with more performance metrics. 

We compare the LAMDA baseline with the API Graph~\cite{api_graph_dataset} dataset and present comprehensive results using four models under AnoShift-style~\cite{anoshift} splits. While a subset of results is highlighted in Table~\ref{tab:model_performance} and Table~\ref{tab:api_graph_performance} in the main paper, we report extended metrics here for completeness.

Across both \texttt{NEAR} and \texttt{FAR} splits, LAMDA consistently exhibits lower scores across all performance metrics compared to API Graph, and notably higher false negative rates (FNR). These trends clearly indicate that LAMDA captures a significantly higher degree of concept drift. Furthermore, the standard deviation across metrics is substantially higher in LAMDA, especially for drifted years, underscoring the dataset’s temporal instability in detection performance---validating the presence of concept drift.

\begin{table}[!t]
\centering
\caption{Performance of models across IID, NEAR, and FAR splits for LAMDA on Baseline (\texttt{VarianceThreshold = 0.001}).}
\label{tab:model_performance}
\scriptsize
\resizebox{\textwidth}{!}{
\begin{tabular}{llcccccccc}
\toprule
\textbf{Split} & \textbf{Model} & \textbf{Accuracy} & \textbf{Precision} & \textbf{Recall} & \textbf{F1} & \textbf{ROC AUC} & \textbf{PR AUC} & \textbf{FPR} & \textbf{FNR} \\
\midrule
\multirow{4}{*}{IID} 
 & LightGBM & \textbf{97.74 $\pm$ 0.35} & 96.74 $\pm$ 0.31 & \textbf{98.26 $\pm$ 0.34} & \textbf{97.49 $\pm$ 0.17} & \textbf{99.55 $\pm$ 0.03} & \textbf{99.50 $\pm$ 0.11} & 2.69 $\pm$ 0.48 & \textbf{1.74 $\pm$ 0.34} \\
 & MLP      & 97.50 $\pm$ 0.44 & 96.91 $\pm$ 0.29 & 97.50 $\pm$ 0.06 & 97.21 $\pm$ 0.12 & 99.48 $\pm$ 0.04 & 99.38 $\pm$ 0.20 & 2.58 $\pm$ 0.85 & 2.50 $\pm$ 0.06 \\
 & SVM      & 95.61 $\pm$ 0.61 & 94.78 $\pm$ 1.41 & 95.18 $\pm$ 0.76 & 94.98 $\pm$ 1.07 & 98.89 $\pm$ 0.28 & 98.75 $\pm$ 0.46 & 4.09 $\pm$ 0.55 & 4.82 $\pm$ 0.76 \\
 & XGBoost  & 97.36 $\pm$ 0.15 & 96.32 $\pm$ 0.70 & 97.80 $\pm$ 0.43 & 97.05 $\pm$ 0.14 & 99.15 $\pm$ 0.16 & 97.68 $\pm$ 1.16 & 2.96 $\pm$ 0.01 & 2.20 $\pm$ 0.43 \\
\midrule
\multirow{4}{*}{NEAR} 
 & LightGBM & \textbf{85.83 $\pm$ 3.96} & \textbf{90.36 $\pm$ 5.21} & \textbf{49.49 $\pm$ 30.82} & \textbf{59.48 $\pm$ 28.20} & 74.05 $\pm$ 23.76 & \textbf{70.18 $\pm$ 27.10} & \textbf{1.85 $\pm$ 0.95} & \textbf{50.51 $\pm$ 30.82} \\
 & MLP      & 83.90 $\pm$ 3.75 & 78.12 $\pm$ 13.98 & 48.05 $\pm$ 30.42 & 56.57 $\pm$ 28.41 & 82.71 $\pm$ 11.19 & 67.94 $\pm$ 24.59 & 3.98 $\pm$ 1.19 & 51.95 $\pm$ 30.42 \\
 & SVM      & 82.08 $\pm$ 3.11 & 72.56 $\pm$ 18.38 & 44.38 $\pm$ 28.88 & 52.91 $\pm$ 28.40 & 75.18 $\pm$ 17.98 & 62.53 $\pm$ 29.82 & 4.71 $\pm$ 0.97 & 55.62 $\pm$ 28.88 \\
 & XGBoost  & 84.59 $\pm$ 3.75 & 86.18 $\pm$ 9.02 & 46.16 $\pm$ 30.94 & 55.84 $\pm$ 29.73 & 77.75 $\pm$ 16.85 & 68.14 $\pm$ 26.55 & 2.14 $\pm$ 0.86 & 53.84 $\pm$ 30.94 \\
\midrule
\multirow{4}{*}{FAR} 
 & LightGBM & 83.94 $\pm$ 10.61 & 74.65 $\pm$ 34.66 & 35.90 $\pm$ 22.97 & 47.24 $\pm$ 27.33 & 78.04 $\pm$ 20.83 & 63.45 $\pm$ 35.80 & 1.30 $\pm$ 0.95 & 64.10 $\pm$ 22.97 \\
 & MLP      & 83.45 $\pm$ 10.74 & \textbf{76.12 $\pm$ 33.39} & 35.60 $\pm$ 20.63 & \textbf{47.59 $\pm$ 25.30} & \textbf{84.04 $\pm$ 11.23} & \textbf{66.16 $\pm$ 34.34} & \textbf{1.14 $\pm$ 0.71} & \textbf{64.40 $\pm$ 20.63} \\
 & SVM      & 80.99 $\pm$ 11.98 & 72.89 $\pm$ 35.60 & 30.07 $\pm$ 16.85 & 41.86 $\pm$ 22.55 & 79.07 $\pm$ 15.06 & 62.27 $\pm$ 34.09 & 1.27 $\pm$ 0.76 & 69.93 $\pm$ 16.85 \\
 & XGBoost  & 82.03 $\pm$ 11.07 & 70.00 $\pm$ 37.26 & 31.89 $\pm$ 20.43 & 42.75 $\pm$ 25.86 & 76.85 $\pm$ 16.49 & 60.33 $\pm$ 35.88 & 1.69 $\pm$ 0.57 & 68.11 $\pm$ 20.43 \\
\bottomrule
\end{tabular}
}
\end{table}

\begin{table}[ht]
\centering
\caption{Performance of models on across IID, NEAR, and FAR splits for API Graph.}
\label{tab:api_graph_performance}
\scriptsize
\resizebox{\textwidth}{!}{
\begin{tabular}{llcccccccc}
\toprule
\textbf{Split} & \textbf{Model} & \textbf{Accuracy} & \textbf{Precision} & \textbf{Recall} & \textbf{F1} & \textbf{ROC AUC} & \textbf{PR AUC} & \textbf{FPR} & \textbf{FNR} \\
\midrule
\multirow{4}{*}{IID}
 & LightGBM & 97.02 $\pm$ 0.00 & \textbf{95.78 $\pm$ 0.00} & 73.44 $\pm$ 0.00 & 83.14 $\pm$ 0.00 & \textbf{98.93 $\pm$ 0.00} & \textbf{94.92 $\pm$ 0.00} & \textbf{0.36 $\pm$ 0.00} & 26.56 $\pm$ 0.00 \\
 & MLP      & 97.35 $\pm$ 0.20 & 94.84 $\pm$ 1.49 & 77.79 $\pm$ 2.85 & 85.43 $\pm$ 1.35 & 94.62 $\pm$ 0.92 & 89.33 $\pm$ 1.75 & 0.47 $\pm$ 0.16 & 22.21 $\pm$ 2.85 \\
 & SVM      & 96.64 $\pm$ 0.00 & 94.12 $\pm$ 0.00 & 70.85 $\pm$ 0.00 & 80.84 $\pm$ 0.00 & 97.27 $\pm$ 0.00 & 90.90 $\pm$ 0.00 & 0.49 $\pm$ 0.00 & 29.15 $\pm$ 0.00 \\
 & XGBoost  & 96.30 $\pm$ 0.00 & 91.57 $\pm$ 0.00 & 69.37 $\pm$ 0.00 & 78.94 $\pm$ 0.00 & 95.93 $\pm$ 0.00 & 89.08 $\pm$ 0.00 & 0.71 $\pm$ 0.00 & 30.63 $\pm$ 0.00 \\
\midrule
\multirow{4}{*}{NEAR}
 & LightGBM & \textbf{94.84 $\pm$ 0.00} & \textbf{94.07 $\pm$ 0.00} & 51.33 $\pm$ 0.00 & 66.42 $\pm$ 0.00 & \textbf{96.28 $\pm$ 0.00} & \textbf{83.45 $\pm$ 0.00} & \textbf{0.36 $\pm$ 0.00} & \textbf{48.67 $\pm$ 0.00} \\
 & MLP      & 94.66 $\pm$ 0.34 & 88.22 $\pm$ 3.83 & 53.67 $\pm$ 4.84 & 66.55 $\pm$ 3.19 & 85.97 $\pm$ 1.25 & 72.24 $\pm$ 2.46 & 0.81 $\pm$ 0.37 & 46.33 $\pm$ 4.84 \\
 & SVM      & 94.01 $\pm$ 0.00 & 81.70 $\pm$ 0.00 & 51.18 $\pm$ 0.00 & 62.93 $\pm$ 0.00 & 90.29 $\pm$ 0.00 & 70.84 $\pm$ 0.00 & 1.26 $\pm$ 0.00 & 48.82 $\pm$ 0.00 \\
 & XGBoost  & 92.76 $\pm$ 0.00 & 77.60 $\pm$ 0.00 & 38.11 $\pm$ 0.00 & 51.12 $\pm$ 0.00 & 92.13 $\pm$ 0.00 & 65.74 $\pm$ 0.00 & 1.21 $\pm$ 0.00 & 61.89 $\pm$ 0.00 \\
\midrule
\multirow{4}{*}{FAR}
 & LightGBM & \textbf{95.31 $\pm$ 0.61} & \textbf{87.48 $\pm$ 2.48} & \textbf{57.41 $\pm$ 6.57} & \textbf{69.07 $\pm$ 4.73} & \textbf{96.11 $\pm$ 0.40} & \textbf{80.94 $\pm$ 1.17} & \textbf{0.84 $\pm$ 0.24} & \textbf{42.59 $\pm$ 6.57} \\
 & MLP      & 94.50 $\pm$ 0.84 & 82.31 $\pm$ 6.28 & 51.14 $\pm$ 7.90 & 62.80 $\pm$ 6.68 & 90.02 $\pm$ 2.91 & 71.38 $\pm$ 5.89 & 1.12 $\pm$ 0.43 & 48.86 $\pm$ 7.90 \\
 & SVM      & 94.33 $\pm$ 0.60 & 77.38 $\pm$ 0.33 & 54.34 $\pm$ 7.27 & 63.56 $\pm$ 5.08 & 92.73 $\pm$ 0.92 & 71.46 $\pm$ 2.43 & 1.60 $\pm$ 0.20 & 45.66 $\pm$ 7.27 \\
 & XGBoost  & 93.70 $\pm$ 0.35 & 75.99 $\pm$ 2.62 & 46.04 $\pm$ 1.92 & 57.29 $\pm$ 1.45 & 87.78 $\pm$ 1.90 & 66.49 $\pm$ 4.80 & 1.49 $\pm$ 0.28 & 53.96 $\pm$ 1.92 \\
\bottomrule
\end{tabular}
}
\end{table}

\subsection{LAMDA Variants and Drift Sensitivity}

\begin{figure}[!t]
    \centering
    \begin{subfigure}[t]{0.48\linewidth}
        \centering
        \includegraphics[width=\linewidth]{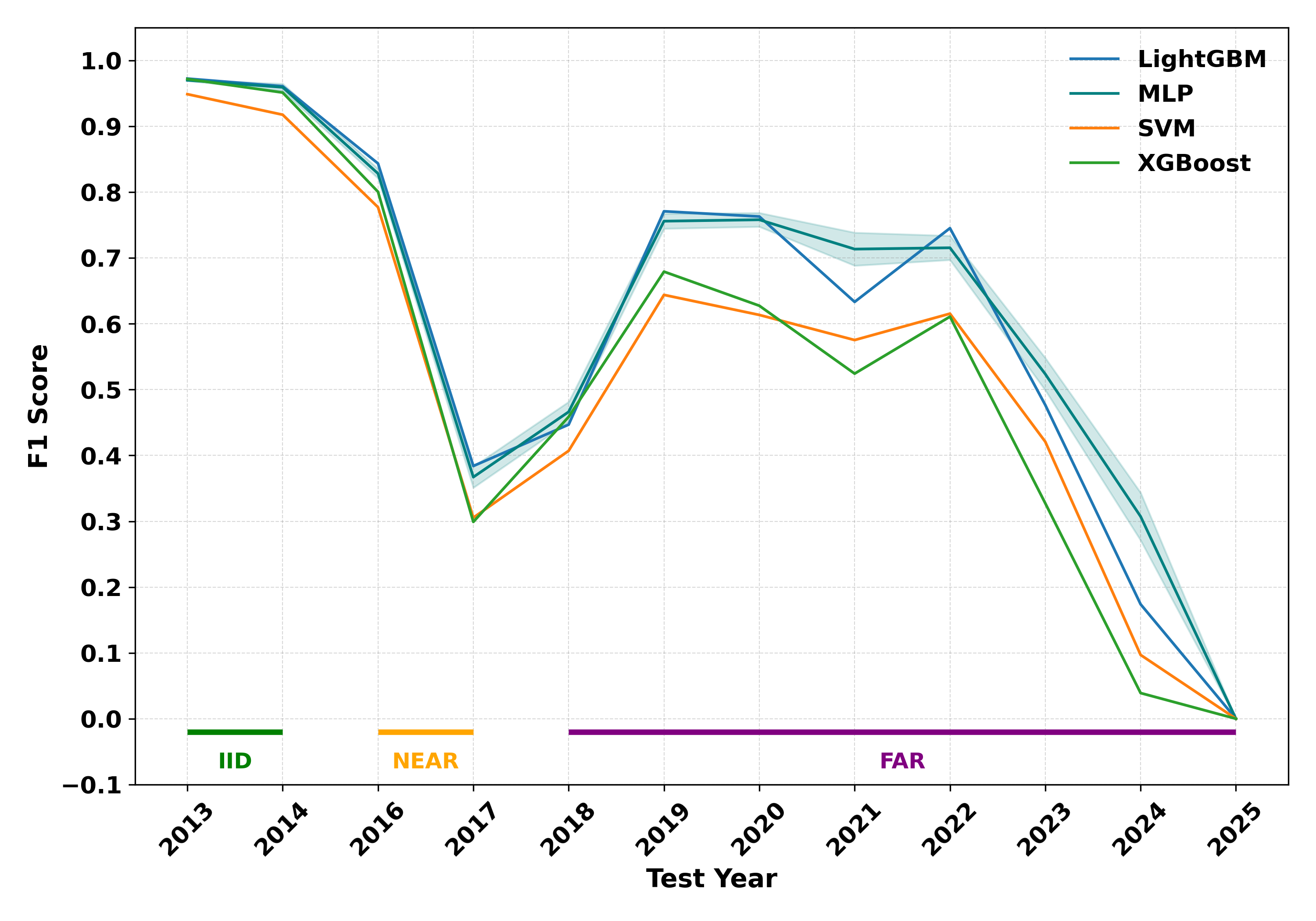} 
        \caption{\texttt{VarianceThreshold} $= 0.01$}
        \label{fig:subfig1}
    \end{subfigure}
    \hfill
    \begin{subfigure}[t]{0.48\linewidth}
        \centering
        \includegraphics[width=\linewidth]{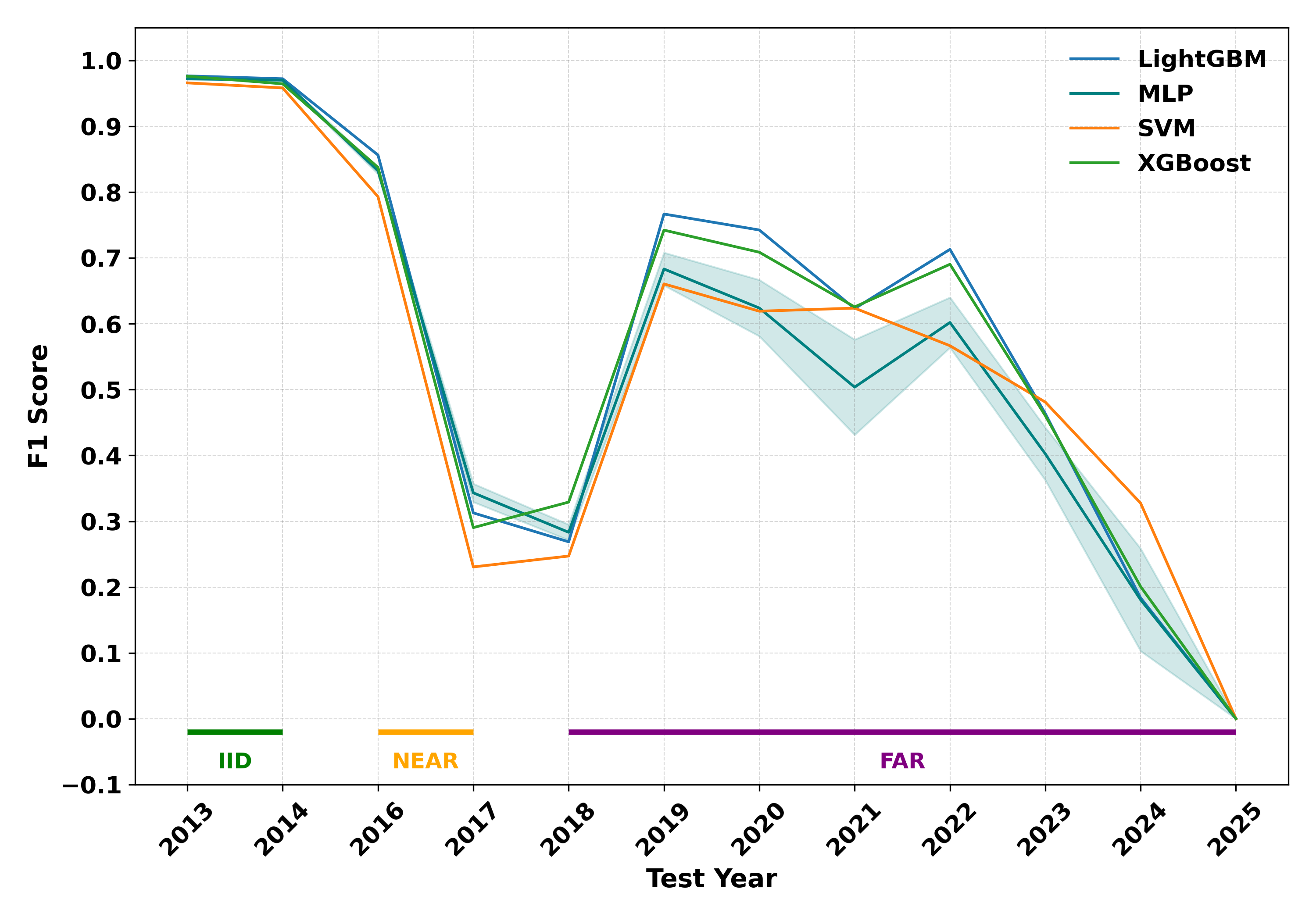} 
        \caption{\texttt{VarianceThreshold} $= 0.0001$}
        \label{fig:subfig2}
    \end{subfigure}
    \caption{F1-scores on different models on based on AnoShift-style split on LAMDA. (a) for \texttt{VarianceThreshold} 0.01 and (b) for \texttt{VarianceThreshold} 0.0001.}
    \label{fig:lamda_variants_side_by_side}
\end{figure}

LAMDA offers flexibility to researchers for Android malware analysis by supporting different feature selection variants. In this section, we evaluate two additional variants of LAMDA. As shown in Table~\ref{tab:model_performance_cleaned} and Table~\ref{tab:model_performance_updated}, we report detailed performance results for the four methods and configurations used in the primary analysis of concept drift with \texttt{VarianceThreshold} of 0.001. The variant using a threshold of 0.01 exhibits a relatively higher average F1-score across NEAR and FAR splits compared to that of the Baseline (\texttt{0.001}) and \texttt{var\_thresh\_0.0001} variants. Figure~\ref{fig:lamda_variants_side_by_side} shows this trend as well, highlighting the comparative performance of the LAMDA variants.


To further understand how these feature selection variants influence drift sensitivity, we focus on the NEAR split with the SVM. For false positive rate (FPR), both the baseline and \texttt{varTh=0.0001} maintain relatively low values ($\sim$4.07~$\pm$~0.18), whereas \texttt{varTh=0.01} shows a sharp increase to 17.09~$\pm$~2.87. This suggests that reducing the feature set too aggressively may misclassify benign as malware. Conversely, the false negative rate (FNR) slightly improves under \texttt{varTh=0.01}, decreasing from $\sim$55.62~$\pm$~28.88 and $\sim$57.72~$\pm$~28.87, baseline and \texttt{varTh=0.0001}, respectively, to 46.22~$\pm$~23.97. This indicates that even with fewer features, the model may still capture certain generalizable malware traits, improving detection of some malicious samples. 

However, for the LightGBM model, this change is accompanied by a drop in precision. While in NEAR region both the baseline and \texttt{varTh=0.0001} variants maintain high precision scores (around 90.36~$\pm$~5.21), the \texttt{varTh=0.01} variant yields a lower precision of 75.03~$\pm$~15.03. This reflects a shift in the methods decision behavior under more aggressive feature selection, emphasizing the importance of balancing dimensionality reduction with predictive consistency.

In summary, while \texttt{varTh=0.01} may occasionally help with generalization under drift, it also amplifies misclassification of benign apps and reduces predictive stability. The baseline and \texttt{varTh=0.0001} offer more shift in data distribution.

\begin{table}[!t]
\centering
\caption{Performance of models across IID, NEAR, and FAR splits for LAMDA variant of \texttt{VarianceThreshold} (0.01).}
\label{tab:model_performance_cleaned}
\scriptsize
\resizebox{\textwidth}{!}{
\begin{tabular}{llcccccccc}
\toprule
\textbf{Split} & \textbf{Model} & \textbf{Accuracy} & \textbf{Precision} & \textbf{Recall} & \textbf{F1} & \textbf{ROC AUC} & \textbf{PR AUC} & \textbf{FPR} & \textbf{FNR} \\
\midrule
\multirow{4}{*}{IID} 
 & {LightGBM} & \textbf{97.08 $\pm$ 0.21} & \textbf{96.40 $\pm$ 0.22} & \textbf{96.95 $\pm$ 1.02} & \textbf{96.67 $\pm$ 0.61} & \textbf{99.42 $\pm$ 0.00} & \textbf{99.33 $\pm$ 0.13} & \textbf{2.94 $\pm$ 0.43} & \textbf{3.05 $\pm$ 1.02} \\
 & MLP      & 96.89 $\pm$ 0.32 & 96.11 $\pm$ 0.37 & 96.84 $\pm$ 1.26 & 96.47 $\pm$ 0.67 & 99.29 $\pm$ 0.08 & 99.15 $\pm$ 0.18 & 3.18 $\pm$ 0.54 & 3.16 $\pm$ 1.26 \\
 & SVM      & 94.13 $\pm$ 0.87 & 92.59 $\pm$ 2.43 & 94.10 $\pm$ 0.84 & 93.33 $\pm$ 1.65 & 97.86 $\pm$ 0.54 & 97.34 $\pm$ 1.15 & 5.83 $\pm$ 0.90 & 5.90 $\pm$ 0.84 \\
 & XGBoost  & 96.65 $\pm$ 0.60 & 95.59 $\pm$ 1.37 & 96.74 $\pm$ 0.80 & 96.16 $\pm$ 1.09 & 99.01 $\pm$ 0.26 & 98.00 $\pm$ 0.36 & 3.45 $\pm$ 0.43 & 3.26 $\pm$ 0.80 \\
\midrule
\multirow{4}{*}{NEAR} 
 & {LightGBM} & \textbf{84.17 $\pm$ 3.77} & \textbf{75.03 $\pm$ 15.03} & \textbf{54.00 $\pm$ 27.34} & \textbf{61.38 $\pm$ 24.22} & \textbf{73.58 $\pm$ 22.25} & \textbf{66.97 $\pm$ 27.37} & \textbf{5.86 $\pm$ 0.94} & \textbf{46.00 $\pm$ 27.34} \\
 & MLP      & 83.39 $\pm$ 3.52 & 73.64 $\pm$ 16.21 & 52.23 $\pm$ 26.78 & 59.77 $\pm$ 24.33 & 79.17 $\pm$ 14.79 & 66.60 $\pm$ 24.33 & 6.09 $\pm$ 1.22 & 47.77 $\pm$ 26.78 \\
 & SVM      & 75.73 $\pm$ 6.47 & 54.52 $\pm$ 25.76 & 53.78 $\pm$ 23.97 & 54.13 $\pm$ 24.86 & 74.88 $\pm$ 14.32 & 58.72 $\pm$ 25.70 & 17.09 $\pm$ 2.87 & 46.22 $\pm$ 23.97 \\
 & XGBoost  & 81.63 $\pm$ 3.39 & 68.78 $\pm$ 20.03 & 47.48 $\pm$ 27.52 & 54.98 $\pm$ 26.43 & 77.32 $\pm$ 14.71 & 62.60 $\pm$ 26.89 & 6.57 $\pm$ 0.68 & 52.52 $\pm$ 27.52 \\
\midrule
\multirow{4}{*}{FAR} 
 & LightGBM & 84.68 $\pm$ 9.54 & 71.15 $\pm$ 35.65 & 39.50 $\pm$ 23.10 & 50.13 $\pm$ 27.27 & 75.80 $\pm$ 21.64 & 61.66 $\pm$ 35.89 & 2.34 $\pm$ 1.72 & 60.50 $\pm$ 23.10 \\
 & {MLP}      & \textbf{85.15 $\pm$ 9.23} & \textbf{76.93 $\pm$ 33.45} & \textbf{41.38 $\pm$ 21.41} & \textbf{53.00 $\pm$ 25.53} & \textbf{87.00 $\pm$ 10.48} & \textbf{68.52 $\pm$ 33.91} & \textbf{1.80 $\pm$ 1.61} & \textbf{58.62 $\pm$ 21.41} \\
 & SVM      & 77.50 $\pm$ 10.05 & 55.00 $\pm$ 32.97 & 36.40 $\pm$ 16.42 & 42.16 $\pm$ 23.50 & 76.19 $\pm$ 12.22 & 52.48 $\pm$ 31.86 & 9.28 $\pm$ 3.56 & 63.60 $\pm$ 16.42 \\
 & XGBoost  & 76.73 $\pm$ 6.52 & 56.84 $\pm$ 36.66 & 34.05 $\pm$ 17.19 & 40.84 $\pm$ 25.02 & 69.95 $\pm$ 15.23 & 51.85 $\pm$ 34.08 & 8.71 $\pm$ 3.04 & 65.95 $\pm$ 17.19 \\
\bottomrule
\end{tabular}
}
\end{table}

\begin{table}[!t]
\centering
\caption{Performance of models across IID, NEAR, and FAR splits for LAMDA variant of \texttt{VarianceThreshold} (0.0001).}
\label{tab:model_performance_updated}
\scriptsize
\resizebox{\textwidth}{!}{
\begin{tabular}{llcccccccc}
\toprule
\textbf{Split} & \textbf{Model} & \textbf{Accuracy} & \textbf{Precision} & \textbf{Recall} & \textbf{F1} & \textbf{ROC AUC} & \textbf{PR AUC} & \textbf{FPR} & \textbf{FNR} \\
\midrule
\multirow{4}{*}{IID} 
 & {LightGBM} & \textbf{97.73 $\pm$ 0.04} & \textbf{96.80 $\pm$ 0.33} & \textbf{98.11 $\pm$ 0.12} & \textbf{97.45 $\pm$ 0.23} & \textbf{99.58 $\pm$ 0.00} & \textbf{99.54 $\pm$ 0.09} & \textbf{2.61 $\pm$ 0.25} & \textbf{1.89 $\pm$ 0.12} \\
 & MLP      & 97.42 $\pm$ 0.21 & 96.44 $\pm$ 0.35 & 97.80 $\pm$ 0.23 & 97.12 $\pm$ 0.15 & 99.31 $\pm$ 0.06 & 99.24 $\pm$ 0.17 & 2.92 $\pm$ 0.41 & 2.20 $\pm$ 0.23 \\
 & SVM      & 96.68 $\pm$ 0.02 & 96.68 $\pm$ 0.08 & 95.77 $\pm$ 0.72 & 96.22 $\pm$ 0.40 & 99.13 $\pm$ 0.20 & 99.13 $\pm$ 0.29 & 2.69 $\pm$ 0.48 & 4.23 $\pm$ 0.72 \\
 & XGBoost  & 97.39 $\pm$ 0.26 & 96.27 $\pm$ 0.69 & 97.83 $\pm$ 0.53 & 97.04 $\pm$ 0.61 & 99.35 $\pm$ 0.08 & 98.92 $\pm$ 0.10 & 3.01 $\pm$ 0.04 & 2.17 $\pm$ 0.53 \\
\midrule
\multirow{4}{*}{NEAR} 
 & {LightGBM} & \textbf{85.55 $\pm$ 3.91} & \textbf{90.24 $\pm$ 5.75} & 48.34 $\pm$ 30.74 & 58.46 $\pm$ 28.66 & 74.58 $\pm$ 23.25 & \textbf{70.47 $\pm$ 26.85} & \textbf{1.70 $\pm$ 0.80} & 51.66 $\pm$ 30.74 \\
 & MLP      & 84.22 $\pm$ 3.38 & 80.23 $\pm$ 13.34 & 49.38 $\pm$ 27.83 & 58.79 $\pm$ 25.81 & 81.65 $\pm$ 13.28 & 69.02 $\pm$ 25.24 & 3.71 $\pm$ 0.95 & 50.62 $\pm$ 27.83 \\
 & SVM      & 81.79 $\pm$ 3.36 & 71.61 $\pm$ 21.55 & 42.28 $\pm$ 28.87 & 51.19 $\pm$ 29.64 & 76.15 $\pm$ 16.45 & 63.54 $\pm$ 29.13 & 4.07 $\pm$ 0.18 & 57.72 $\pm$ 28.87 \\
 & XGBoost  & 84.58 $\pm$ 3.41 & 86.82 $\pm$ 6.81 & 46.86 $\pm$ 30.72 & 56.40 $\pm$ 28.84 & 77.29 $\pm$ 18.06 & 69.61 $\pm$ 25.51 & 2.52 $\pm$ 1.41 & 53.14 $\pm$ 30.72 \\
\midrule
\multirow{4}{*}{FAR} 
 & {LightGBM} & \textbf{83.96 $\pm$ 10.66} & \textbf{75.34 $\pm$ 34.34} & \textbf{35.62 $\pm$ 23.11} & \textbf{47.02 $\pm$ 27.47} & 77.98 $\pm$ 21.70 & \textbf{64.21 $\pm$ 35.59} & \textbf{1.18 $\pm$ 0.85} & \textbf{64.38 $\pm$ 23.11} \\
 & MLP      & 80.88 $\pm$ 11.82 & 71.93 $\pm$ 36.09 & 29.64 $\pm$ 17.51 & 40.99 $\pm$ 23.11 & 82.52 $\pm$ 12.80 & 63.53 $\pm$ 35.27 & 1.47 $\pm$ 1.11 & 70.36 $\pm$ 17.51 \\
 & SVM      & 81.23 $\pm$ 11.94 & 73.48 $\pm$ 34.34 & 32.40 $\pm$ 16.44 & 44.08 $\pm$ 21.95 & 76.94 $\pm$ 16.51 & 63.06 $\pm$ 32.51 & 1.31 $\pm$ 0.86 & 67.60 $\pm$ 16.44 \\
 & XGBoost  & 83.47 $\pm$ 10.41 & 73.53 $\pm$ 35.83 & 35.43 $\pm$ 20.83 & 46.97 $\pm$ 25.75 & 77.49 $\pm$ 19.29 & 63.56 $\pm$ 34.89 & 1.32 $\pm$ 0.67 & 64.57 $\pm$ 20.83 \\
\bottomrule
\end{tabular}
}
\end{table}

\section{Behind the Scenes: Practical Challenges in LAMDA Creation}
\label{appendix:practical-issues}

\subsection{Administrative Challenges}
Downloading large volumes of real-world malware presents significant cybersecurity risks within any institutional environment. During our data collection process, the downloading and unpacking of live malware samples triggered internal threat detection systems, as automated security tools flagged these activities as potential breaches. To mitigate these risks, we implemented strict containment policies, such as disabling execution permissions. Additionally, we worked closely with the university’s cybersecurity team to obtain the necessary approvals and ensure compliance with all relevant security policies. We maintained continuous communication with them throughout the process to ensure proper coordination and promptly address any emerging issues.

\subsection{Technical Challenges}
We faced several technical constraints during the sample collection and processing pipeline. The AndroZoo platform imposes strict download rate limits, allowing only 40 concurrent downloads per user. As a result, we had to be extremely cautious to avoid violating their terms and conditions. Unfortunately, accidental oversights on our part led to temporary request blocks which disrupted the collection process.  Similarly, the VirusTotal API has strict rate limits, which can significantly slow down the retrieval of metadata. Additionally, a considerable number of APKs failed to decompile successfully using Apktool. These failures were often due to obfuscation, corrupted files, or non-standard packaging formats. So, we had to perform multiple rounds of sampling to reach our target number of usable samples.


\section{Effect of Label Noise in Training Data}
\label{appendix:label-noise}

\begin{table}[!t]
\centering
\caption{Effect of thresholding on sample counts and relative percentage change (w.r.t. threshold 4).}
\label{tab:threshold_impact}
\resizebox{\linewidth}{!}{
\begin{tabular}{ccccc}
\toprule
\textbf{Threshold} & \textbf{Benign Samples} & \textbf{Malware Samples} & \textbf{\% Change (Benign)} & \textbf{\% Change (Malware)} \\
\midrule
1  & 638{,}475 & 369{,}906 & 0.0\%  & 0.0\%   \\
2  & 638{,}475 & 369{,}906 & 0.0\%  & 0.0\%   \\
3  & 638{,}475 & 369{,}906 & 0.0\%  & 0.0\%   \\
\rowcolor{gray!20} 4  & 638{,}475 & 369{,}906 & 0.0\%  & 0.0\%   \\ 
5  & 638{,}475 & 324{,}927 & 0.0\%  & \textcolor{red}{$\downarrow$ 12.16\%} \\
6  & 638{,}475 & 281{,}824 & 0.0\%  & \textcolor{red}{$\downarrow$ 23.81\%} \\
7  & 638{,}475 & 241{,}690 & 0.0\%  & \textcolor{red}{$\downarrow$ 34.66\%} \\
8  & 638{,}475 & 206{,}644 & 0.0\%  & \textcolor{red}{$\downarrow$ 44.14\%} \\
9  & 638{,}475 & 177{,}707 & 0.0\%  & \textcolor{red}{$\downarrow$ 51.96\%} \\
10 & 638{,}475 & 155{,}376 & 0.0\%  & \textcolor{red}{$\downarrow$ 58.00\%} \\
11 & 638{,}475 & 138{,}041 & 0.0\%  & \textcolor{red}{$\downarrow$ 62.68\%} \\
12 & 638{,}475 & 123{,}783 & 0.0\%  & \textcolor{red}{$\downarrow$ 66.53\%} \\
13 & 638{,}475 & 111{,}350 & 0.0\%  & \textcolor{red}{$\downarrow$ 69.89\%} \\

\bottomrule
\end{tabular}
}
\end{table}

Label noise in Android malware family classification, particularly when using the Drebin~\cite{arp2014drebin} feature set can significantly impact model performance due to ambiguous and overlapping feature representations. As demonstrated in recent work~\cite{oyen2022robustness}, the robustness of classification models depends not only on the amount of label noise but also on its distribution within the feature space. Specifically, feature-dependent label noise, where the probability of a label flip is contingent on the position of a sample in feature space, can cause a substantial drop in accuracy, even at low noise levels. 

This is especially relevant for Drebin features, where different malware families may share static features like permissions ($x_1 = \texttt{INTERNET}$, $x_2 = \texttt{SEND\_SMS}$), API calls ($x_3 = \texttt{getDeviceId}$), and hardware access ($x_4 = \texttt{READ\_PHONE\_STATE}$). Samples with minimal or ambiguous patterns (e.g., $x_5 = \texttt{ACCESS\_NETWORK\_STATE}$ and $x_6 = \texttt{RECEIVE\_BOOT\_COMPLETED}$) are likely to fall near decision boundaries, increasing the risk of mislabeling. Such feature-dependent noise is more detrimental than uniform or class-dependent noise and warrants careful consideration in malware classification tasks.

\begin{figure}[!t]
    \centering
    \includegraphics[width=\textwidth]{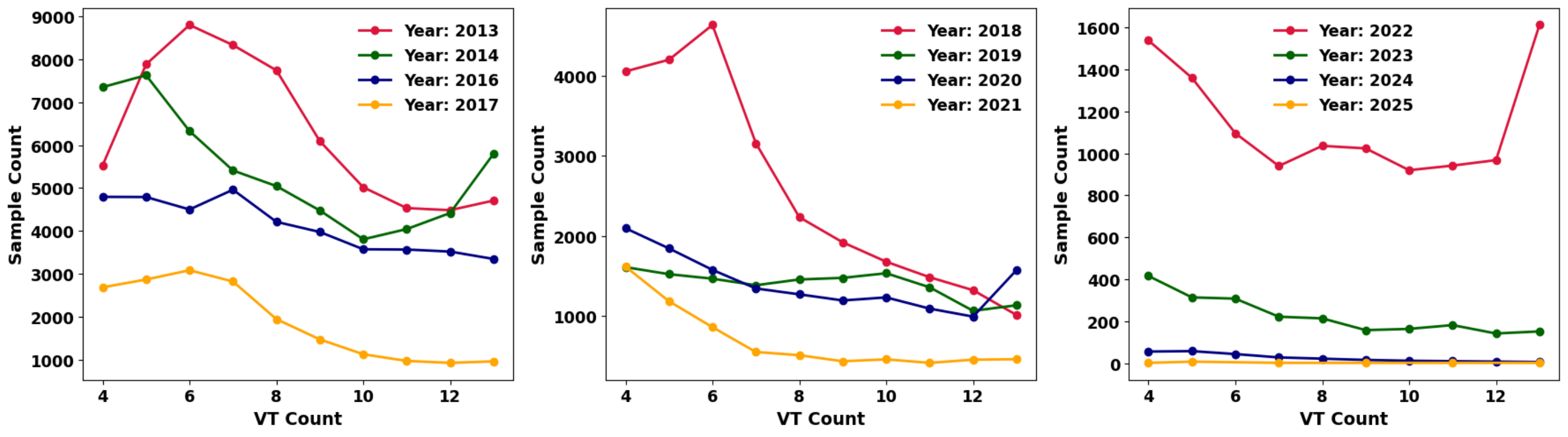}
    \caption{Sample Count Year-wise for Each VT Detection Threshold.}
    \label{fig:vt_detection_sample_count}
\end{figure}

\begin{figure}[!t]
    \centering
    \includegraphics[width=\textwidth]{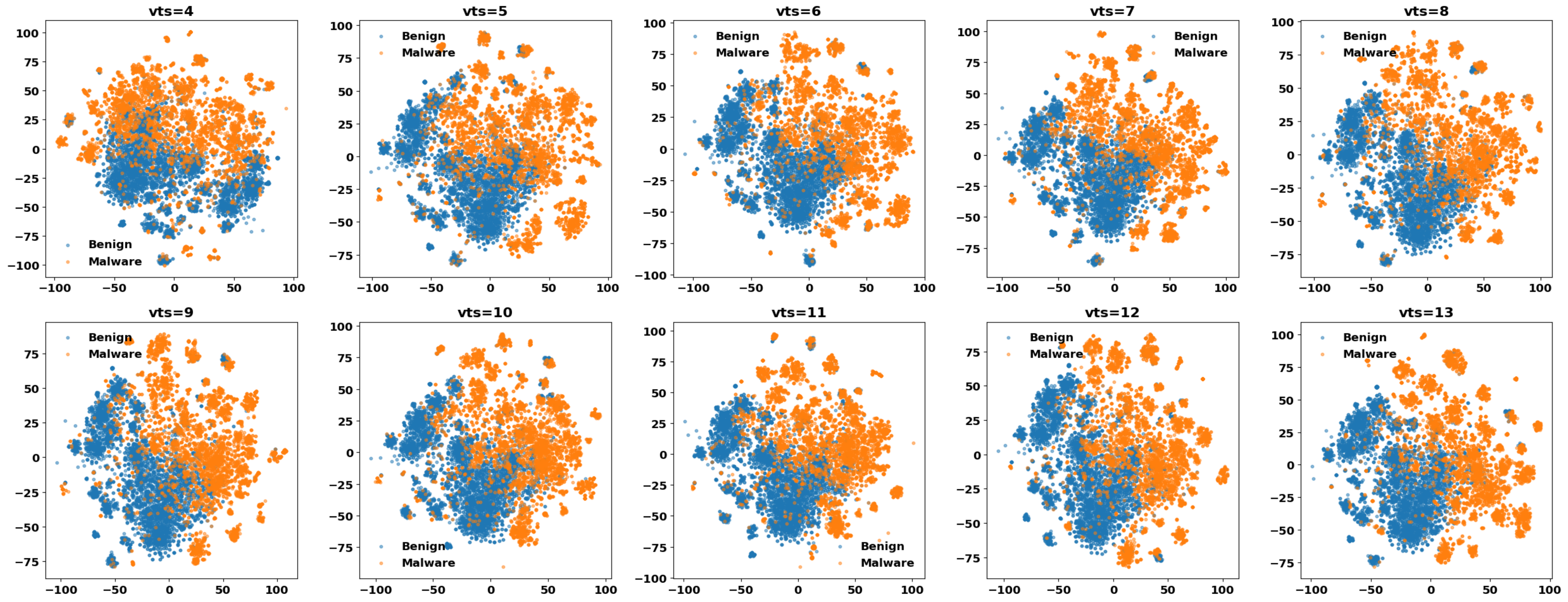}
    \caption{t-SNE visualization of benign and malware samples at varying Virus Total (VT) detection threshold.}
    \label{fig:vt_detection_tsne_plot}
\end{figure}

Table~\ref{tab:threshold_impact} shows the effect of increasing the VirusTotal (VT) detection~\cite{virustotal} threshold on malware labeling in the LAMDA dataset. According to previous studies~\cite{tesseract,continual-learning-malware, cade}, a sample is considered benign if \texttt{vt\_detection} = 0, and labeled as malware if \texttt{vt\_detection} $\geq$ 4. As seen from the Benign sample column in the Table~\ref{tab:threshold_impact}, the number of benign samples remains unchanged across all thresholds, since the benign definition is fixed and independent of the malware thresholding rule. However, the number of malware samples decreases significantly as the threshold increases from 4 to 13. For instance, at a threshold of 10, the number of malware samples drops by 58\% compared to the baseline at threshold 4. This trend continues, reaching a 69.89\% reduction at threshold 13. These results demonstrate that requiring stronger agreement among antivirus engines (i.e., a higher threshold) leads to more conservative malware labeling, effectively excluding a substantial portion of potentially malicious samples. While this may improve the confidence in the labeled malware, it also drastically reduces dataset coverage. Therefore, the choice of VT threshold directly impacts the balance between label precision and data availability, and threshold 4 provides a practical trade-off commonly adopted in existing literature~\cite{droidevolver,tesseract,malcl}.



Figure~\ref{fig:vt_detection_sample_count} illustrates the distribution of malware sample counts across different VirusTotal (VT) threshold~\cite{virustotal} values for each year from 2013 to 2025 (except 2015). The VT count, plotted on the $x$-axis, represents the number of antivirus (AV) engines that flagged a sample as malicious, serving as a metric for detection consensus or confidence. A consistent trend is observed across all years --- as the VT threshold increases, the number of flagged samples decreases. This suggests that only a small fraction of malware samples achieve strong consensus among AV engines, while the majority are detected by relatively few engines. The sample count is highest between VT counts of 5 to 7, especially in earlier years such as 2013--2017, indicating a moderate level of agreement in those periods. In contrast, from 2022 onward, the overall volume of detected samples decreases sharply, and the detections are largely concentrated in the lower VT ranges, which may reflect advancements in malware evasion techniques or shifts in detection criteria. These observations justify the use of a VT threshold. Using a higher threshold (e.g., $\geq$10) may lead to overly conservative labeling with potential false negatives, while lower thresholds increase coverage but may introduce noise. Thus, this temporal analysis provides critical insight into threshold selection and highlights the evolving nature of malware detection over time.

Figure~\ref{fig:vt_detection_tsne_plot} shows the t-SNE projections across varying VirusTotal (VT) detection~\cite{virustotal} thresholds. At lower thresholds (e.g., VT~$\geq$~4 to 6), there is significant overlap between the malware and benign clusters, indicating that many samples labeled as malware may exhibit similar characteristics to benign samples. This suggests that lower thresholds capture a broader range of potentially ambiguous or borderline malicious behaviors. As the VT detection count increases (e.g., VT~$\geq$~10), the overlap diminishes, and malware samples become more distinct and spatially separated from benign samples in the embedded space. This indicates that higher-threshold malware samples possess more distinguishable feature representations, likely reflecting stronger and more consistent malicious behaviors detected by a greater number of antivirus engines. Furthermore, the density of malware samples decreases as the threshold rises, aligning with the observed reduction in malware counts from the dataset.

\begin{figure}[!t]
    \centering
    \includegraphics[width=\linewidth]{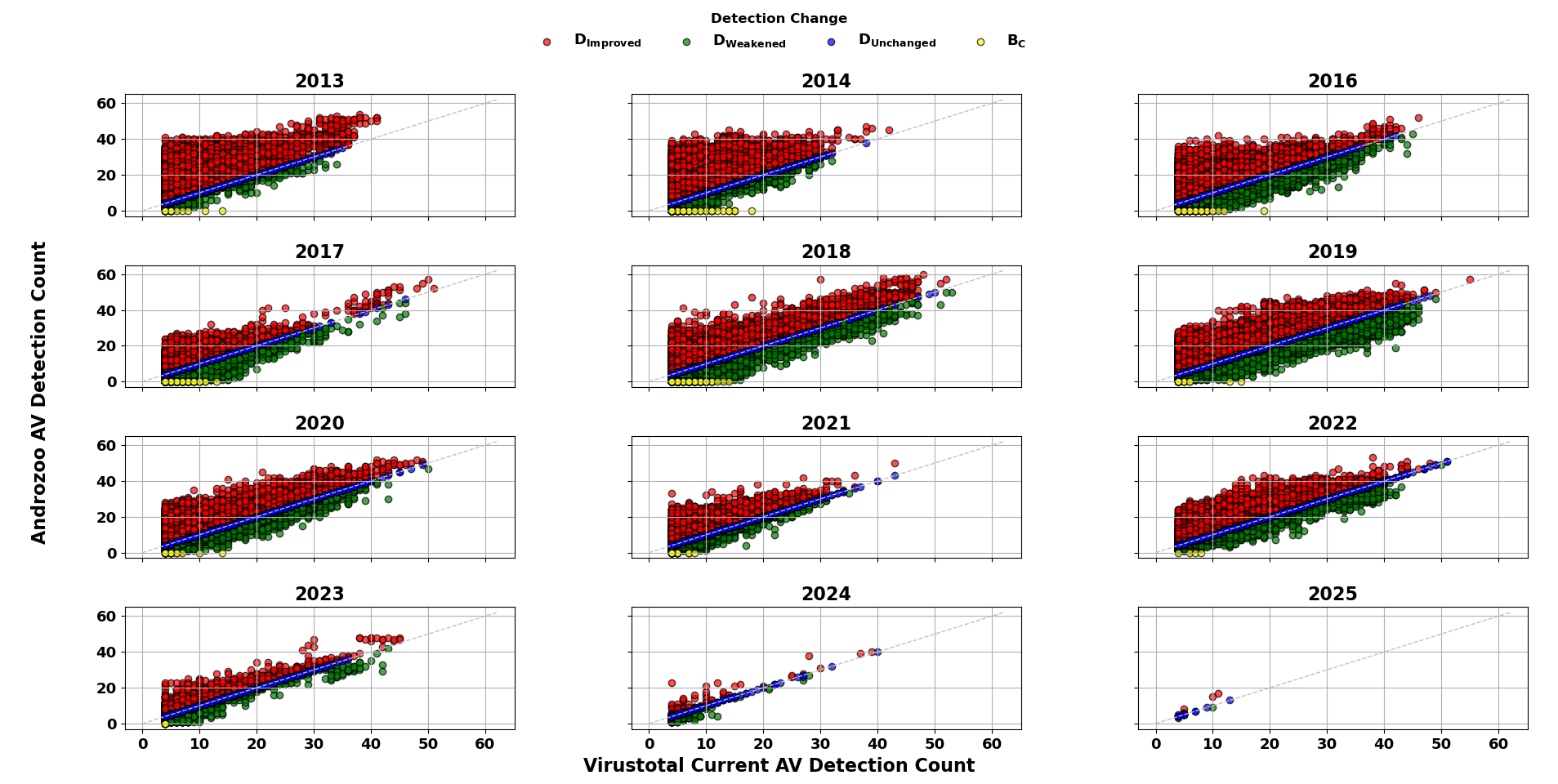}
    \caption{Malware AV detection drift over the years Virustotal vs AndroZoo Metadata. {\bf $B{_C}$}: Currently Labeled as Benign, {\bf $D_{Improved}$}: Improved Detection, {\bf $D_{Weakened}$}: Weakened Detection, {\bf $D_{Unchanged}$}: Unchanged Detection.}
    \label{fig:detection_drift_by_year}
\end{figure}

\begin{table}[ht]
\centering
\caption{F1 scores of the baseline malware detectors with varying VT thresholds.}
\label{tab:vt_threshold_f1}
\scriptsize
\begin{tabular}{llcccccccccc}
\toprule
\textbf{Split} & \textbf{Model} & \textbf{VT=4} & \textbf{VT=5} & \textbf{VT=6} & \textbf{VT=7} & \textbf{VT=8} & \textbf{VT=9} & \textbf{VT=10} & \textbf{VT=11} & \textbf{VT=12} & \textbf{VT=13} \\
\midrule
\multirow{4}{*}{IID}
& LightGBM & 0.8515 & 0.8566 & 0.8148 & 0.8397 & 0.8172 & 0.8511 & 0.8574 & 0.8364 & 0.6889 & 0.8385 \\
& MLP      & 0.8364 & 0.8173 & 0.8166 & 0.8319 & 0.8393 & 0.8232 & 0.8102 & 0.8628 & 0.7536 & 0.8623 \\
& SVM      & 0.7898 & 0.7954 & 0.7605 & 0.7543 & 0.8153 & 0.7492 & 0.7598 & 0.7353 & 0.6559 & 0.7462 \\
& XGBoost  & 0.8456 & 0.8028 & 0.8298 & 0.8255 & 0.8119 & 0.8370 & 0.8273 & 0.7844 & 0.7324 & 0.7538 \\
\midrule
\multirow{4}{*}{NEAR}
& LightGBM & 0.2423 & 0.2008 & 0.2308 & 0.1994 & 0.2470 & 0.1962 & 0.1956 & 0.2178 & 0.1639 & 0.2133 \\
& MLP      & 0.1971 & 0.1894 & 0.2191 & 0.2201 & 0.2223 & 0.2275 & 0.2249 & 0.1771 & 0.1867 & 0.2098 \\
& SVM      & 0.1330 & 0.1512 & 0.1569 & 0.1763 & 0.1971 & 0.2712 & 0.1386 & 0.1381 & 0.1261 & 0.1865 \\
& XGBoost  & 0.2221 & 0.2064 & 0.2703 & 0.2788 & 0.2761 & 0.1937 & 0.2788 & 0.2361 & 0.2563 & 0.2241 \\
\midrule
\multirow{4}{*}{FAR}
& LightGBM & 0.1284 & 0.1477 & 0.1559 & 0.1039 & 0.0852 & 0.1562 & 0.1551 & 0.0935 & 0.0511 & 0.2306 \\
& MLP      & 0.1284 & 0.2716 & 0.0918 & 0.0721 & 0.1362 & 0.1873 & 0.1954 & 0.0841 & 0.0912 & 0.0947 \\
& SVM      & 0.2393 & 0.2272 & 0.1503 & 0.0655 & 0.0938 & 0.1104 & 0.1322 & 0.0540 & 0.0508 & 0.0706 \\
& XGBoost  & 0.1560 & 0.3513 & 0.2926 & 0.2613 & 0.1971 & 0.1787 & 0.2133 & 0.1067 & 0.1385 & 0.1452 \\
\bottomrule
\end{tabular}
\end{table}

\begin{figure}[ht]
    \centering
    \includegraphics[width=0.95\textwidth]{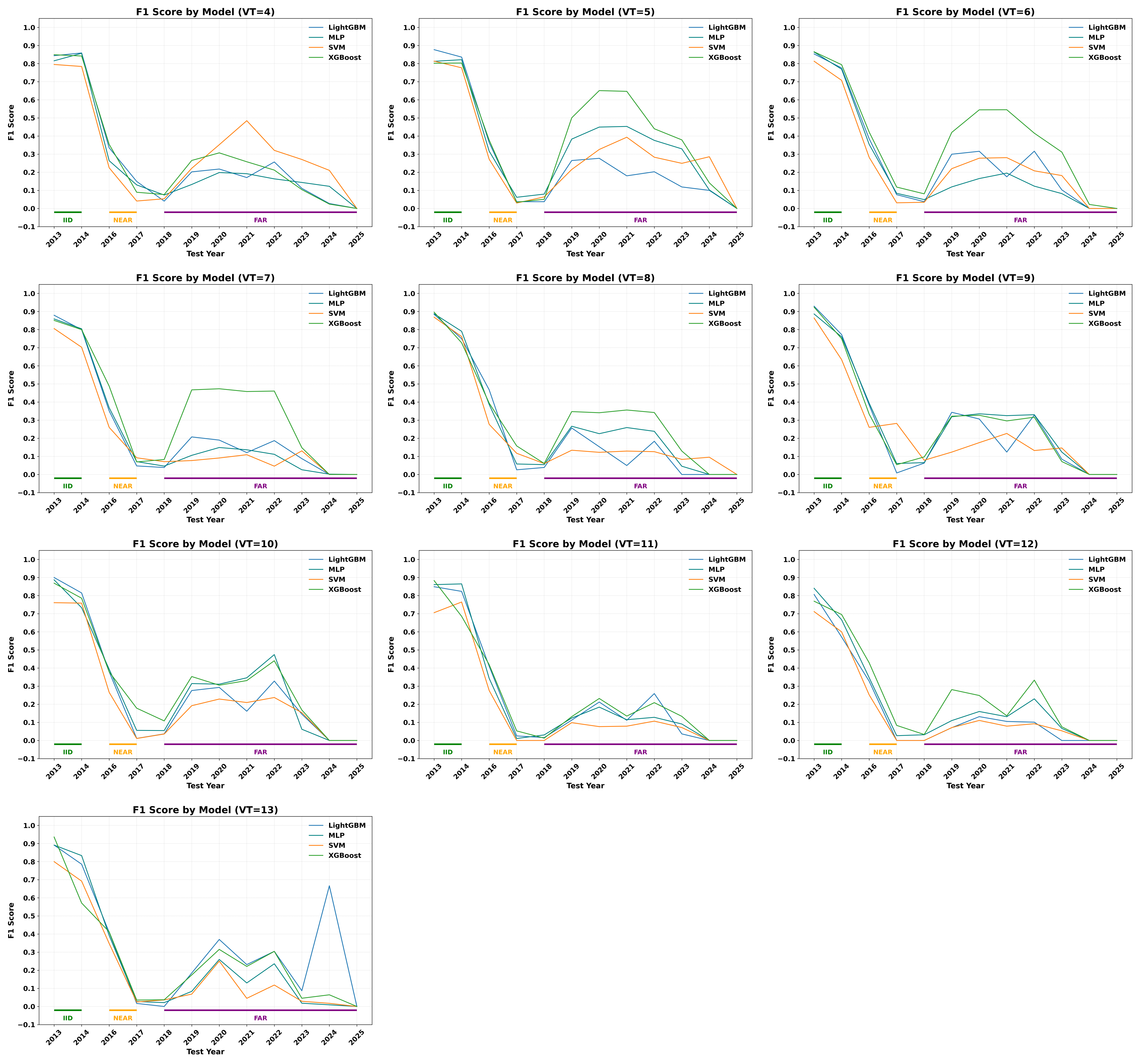}
    \caption{Combined F1 score plots for VT thresholds 4 to 13.}
    \label{fig:combined-f1-scores}
\end{figure}

\begin{figure*}[!t]
    \centering
    \begin{subfigure}[t]{0.32\textwidth}
        \centering
        \includegraphics[width=\linewidth]{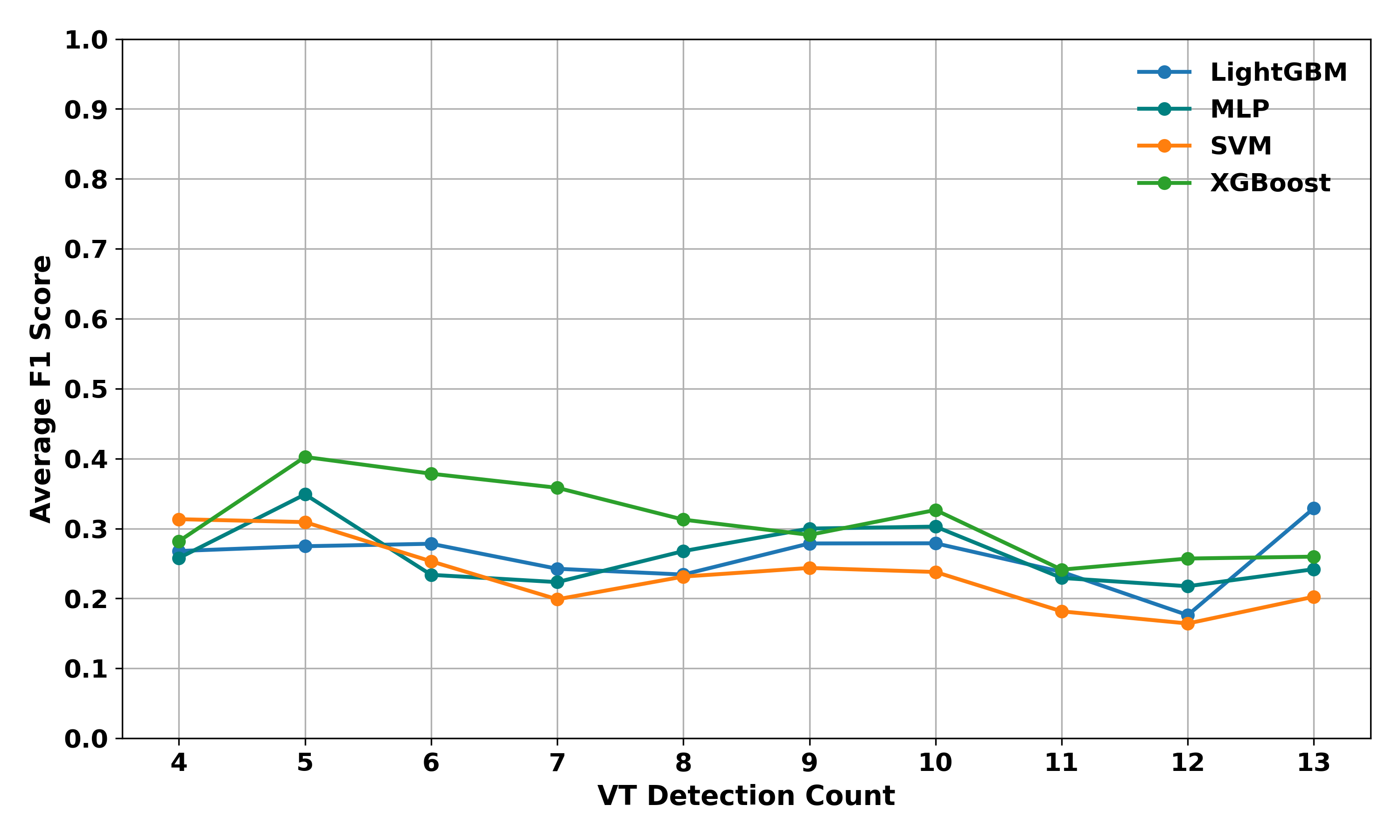}
        \caption{}
        \label{fig:avg_f1_vt}
    \end{subfigure}
    \hfill
    \begin{subfigure}[t]{0.32\textwidth}
        \centering
        \includegraphics[width=\linewidth]{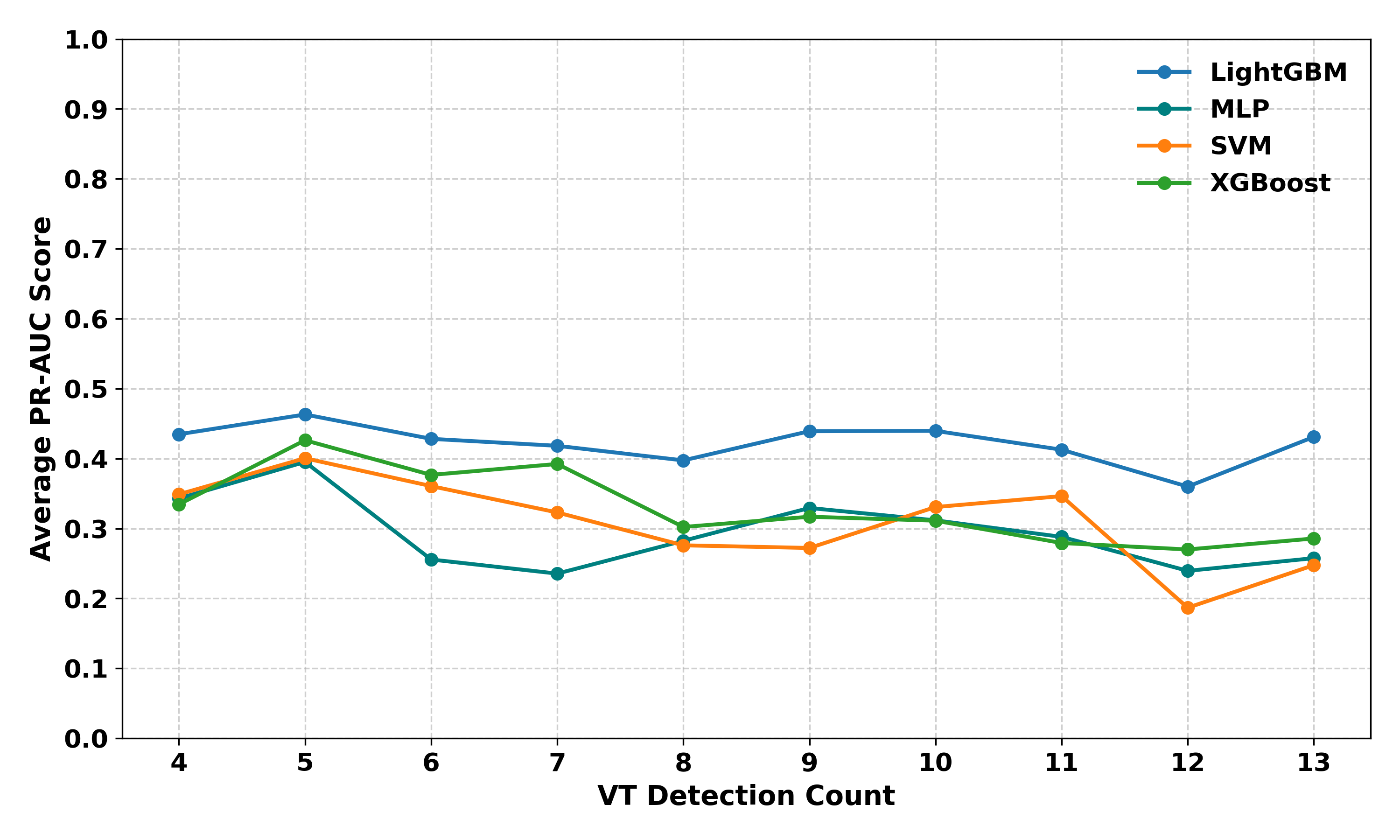}
        \caption{}
        \label{fig:avg_pr_auc_vt}
    \end{subfigure}
    \hfill
    \begin{subfigure}[t]{0.32\textwidth}
        \centering
        \includegraphics[width=\linewidth]{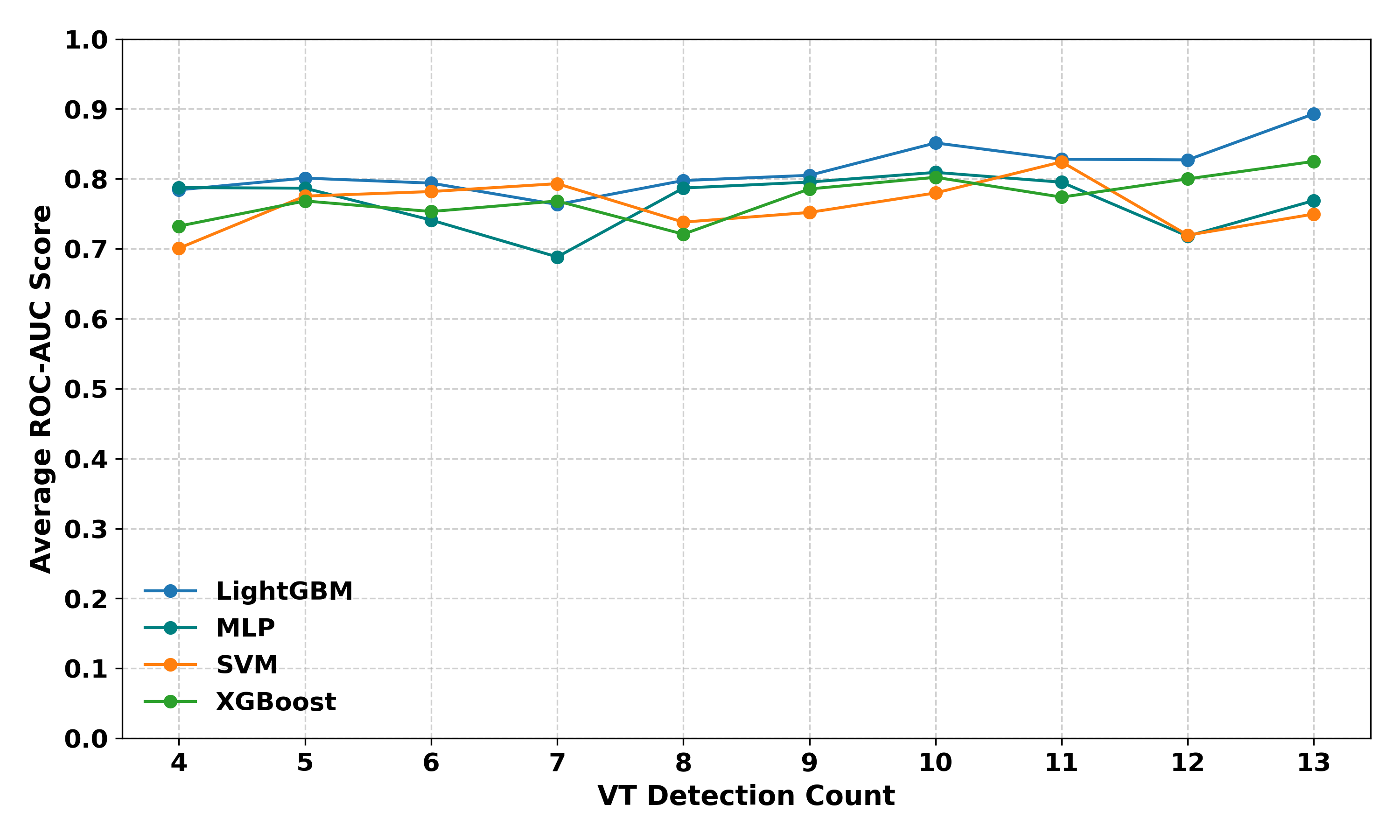}
        \caption{}
        \label{fig:avg_roc_auc_vt}
    \end{subfigure}
    \caption{Comparison of model performance across VirusTotal detection thresholds for (a) F1-score, (b) PR-AUC, and (c) ROC-AUC.}
    \label{fig:avg_vt_score_comparison}
\end{figure*}

To assess the impact of label noise on malware detection, we conduct a set of experiment varying the VirusTotal (VT) labeling Threshold. We first create a set of LAMDA datasets, where in each dataset contain \emph{all benign samples} and \emph{only malware samples with a specific VT labeling count}. For example, for the first dataset, we keep all benign samples and those malware samples that were flagged exactly by $4$ antivirus engines (VT=4). In the next dataset, we include malware samples with VT=5, and so on, up to VT=13. This resulted in creating ten separate LAMDA dataset variants, each reflecting a different level of confidence in the AV engines.

Next, We evaluate standard malware detectors (LightGBM, MLP, SVM, and XGBoost) on these datasets using the \textit{AnoShift}-style splits, which simulate temporal concept drift by training on IID split and tested on NEAR and FAR splits. Each malware detector's performance is evaluated using F1-score metric.

Table~\ref{tab:vt_threshold_f1} presents performance details of the baseline malware detectors using F1-scores metric with varying VT count. We made the following observations. LightGBM with VT=4, we observe an F1-score of $0.8515$ under the IID split, however it drops significantly to $0.2423$ on NEAR and to $0.1284$ on FAR splits. This decline highlights the degradation of malware detector over time. 

Figure~\ref{fig:avg_vt_score_comparison} illustrates the average F1, PR-AUC, and ROC-AUC scores of the baseline malware detectors with varying VT labeling threshold. While Figure~\ref{fig:avg_f1_vt} is the summarization of Figure~\ref{fig:combined-f1-scores}, subplots (a), (b) and (c) illustrate how performance metrics vary when the VT threshold is set to different values. Across all these three evaluation metrics, we observe only minor differences in baseline malware detectors performance. This suggests that, varying the VT labeling threshold has minimal impact on baseline accuracy. This result suggests that the primary causes of performance degradation in our main experiments may not be the labeling noise from VT, but rather factors such as temporal concept drift and class imbalance. 
The observed performance degradation is more likely attributed to the distributional shifts over time, reinforcing the relevance of concept drift in real-world malware detection scenarios.


\section{Label Drift Across Years Based on VirusTotal Label Changes}
\label{appendix:label-drift}
Prior works ~\cite{zhu2020benchmarking,zhu2020measuring} show that a significant portion of malware samples are observed to changes their labels over time, with some initially labeled as malicious later reclassified as benign, and vice versa. Based on these studies, we analyze how sample labels have evolved within our dataset, using VirusTotal and AndroZoo as reference points. Table~\ref{tab:vt_detection_summary} summarizes how malware sample labels have changed over time according to VirusTotal and AndroZoo metadata. The table depicts the yearly statistics from 2013 to 2025 (excluding 2015), including the number of samples now considered benign, and how many show improved, weakened, or unchanged detection. The \textit{significant drop} and \textit{significantly increased} columns highlight cases with drastic shifts (more than 50\%) in detection confidence. 

\begin{table*}[!t]
\centering
\caption{This table summarizes label drift for Android malware samples, highlighting shifts in detection status over time. {\bf TS}: Total \#of Malware Samples, {\bf $B{_C}$}: Currently Labeled as Benign, {\bf $\%B{_C}$}: Percentage of Total Malware Samples Currently Labeled as Benign. {\bf $D_{Improved}$}: Improved Detection, {\bf $D_{Weakened}$}: Weakened Detection, {\bf $D_{Unchanged}$}: Unchanged Detection, {\bf $DS_{Drop}$}: Significant Drop of Detection Count, \textbf{$DS_{Improve}$}: Detection Count Significantly Increased.}

\label{tab:vt_detection_summary}
\resizebox{\textwidth}{!}{
\begin{tabular}{c|c|c|c|c|c|c|c|c}
\toprule
\textbf{Year} & \textbf{TS} & \textbf{$B{_C}$} & \textbf{$\%B{_C}$} & \textbf{$D_{Improved}$} & \textbf{$D_{Weakened}$} & \textbf{$D_{Unchanged}$} & \textbf{$DS_{Drop}$} & \textbf{$DS_{Increase}$} \\
\midrule
2013 & 44383 & 24 & 0.05 & 40436 & 439 & 3484 & 85 & 34481 \\ \hline
2014 & 45756 & 345 & 0.75 & 37108 & 1554 & 6749 & 863 & 27239 \\ \hline
2016 & 45134 & 177 & 0.39 & 26963 & 7485 & 10509 & 1160 & 13581 \\ \hline
2017 & 21359 & 1108 & 5.19 & 7765 & 10289 & 2197 & 5061 & 3362 \\ \hline
2018 & 39350 & 1242 & 3.16 & 17561 & 15346 & 5201 & 7304 & 7600 \\ \hline
2019 & 41585 & 22 & 0.05 & 22905 & 9294 & 9364 & 467 & 7518 \\ \hline
2020 & 46355 & 25 & 0.05 & 20755 & 3931 & 21644 & 294 & 8001 \\ \hline
2021 & 35627 & 23 & 0.06 & 10385 & 4482 & 20737 & 176 & 2531 \\ \hline
2022 & 41648 & 4 & 0.01 & 10445 & 3629 & 27570 & 121 & 2719 \\ \hline
2023 & 7892 & 15 & 0.19 & 1763 & 1979 & 4135 & 592 & 416 \\ \hline
2024 & 794 & 0 & 0.00 & 74 & 319 & 401 & 79 & 19 \\ \hline
2025 & 23 & 0 & 0.00 & 6 & 2 & 15 & 0 & 2 \\ 

\bottomrule
\end{tabular}
}
\end{table*}

In addition to the table, we include a visualization of how detection counts have changed over time. As shown in Figure~\ref{fig:detection_drift_by_year}, it clearly illustrates how metadata from prior state-of-the-art diverges from current data. This drift contributes to label uncertainty; for instance, LAMDA includes approximately 2,985 samples in Table~\ref{tab:vt_detection_unknown} as unknown family. LAMDA captures this real-world variability, highlighting the importance of regularly updating metadata to minimize label noise and maintain the reliability of malware analysis.

To the best of our knowledge, LAMDA is the only public android malware dataset that tracks label drift, showing how “malicious” and “benign” verdicts shift as antivirus engines advance. By updating the metadata for over a million of samples, it allows researchers to avoid relying on outdated labels from prior works. LAMDA also highlights a critical issue in malware research: static labels can quickly become obsolete, and updating metadata over time is must.

\section{Scalability of LAMDA}
\label{appendix:scalability}

To support long-term use and extensibility, we have designed \textsc{LAMDA} with scalability in mind. In the context of LAMDA, scalability refers to the extensibility of the dataset—specifically, its ability to be easily expanded with new samples. We have published three variants of LAMDA on the \texttt{HuggingFace} repository, each supporting a different \texttt{VarianceThreshold} configuration. The dataset creation process begins by splitting the static feature files (i.e., \texttt{.data} files extracted from each APK) into stratified train and test splits. From the training split, we collect the global set of all unique tokens (i.e., features), encode both train and test samples into binary vectors in this raw feature space, and apply \texttt{VarianceThreshold} to select high-variance features from the training data. The same selected features are then applied to the test data using the saved threshold \texttt{object}.

We publish the following artifacts to facilitate scalability: raw feature matrices (before thresholding), reduced feature matrices (after thresholding), and the serialized \texttt{VarianceThreshold} object (in \texttt{joblib} format). Using these resources and the accompanying codebase, researchers can seamlessly extend \textsc{LAMDA} by collecting newer APKs, extracting static features, encoding them, and applying the same thresholding object to map them into \textsc{LAMDA}’s feature space. While it is not feasible to add new samples to the training set—because doing so would alter the global vocabulary and invalidate the original thresholding, researchers can add test-time samples for evaluation. This supports drift detection on newer and future malware variants without requiring retraining. Thus, \textsc{LAMDA} enables reproducible research and practical testing of detection models against evolving threats.



\section{Concept Drift Adaptation on LAMDA}
\label{appendix:comparison-sota}

Concept drift adaptation remains a significant challenge in Android malware detection. As Android malware evolves, detection models must adapt to shifting feature distributions and the emergence of new malware. LAMDA captures this evolution over a 12-year span, offering a realistic and temporally diverse benchmark for evaluating concept drift adaptation techniques.

We evaluate CADE~\cite{cade}, a concept drift adaptation method on the API Graph dataset~\cite{api_graph_dataset} and LAMDA. However, API Graph only spans 2012 to 2018 and lacks sufficient representation of recent malware evolution. To assess robustness under more severe and recent drift, we compare this method on both datasets.

\begin{figure}[ht]
    \centering
    \begin{tabular}{ccc}
        \includegraphics[width=0.3\textwidth]{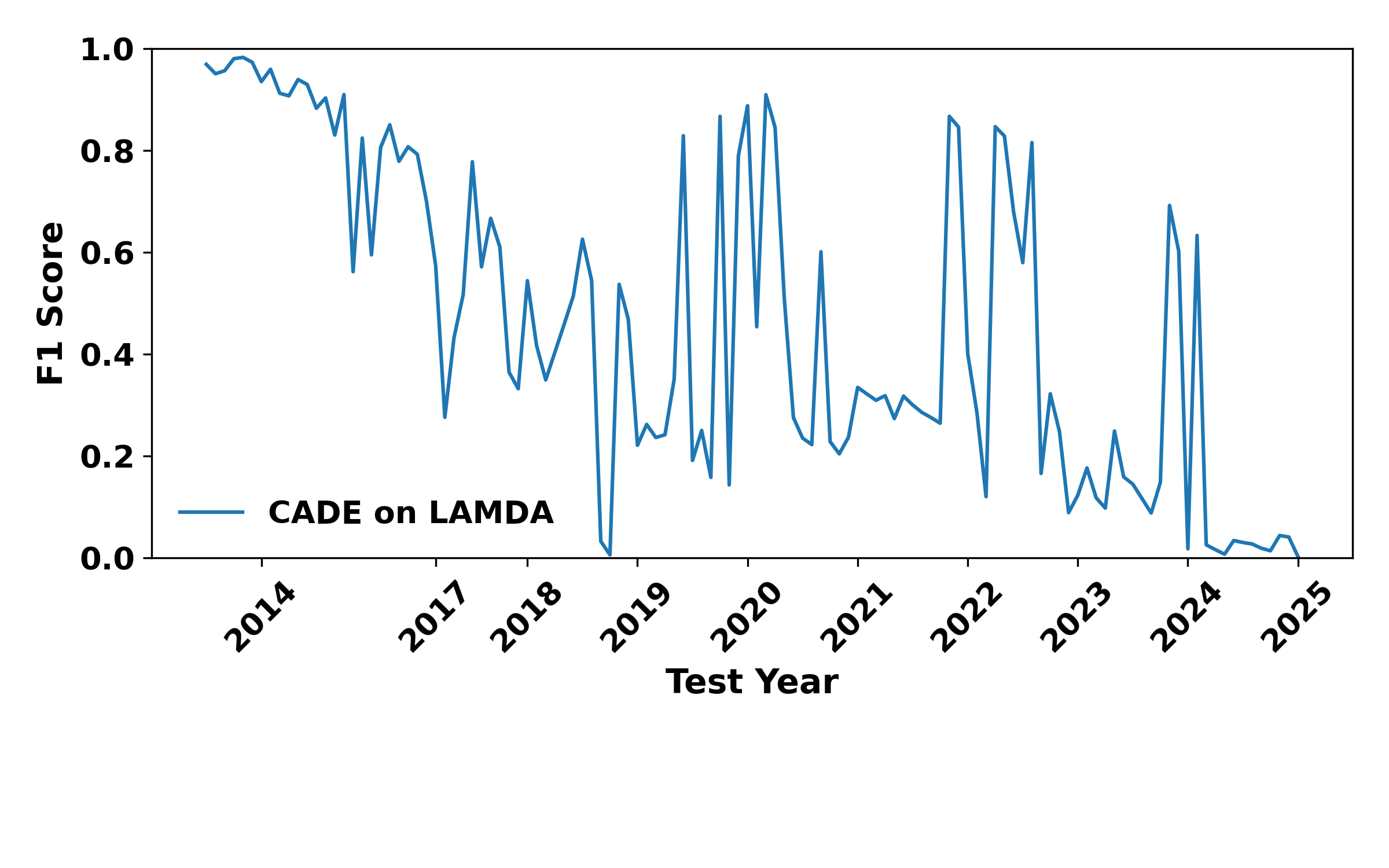} &
        \includegraphics[width=0.3\textwidth]{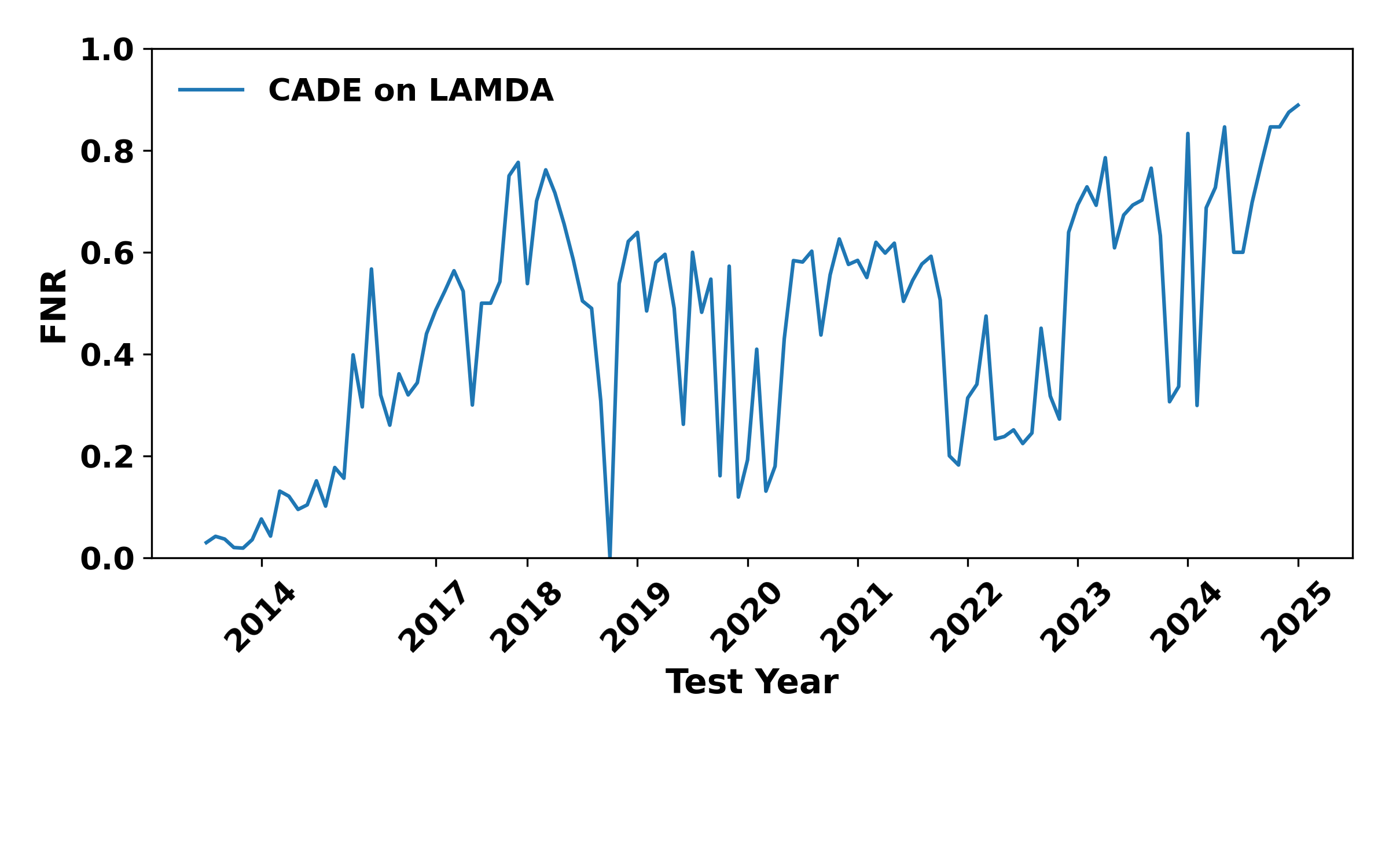} &
        \includegraphics[width=0.3\textwidth]{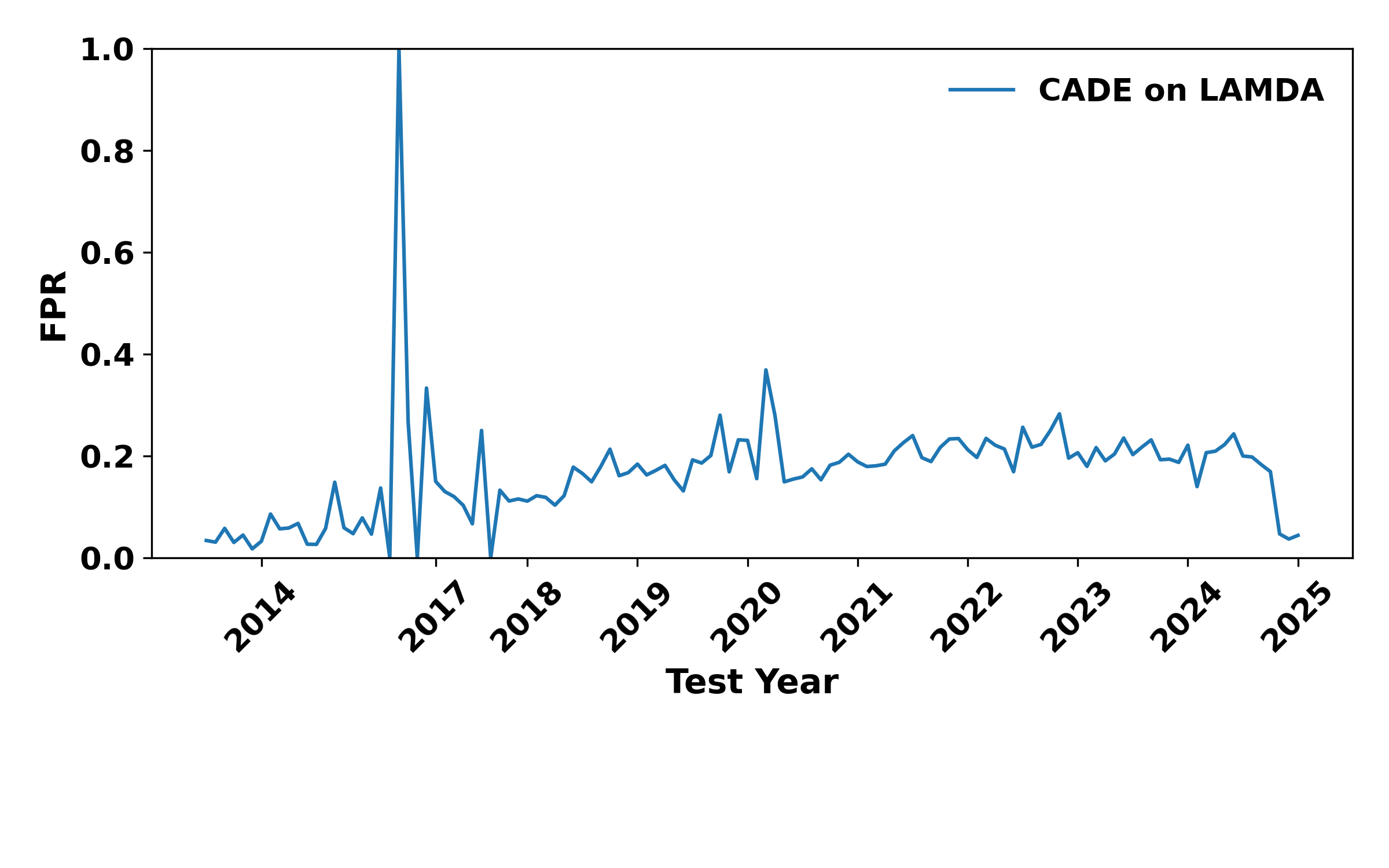} \\
        \includegraphics[width=0.3\textwidth]{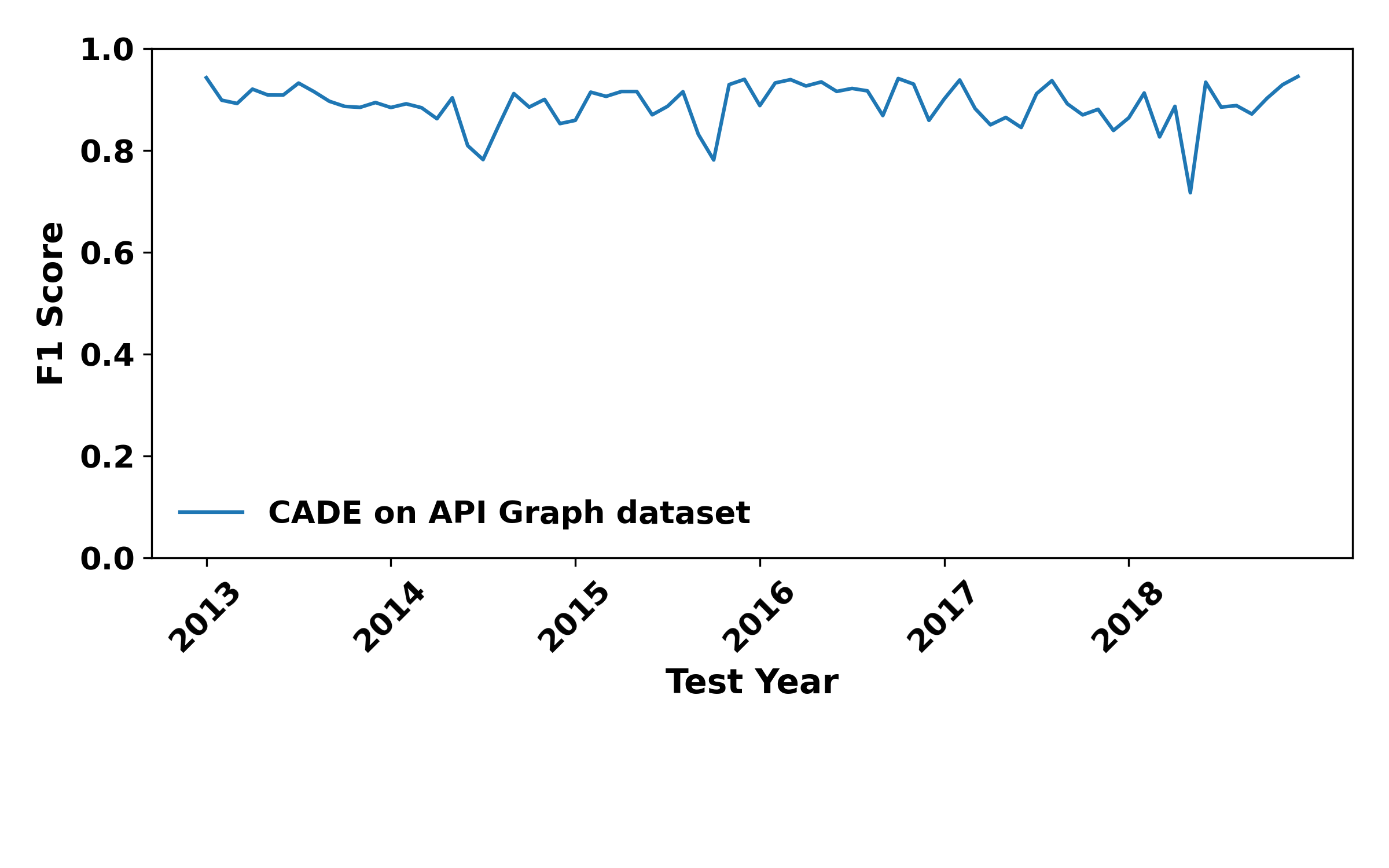 } &
        \includegraphics[width=0.3\textwidth]{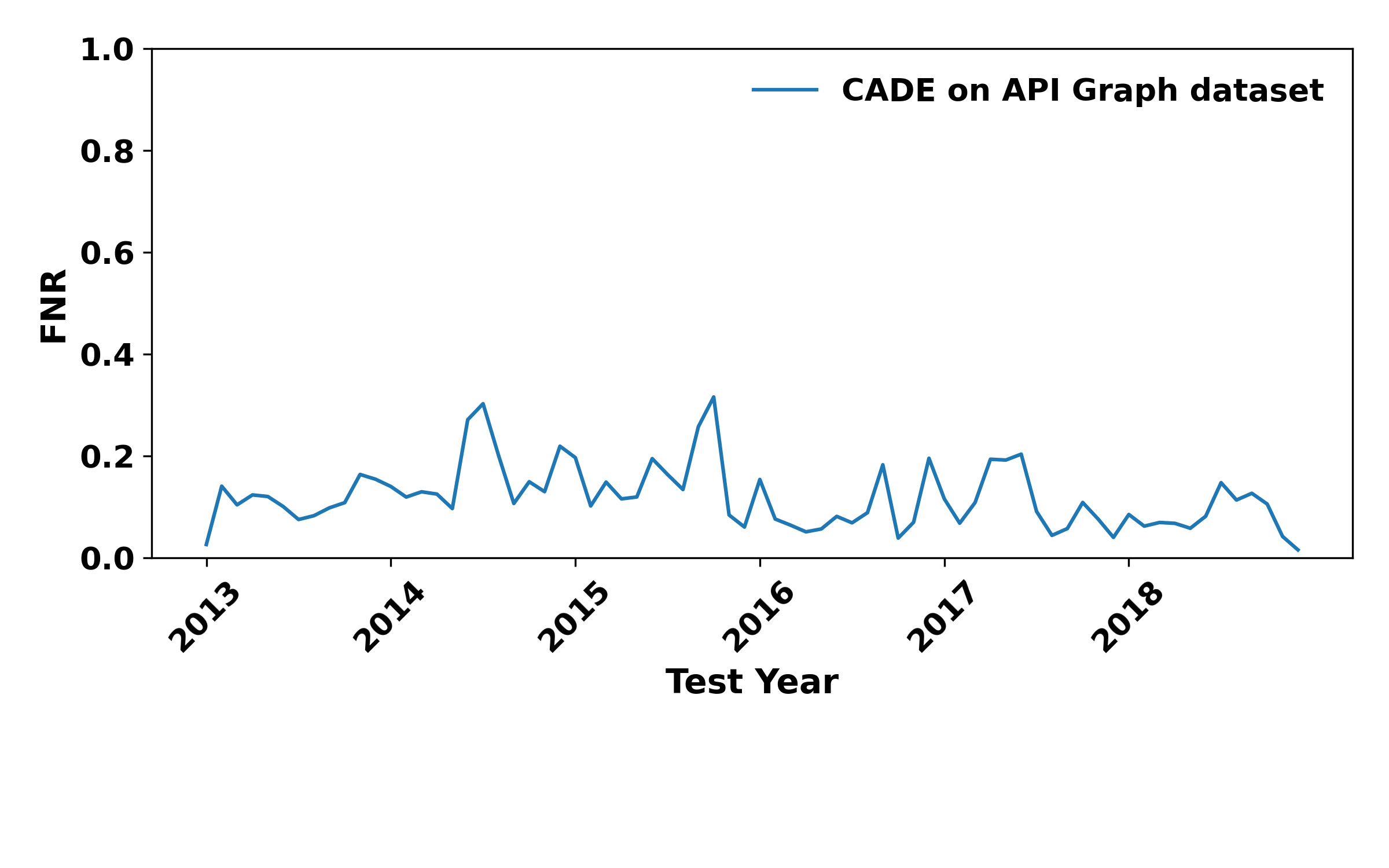} &
        \includegraphics[width=0.3\textwidth]{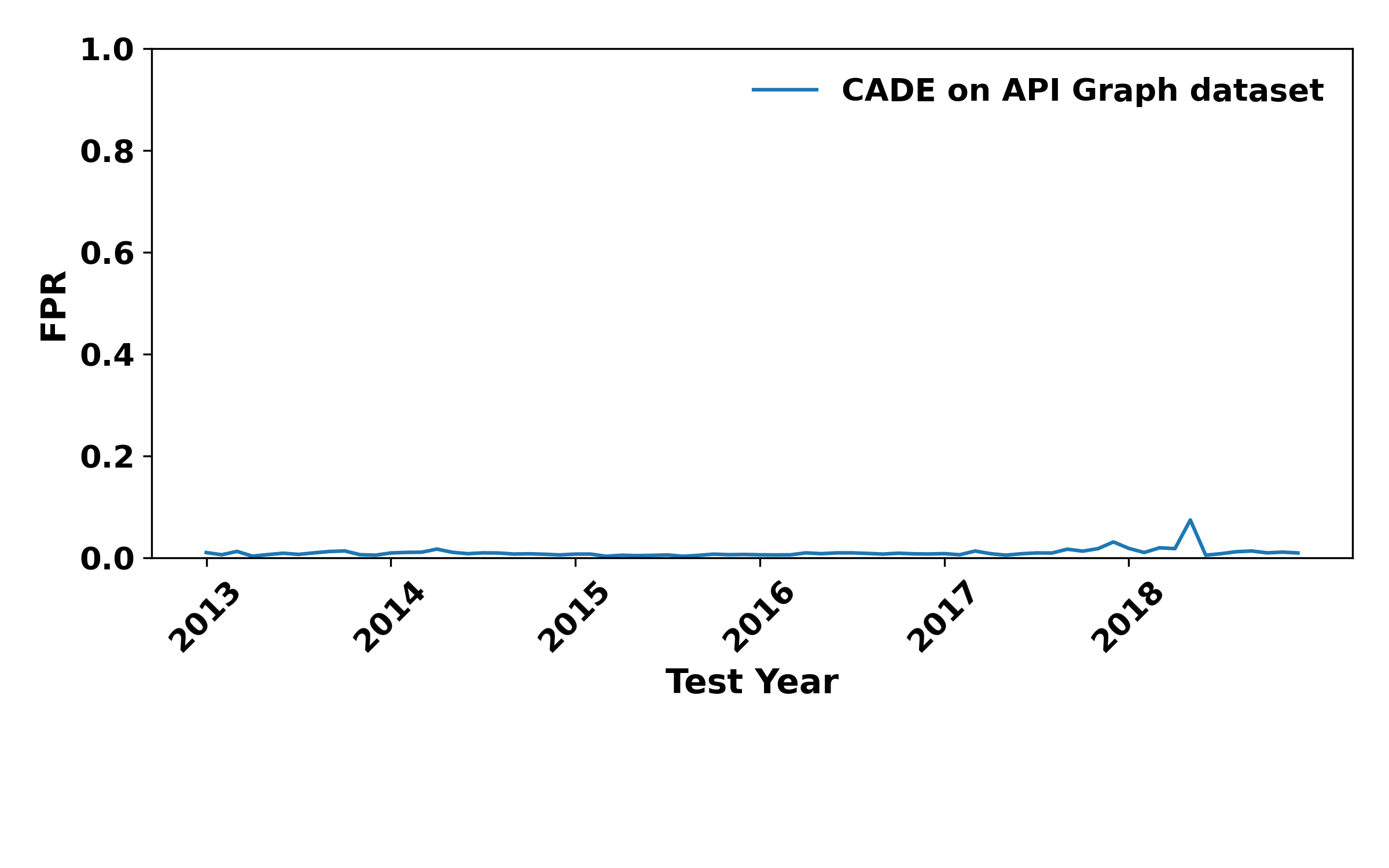} \\
    \end{tabular}
    \vspace{-0.4cm}
    \caption{Comparison with F1-score, False negative rate (FNR) and False positive rate (FPR) of CADE on LAMDA and API Graph dataset.}
    \label{fig:cade_f1_fnr_fpr_comparison}
\end{figure}

Figure~\ref{fig:cade_f1_fnr_fpr_comparison} demonstrates that while CADE~\cite{cade} performs strongly on the API Graph dataset, its effectiveness declines when applied to LAMDA. On API Graph, CADE achieves an impressive F1-score of 0.8904 with a low false negative rate (FNR) of 0.1191 and false positive rate (FPR) of 0.0101, reflecting its strong capability in handling moderate concept drift. However, when evaluated on LAMDA that spans a longer temporal range and introduces more severe drift—CADE's performance deteriorates significantly, with the F1-score dropping to 0.4407, the FNR rising to 0.4734, and the FPR increasing to 0.1729. These results highlight that existing adaptation methods struggle to generalize under the more challenging and realistic drift scenarios captured by LAMDA, emphasizing the need for more robust drift-aware detection techniques.


 
\begin{table}[!t]
\centering
\caption{Performance comparison of 
CADE~\cite{cade} 
on API Graph and LAMDA datasets. Reported as average of each metric.}
\label{tab:chen_cade_comparison}
\begin{tabular}{llccc}
\toprule
\textbf{Dataset} & \textbf{Method} & \textbf{F1-score} & \textbf{FNR} & \textbf{FPR} \\
\midrule
API Graph & CADE    & 0.8904 & 0.1191 & 0.0101 \\
LAMDA & CADE    & 0.4407 & 0.4734 & 0.1729 \\
\bottomrule
\end{tabular}
\end{table}

These findings reinforce the necessity of LAMDA to study longitudinal and modern threats, and provide a stronger foundation for evaluating concept drift adaptation methods in real-world malware detection settings.

\section{Temporal Analysis of SHAP-Based Explanation Drift (Top 1000 Features)}
\label{appendix:explanation-drift}

\begin{figure}
\centering
\begin{minipage}{\textwidth}
    \centering
        \begin{subfigure}{0.45\textwidth}
        \centering
        \includegraphics[width=\textwidth]{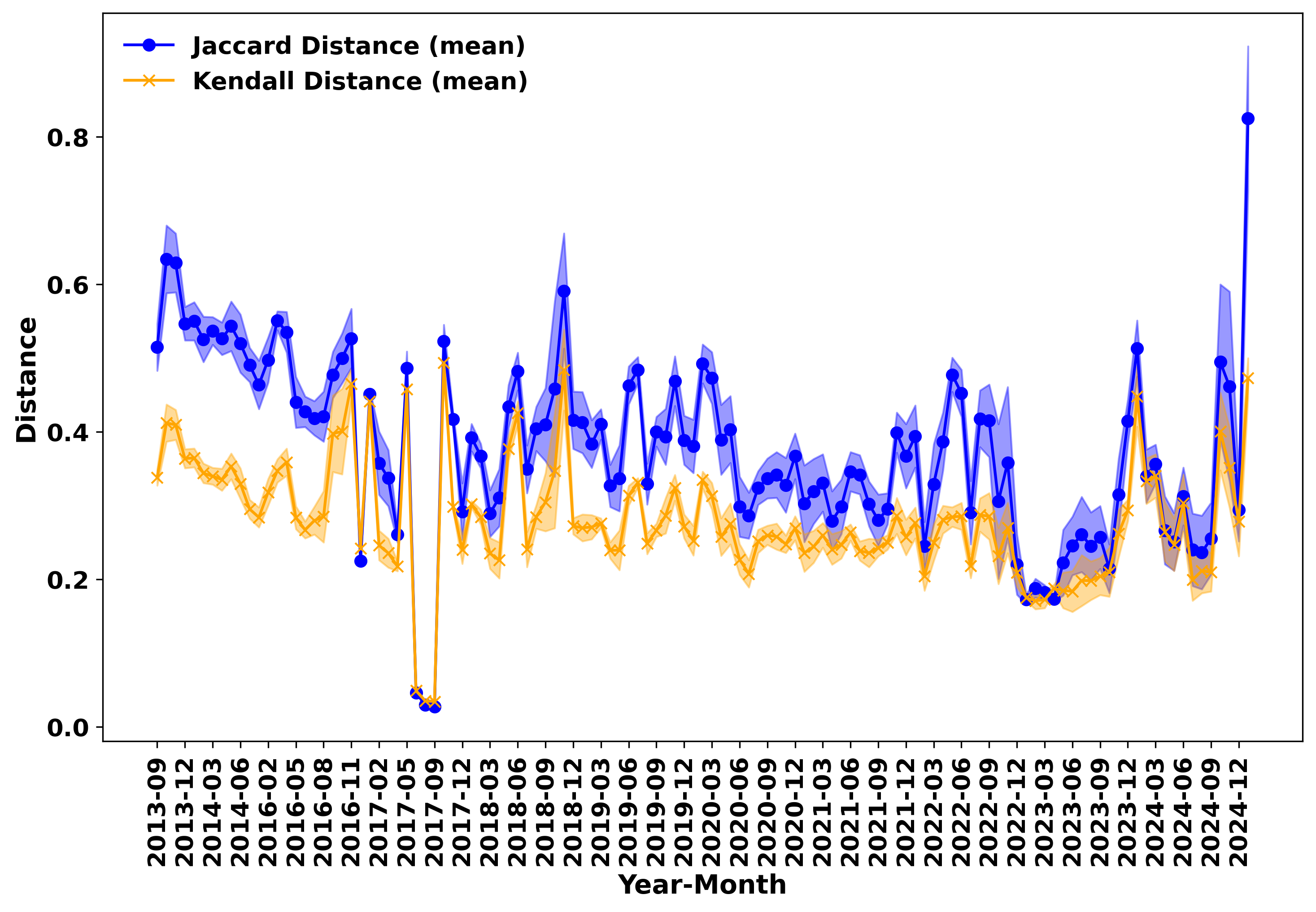}
        \caption{Analysis on LAMDA dataset from 2013 to 2024.}
        \label{fig:explanation-drift-analysis-on-lamda-dataset-1000-features}
    \end{subfigure}
    \hfill
    \begin{subfigure}{0.45\textwidth}
        \centering
        \includegraphics[width=\textwidth]{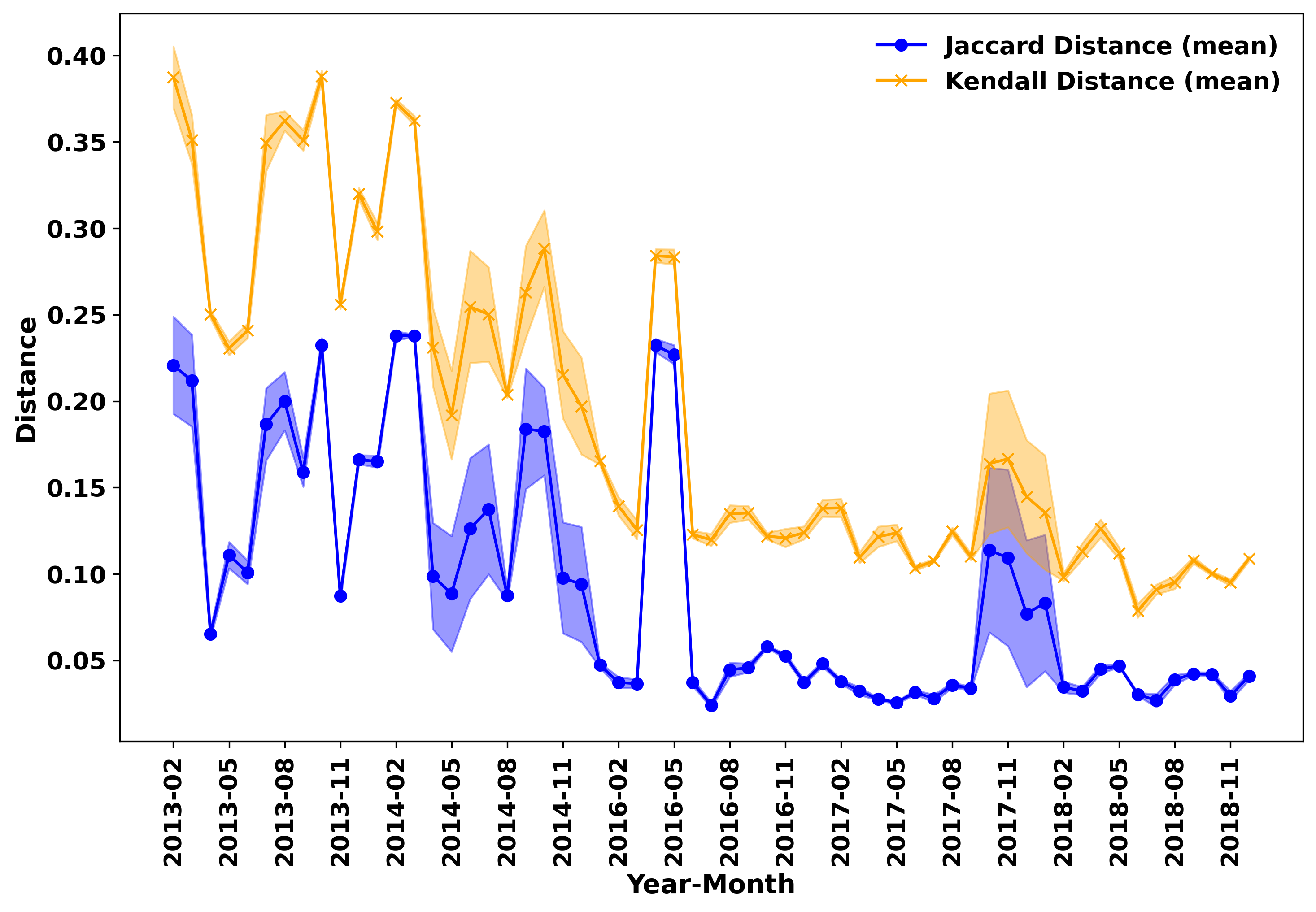}
        \caption{Analysis on APIGraph dataset from 2013 to 2018.}
        \label{fig:explanation-drift-analysis-on-apigraph-dataset-1000-features}
    \end{subfigure}
    \hfill
\end{minipage}
\caption{Monthly explanation drift measured by Jaccard and Kendall distances between top-1000 SHAP features.}
\label{fig:shap-based-monthly-explanation-drift-analysis-1000-features}
\vspace{-0.4cm}
\end{figure}


\subsection{Analysis Setting}
For the analysis of explanation drift using the top-1000 SHAP features, we adopt the same experimental configuration as the top-100 feature setting. Specifically, we use an MLP-based malware classifier following the architecture in Chen-AL.~\cite{chen2023continuous}, and compute SHAP values using the KernelExplainer with 100 background samples and 100 test samples per month. Each monthly run is repeated 5 times to ensure statistical stability. The only difference in this setting is that distances are computed based on the top-1000 ranked SHAP features instead of the top-100.

\subsection{Jaccard and Kendall Distance Analysis}
To validate the effectiveness of our proposed dataset LAMDA we also compute Jaccard and Kendall distances across monthly SHAP attributions using the top-1000 most important features. As shown in Figure~\ref{fig:explanation-drift-analysis-on-lamda-dataset-1000-features}, the LAMDA dataset exhibits moderate to high Jaccard distance values with frequent spikes, especially in early 2017, 2018, and 2024, indicating substantial shifts in the broader set of influential features. The Kendall distances remain relatively stable yet fluctuate noticeably in the same periods, suggesting not only changes in which features matter, but also in their relative rankings. This behavior implies strong temporal variability in the model’s decision logic driven by evolving data patterns. Conversely, the APIGraph dataset, illustrated in Figure~\ref{fig:explanation-drift-analysis-on-apigraph-dataset-1000-features}, demonstrates much lower overall distance values, with both Jaccard and Kendall distances remaining mostly below 0.3. This pattern indicates limited month-to-month variability in explanation behavior, reflecting a more static and homogeneous data distribution over time~\cite{api_graph_dataset}. These trends further highlight the superior effectiveness of the LAMDA dataset over APIGraph, as SHAP explanations~\cite{shap} more clearly capture the dynamic shifts in feature importance and explanation drift across time in LAMDA, reflecting its richer temporal variability and evolving behavior patterns.

\subsection{LAMDA and APIGraph Comparative Analysis}
From the 1000-feature analysis, it is evident that the LAMDA dataset presents a more dynamic and variable model explanation pattern than the APIGraph dataset. The broader feature set allows us to detect finer-grained shifts in the importance landscape, and the higher variance and frequent spikes in Jaccard and Kendall distances for LAMDA (Figure~\ref{fig:explanation-drift-analysis-on-lamda-dataset-1000-features}) suggest that it better reflects evolving malware behaviors and concept drift. In contrast, the APIGraph dataset (Figure~\ref{fig:explanation-drift-analysis-on-apigraph-dataset-1000-features}) maintains a relatively stable feature importance distribution across time, with limited explanation diversity. These findings reinforce the conclusion that LAMDA is more suitable for benchmarking adaptive models that need to operate effectively under changing threat environments and shifting decision boundaries.

\section{Continual Learning on LAMDA}
\label{appendix:cl-exps}


In real-world settings, a large number of new benign and malicious Android applications are introduced each year. As a result, both benign (e.g., due to changes in user demands, Android APIs, security practices) and malicious (e.g., the emergence of novel malware variants) behaviors evolve over time, leading to concept drift. This makes it challenging for static machine learning models to maintain reliable performance over time without retraining regularly. However, complete retraining of past data becomes impractical due to the massive volume of Android applications released daily and the high computational cost associated with retraining. On the other hand, training solely on recent data often leads to catastrophic forgetting ~\cite{DeLange2019ACL,continual-learning-malware}, where previously acquired knowledge is overwritten or lost. In such a situation, continual learning (CL) offers a compelling solution by enabling the detection models to adapt incrementally to new benign and malware applications without the need to retrain with all past data ~\cite{malcl,continual-learning-malware,ghiani2025understanding}. However, some CL techniques may require access to a small subset of past data.

\subsection{LAMDA for Continual Learning}
LAMDA can be a natural choice for benchmarking CL due to several key properties of its design and structure:

\begin{enumerate}[label=\Roman*.]
    \item \textbf{Temporal granularity:} It spans over a decade (2013--2025, excluding 2015) with available both monthly and yearly splits, allowing custom CL as per need.
    \item \textbf{Concept drift:} As shown in Section~4, LAMDA exhibits significant distributional changes over time, both in feature and label space.
    
    \item \textbf{Flexible task construction:}
    \begin{itemize}[label=--]
        \item \textbf{Domain-IL:} Using yearly data splits while maintaining a consistent malware or benign labeling.
        \item \textbf{Class-IL:} Leveraging AVClass2-labeled malware families to incrementally expand the label space. 
    \end{itemize}
    
    \item \textbf{Real world relevance:} LAMDA is derived from real-world Android APKs and VirusTotal reports, introducing authentic drift and noise.
\end{enumerate}

We evaluate CL on the LAMDA benchmark using two established baselines, {\em Naive} (i.e., None) and {\em Joint}, inspired by the prior work~\cite{continual-learning-malware,ghiani2025understanding}. 
Additionally, we include {\em Replay} (i.e., Experience Replay)~\cite{er}, a state-of-the-art memory replay based CL method, configured with a buffer size of 200 samples per experience. The Naive baseline trains the model sequentially on each experience or task without any access to past data, while the Joint baseline retrains the model from scratch using the cumulative data observed up to the current experience or task.

These baselines are tested under two settings: Domain Incremental Learning ({\em Domain-IL}), which involves binary malware \textit{vs} benign classification across yearly tasks, and Class Incremental Learning ({\em Class-IL}), where each task introduces new malware families to classify~\cite{van2022three,continual-learning-malware}. Due to the lack of available prior work that can assign a single behavioral label, we didn't consider Task Incremental Learning ({\em Task-IL}) in our experimental setups. 

We define each experience or task in the Domain-IL experiment as all samples (both benign and malicious) collected within a specific calendar year (e.g., 2013, 2014, ..., 2025). However, for the Class-IL experiments, each experience or task consists of only the malware samples collected during the corresponding year. 

\subsection{Continual Learning Experimental Setup}

\paragraph{Domain-IL.} In this setting, each experience or task corresponds to samples collected during a specific year (i.e., 2013, 2014, ... and so on). The model is designed to continuously learn to distinguish between malware and benign samples as the data distribution evolves over time. We use the \texttt{Baseline} variant of our published dataset, treating each year as a separate task in the learning sequence. The objective is for the model to adapt and maintain accurate binary classification performance despite the temporal distribution shifts.

\paragraph{Class-IL.}
In this setting, we utilize a different dataset derived from the \texttt{Baseline} variant of our published dataset. We selected only those malware families that contained more than 10 samples in the test set, resulting in a total of 154 families for our experiment. Consequently, we excluded the year 2025 from our experiments, as no family in that split met the minimum sample threshold. Additionally, we omit standard class incremental learning~\cite{malcl,continual-learning-malware}, where entirely new classes are introduced in each experience. This approach does not reflect how malware appears in real-world scenarios, malicious samples often come from a mix of previously seen and new families. This claim is supported by the analysis presented in Table~\ref{tab:lamda_metadata}. The model is expected to learn incrementally to classify samples across all malware families encountered.

\paragraph{Model Architecture.} We use a shared base architecture, a multi-layer perceptron (MLP) for both Class-IL and Domain-IL settings, consisting of four hidden layers --- 512, 384, 256, 128, with ReLU activation. However, Task-specific heads are added to support each learning scenario. For Class-IL, we add a single linear layer outputting logits for all classes and train with categorical cross-entropy loss. For Domain-IL, we use a two-layer MLP head (100 units each, with dropout p=0.2) and a final sigmoid output, trained with binary cross-entropy. All networks are optimized with SGD (learning-rate 0.01, momentum 0.9, weight-decay 0.000001).

\paragraph{Evaluation Metrics.} We evaluate classification performance using F1 score, which is the harmonic mean of precision and recall. It provides a balanced measure of a model's predictions, particularly important in {\em imbalanced} datasets. Following the prior work ~\cite{ghiani2025understanding}, and we compute the F1 score after training on the $k$-th experience using two complementary evaluation modes:

\begin{itemize}
    \item \textbf{Backward Transfer Performance}: We measures the model's ability to retain knowledge from previous tasks. After training on experience $k$, we compute the F1 scores on all previously seen experiences ($\leq k$). This helps quantify the extent of {\em catastrophic forgetting} (CF).

    \item \textbf{Forward Transfer Performance}: We measures the model's ability to generalize to future, unseen tasks. After training on experience $n$, we compute the F1 score on all future experiences ($> k$). This indicates how well the model's current knowledge transfers to upcoming distributions.
\end{itemize}


\subsection{Continual Learning Experimental Results}

\begin{figure}[!t]
\centering
\begin{minipage}[c]{0.49\linewidth}
\centering
\includegraphics[width=0.9\linewidth]{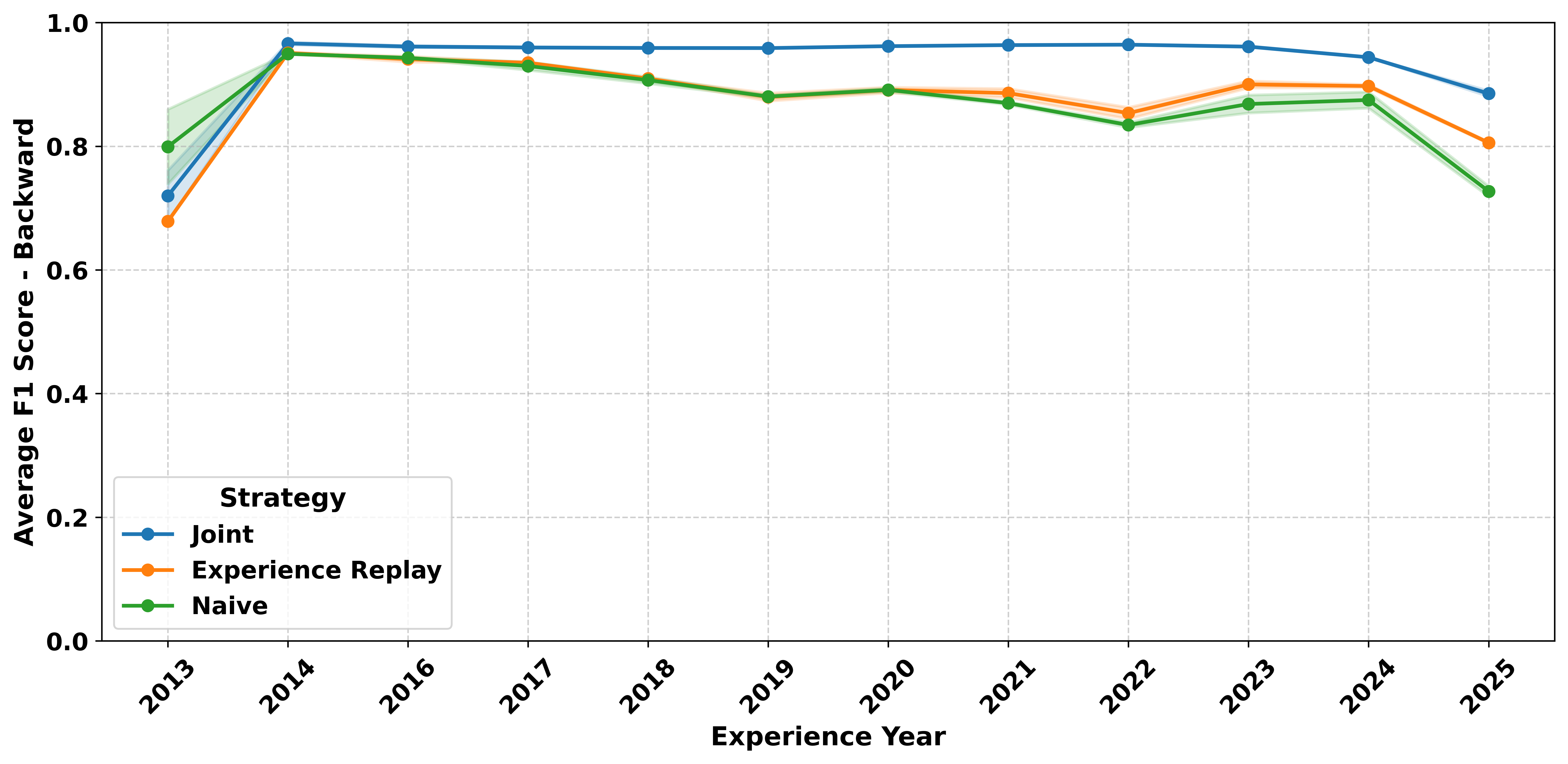}
\vskip -0.25cm 
\caption{F1 Score in Domain-IL (Forward)}
\label{fig:domain_il_backward_avg_f1}
\end{minipage}%
\hfill
\begin{minipage}[c]{0.49\linewidth}
\includegraphics[width=0.9\linewidth]{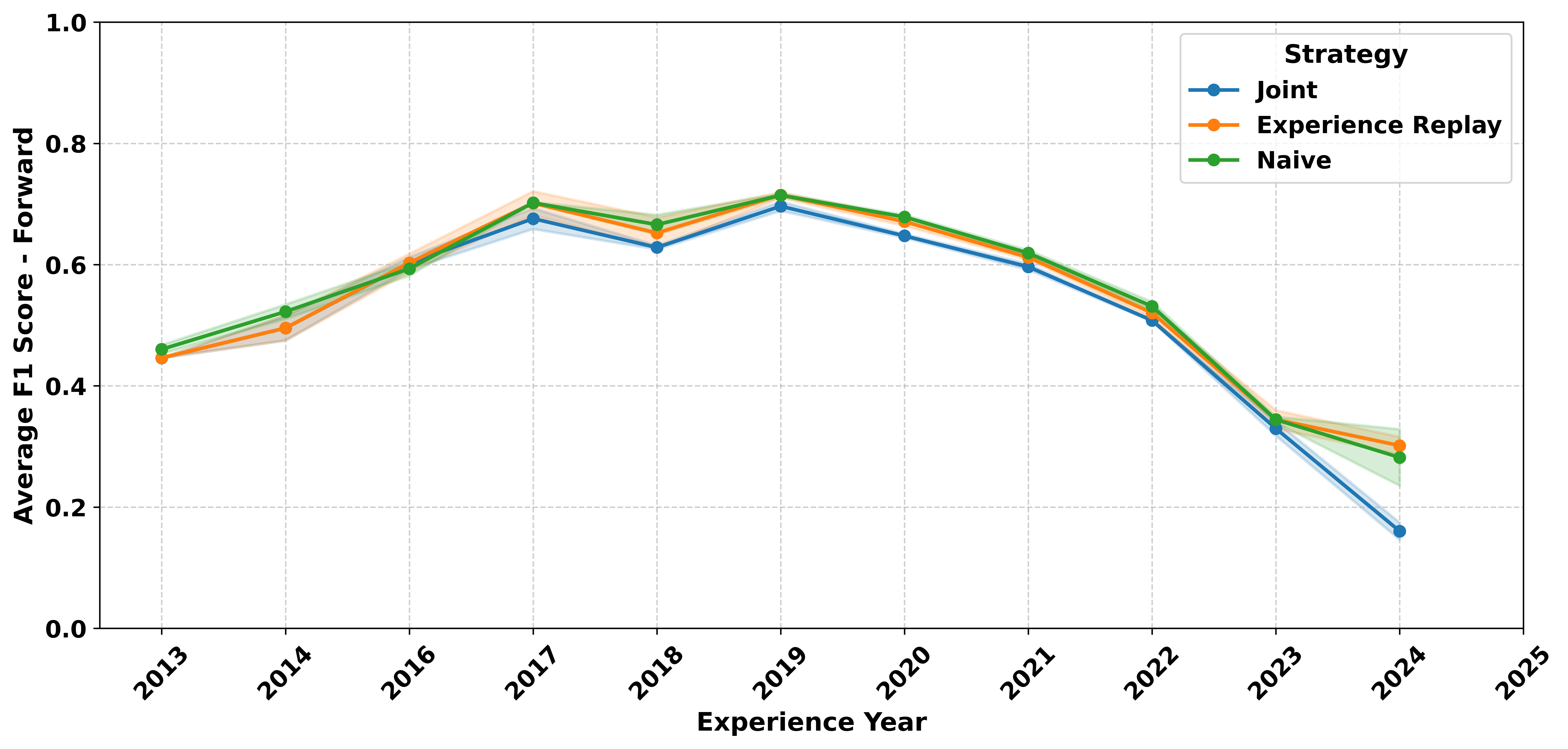}
\vskip -0.25cm
\caption{F1 Score in Domain-IL (Backward)}
\label{fig:domain_il_forward_avg_f1}
\end{minipage}
\end{figure}

\begin{figure}[!t]
\centering
\begin{minipage}[c]{0.49\linewidth}
\centering
\includegraphics[width=0.9\linewidth]{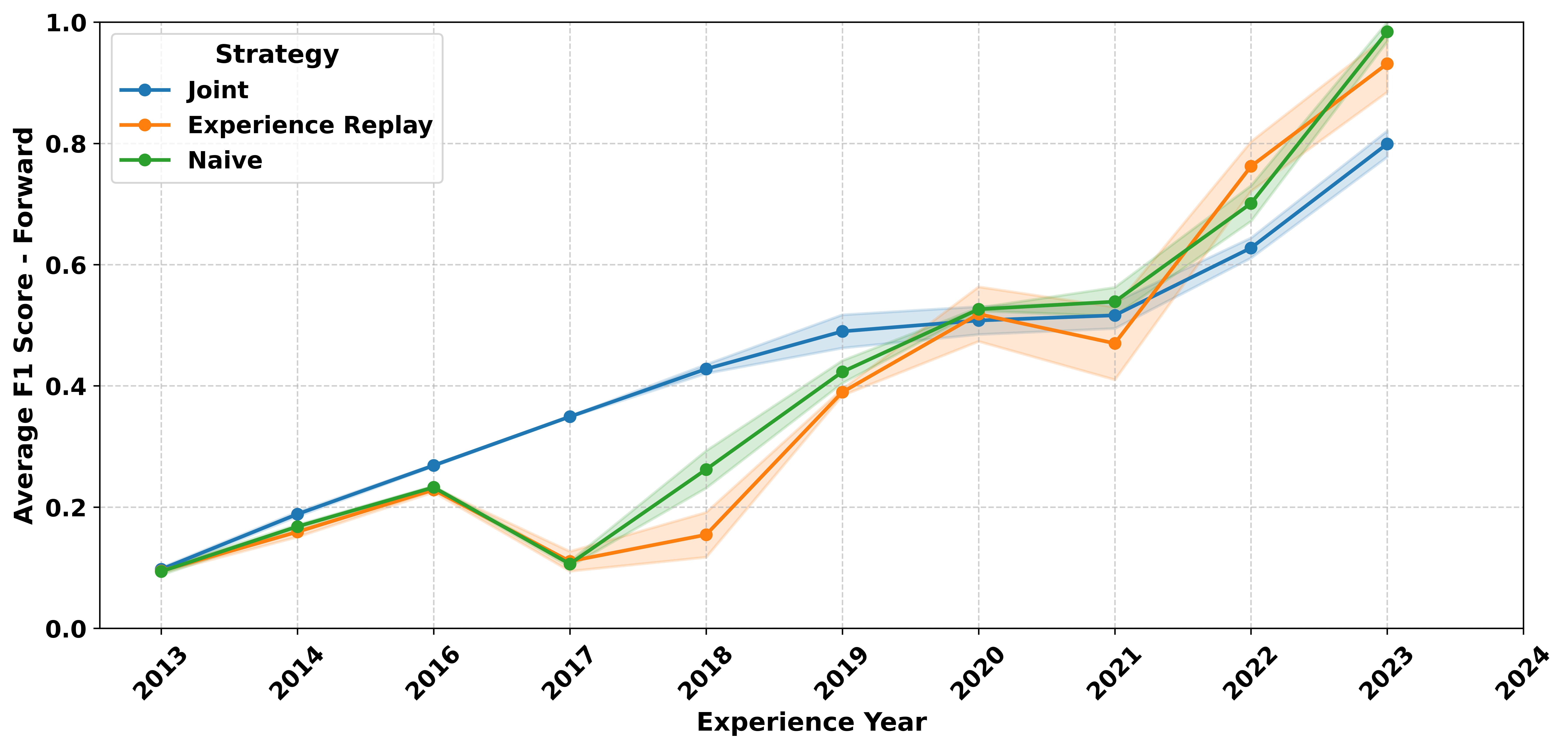}
\vskip -0.25cm 
\caption{F1 Score in Class-IL (Forward)}
\label{fig:class_il_forward_avg_f1}
\end{minipage}%
\hfill
\begin{minipage}[c]{0.49\linewidth}
\includegraphics[width=0.9\linewidth]{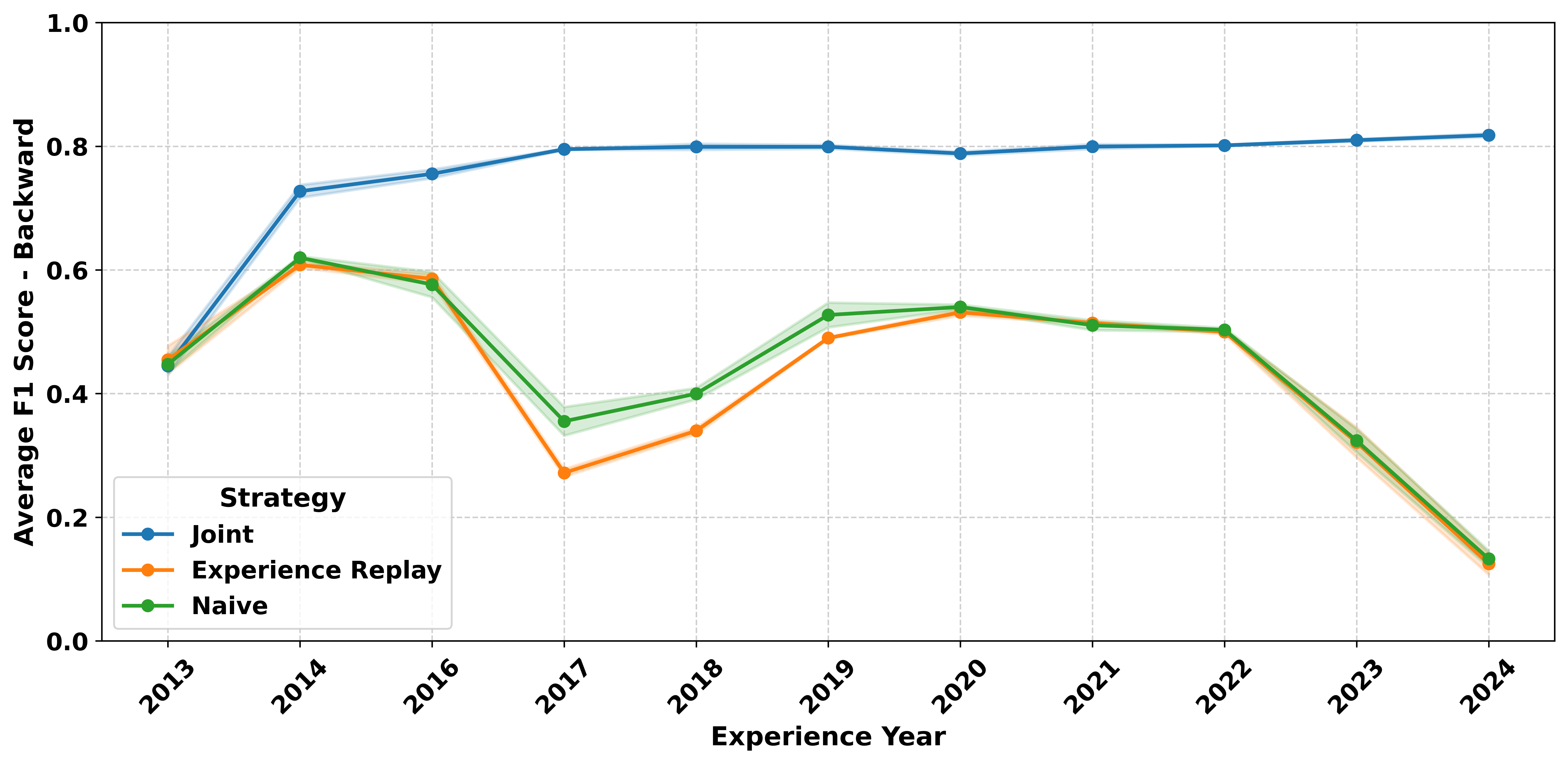}
\vskip -0.25cm
\caption{F1 Score in Class-IL (Backward)}
\label{fig:class_il_backward_avg_f1}
\end{minipage}
\end{figure}

Figures~\ref{fig:domain_il_backward_avg_f1}, \ref{fig:domain_il_forward_avg_f1}, \ref{fig:class_il_forward_avg_f1}, and \ref{fig:class_il_backward_avg_f1} demonstrate the effectiveness of the CL methods in evaluating in realistic scenarios using LAMDA benchmark. In the Class-IL setting, we observe strong signs of catastrophic forgetting, especially in the Naive and Replay (\textit{Experience Replay}) strategies. Backward F1 scores drop sharply after certain years, showing that learning new classes without retaining the previous knowledge leads to forgetting. Joint retains high performance as expected due to its exposure to all the previous data. In the Domain-IL setting, we observe that forgetting is relatively limited due to the fixed set of classes (\textit{malware} or \textit{benign}). Although the data distribution evolves over time, which leads to all strategies experiencing a gradual decline in forward F1 scores as they fail to adapt to new distributions. 
Additionally, we report the average F1 scores of LAMDA across all tasks under the Class-IL and Domain-IL scenarios in Tables~\ref{tab:f1_dual_column_class} and~\ref{tab:f1_dual_column}. In the Class-IL setting (Table~\ref{tab:f1_dual_column_class}), the Joint strategy consistently achieves the highest performance, as expected, due to its access to the full dataset during training. However, this advantage also implies the need for significantly higher computational resources which makes it less practical for real-world settings. The Naive and Replay strategies perform considerably worse, which was also expected as the continual introduction of new classes. In contrast, the Domain-IL results (Table~\ref{tab:f1_dual_column}) show generally higher and more stable F1 scores across all strategies. Since the label space remains fixed over time, both Replay and even the Naive strategy perform reasonably well. This observation suggests that the primary challenge in Domain-IL is not always forgetting, but rather adapting to distributional shifts in the data.

These results highlight LAMDA’s ability to capture both key challenges in continual learning: class expansion and distributional shift. As such, LAMDA serves as a realistic and challenging benchmark that supports future research in continual learning.

\begin{table}[!ht]
\centering
\caption{Average F1 scores of Domain-IL across all experiences or tasks for LAMDA.}
\label{tab:f1_dual_column}
\scriptsize
\begin{tabular}{lll|@{\hskip 0.5cm}lll}
\toprule
\textbf{Year} & \textbf{Strategy} & \textbf{Average F1 Score} & \textbf{Year} & \textbf{Strategy} & \textbf{Average F1 Score} \\
\midrule

\multirow{3}{*}{2013} & Naive  & 48.86 $\pm$ 1.15   & 2020 & Naive  & 80.26 $\pm$ 0.27 \\
     & Joint  & 46.89 $\pm$ 0.33            &      & Joint  & 83.08 $\pm$ 0.13 \\
     & Replay & 46.56 $\pm$ 0.00            &      & Replay & 79.93 $\pm$ 0.27 \\
\midrule
2014 & Naive  & 59.36 $\pm$ 0.99            & 2021 & Naive  & 78.61 $\pm$ 0.33 \\
     & Joint  & 57.38 $\pm$ 1.67            &      & Joint  & 84.12 $\pm$ 0.18 \\
     & Replay & 57.13 $\pm$ 1.71            &      & Replay & 79.46 $\pm$ 0.55 \\
\midrule
2016 & Naive  & 68.07 $\pm$ 0.82            & 2022 & Naive  & 75.86 $\pm$ 0.48 \\
     & Joint  & 69.21 $\pm$ 0.78            &      & Joint  & 85.02 $\pm$ 0.03 \\
     & Replay & 68.75 $\pm$ 1.20            &      & Replay & 77.02 $\pm$ 0.52 \\
\midrule
2017 & Naive  & 77.79 $\pm$ 0.31            & 2023 & Naive  & 78.10 $\pm$ 1.28 \\
     & Joint  & 77.05 $\pm$ 1.13            &      & Joint  & 85.59 $\pm$ 0.23 \\
     & Replay & 77.93 $\pm$ 1.27            &      & Replay & 80.74 $\pm$ 0.73 \\
\midrule
2018 & Naive  & 76.64 $\pm$ 1.06            & 2024 & Naive  & 82.54 $\pm$ 0.81 \\
     & Joint  & 76.61 $\pm$ 0.15            &      & Joint  & 87.86 $\pm$ 0.11 \\
     & Replay & 75.94 $\pm$ 1.46            &      & Replay & 84.77 $\pm$ 0.11 \\
\midrule
2019 & Naive  & 79.74 $\pm$ 0.16            & 2025 & Naive  & 72.71 $\pm$ 0.83 \\
     & Joint  & 82.76 $\pm$ 0.41            &      & Joint  & 88.52 $\pm$ 0.46 \\
     & Replay & 79.72 $\pm$ 0.09            &      & Replay & 80.57 $\pm$ 0.15 \\
\bottomrule
\end{tabular}
\end{table}

\begin{table}[!ht]
\centering
\caption{Average F1 scores of Class-IL across all experiences or tasks for LAMDA.}
\label{tab:f1_dual_column_class}
\scriptsize
\begin{tabular}{lll|@{\hskip 0.5cm}lll}
\toprule
\textbf{Year} & \textbf{Strategy} & \textbf{Average F1 Score} & \textbf{Year} & \textbf{Strategy} & \textbf{Average F1 Score} \\
\midrule

\multirow{3}{*}{2013} & Naive  & 12.60 $\pm$ 0.65   & 2020 & Naive  & 53.50 $\pm$ 0.18 \\
     & Joint  & 12.89 $\pm$ 0.37            &      & Joint  & 68.63 $\pm$ 0.67 \\
     & Replay & 12.75 $\pm$ 0.48            &      & Replay & 52.67 $\pm$ 1.34 \\
\midrule
2014 & Naive  & 25.01 $\pm$ 0.28            & 2021 & Naive  & 51.83 $\pm$ 0.87 \\
     & Joint  & 28.64 $\pm$ 0.49            &      & Joint  & 72.23 $\pm$ 0.46 \\
     & Replay & 24.07 $\pm$ 0.74            &      & Replay & 50.21 $\pm$ 1.51 \\
\midrule
2016 & Naive  & 32.64 $\pm$ 0.72            & 2022 & Naive  & 53.91 $\pm$ 0.42 \\
     & Joint  & 40.14 $\pm$ 0.22            &      & Joint  & 76.98 $\pm$ 0.36 \\
     & Replay & 32.60 $\pm$ 0.36            &      & Replay & 54.74 $\pm$ 0.43 \\
\midrule
2017 & Naive  & 19.68 $\pm$ 1.12            & 2023 & Naive  & 38.39 $\pm$ 1.84 \\
     & Joint  & 51.14 $\pm$ 0.09            &      & Joint  & 80.88 $\pm$ 0.25 \\
     & Replay & 16.92 $\pm$ 1.21            &      & Replay & 37.65 $\pm$ 2.46 \\
\midrule
2018 & Naive  & 32.47 $\pm$ 2.06            & 2024 & Naive  & 13.27 $\pm$ 1.33 \\
     & Joint  & 59.66 $\pm$ 0.31            &      & Joint  & 81.79 $\pm$ 0.29 \\
     & Replay & 23.86 $\pm$ 2.22            &      & Replay & 12.49 $\pm$ 1.76 \\
\midrule
2019 & Naive  & 47.98 $\pm$ 1.71            &  &   &  \\
     & Joint  & 65.86 $\pm$ 1.20            &      &   &  \\
     & Replay & 44.43 $\pm$ 0.26            &      &  &  \\
\bottomrule
\end{tabular}
\end{table}

\section{Computational Resources for LAMDA generation}
\label{appendix:computation}

All dataset processing and experiments for LAMDA were conducted on a high-performance compute server with the following configuration:

\begin{itemize}
    \item \textbf{CPU}: Dual-socket \texttt{Intel Xeon Gold 6430} with a total of 128 logical cores (64 physical cores, 2 threads per core).
    \item \textbf{Memory}: 1 TB RAM, with approximately 810 GB available during runtime.
    \item \textbf{GPU}: 4$\times$ NVIDIA H100 NVL GPUs with 95.8 GB memory per GPU. Experiments were conducted under \texttt{CUDA 12.8} and driver version \texttt{570.124.06}. 
\end{itemize}

This infrastructure enabled us to efficiently process over 1 million APKs, large-scale temporal benchmarking over 12 years of Android malware data.

\section{Dataset Documentation}
\label{appendix:datasetdocument}

\subsection{Hosted URLs}

\paragraph{DOI.} \url{https://doi.org/10.57967/hf/5563}

\paragraph{Hugging Face.} \url{https://huggingface.co/datasets/IQSeC-Lab/LAMDA}.

\paragraph{Croissant.} \url{https://huggingface.co/api/datasets/IQSeC-Lab/LAMDA/croissant} 

\paragraph{GitHub Code Access.} \url{https://github.com/IQSeC-Lab/LAMDA}

\paragraph{Project Page.} \url{https://iqsec-lab.github.io/LAMDA/}

\subsection{Dataset Curation and Preprocessing Methodology}
\begin{itemize}
    \item \textbf{Dataset Construction:} A corpus of over one million Android Package Kits (APKs), spanning the years 2013 to 2025 with the exclusion of 2015, is compiled from the AndroZoo repository~\cite{androzoo,androzooMetadata}. A 20\% overhead hases is included in the downloading process to account for download and decompilation failures. The collected APKs are systematically organized into year-specific directories, with subdirectories designated for malware (\texttt{[year]/malware/}) and benign applications (\texttt{[year]/benign/}).
    
    \item \textbf{Label Assignment:} Binary classification labels is assigned based on the output of VirusTotal (VT) analysis reported in the AndroZoo repository~\cite{androzoo,androzooMetadata}:
    \begin{itemize}
        \item \textbf{Benign:} $\text{vt\_detection} = 0$
        \item \textbf{Malware:} $\text{vt\_detection} \geq 4$
        \item \textbf{Uncertain:} $\text{vt\_detection} \in [1, 3]$ (discarded)
    \end{itemize}

    \item \textbf{Malware Family Labeling:} 
     AVClass2~\cite{sebastian2016avclass} is used to standardize malware family labels using VirusTotal reports. Labels are linked to APKs using \texttt{SHA256} hashes to support multi-class and temporal malware analysis.
    

    \item \textbf{Static Feature Extraction based on Drebin:} Each APK is decompiled using \texttt{apktool}~\cite{apktool} to extract static features:
    \begin{itemize}
        \item From \texttt{AndroidManifest.xml}: permissions, components (activities, services, receivers), hardware features, intent filters
        \item From \texttt{smali} code: restricted/suspicious API calls, hardcoded URLs/IPs
    \end{itemize}

    \item \textbf{Vectorization \& Preprocessing:} Extracted features are vectorized into high-dimensional binary vectors using a bag-of-tokens approach. A global vocabulary ($\sim$9.69M tokens) was constructed. Dimensionality was reduced using \texttt{VarianceThreshold} (threshold = 0.001), resulting in 4,561 final features.
    
    \item \textbf{Data Splitting:} Each year's data is split using stratified sampling:
    \begin{itemize}
        \item \textbf{Training:} 80\%
        \item \textbf{Testing:} 20\%
    \end{itemize}
    Class balance is maintained within each split.

    \item \textbf{Storage \& Format:} Final dataset is saved in both \texttt{.npz} (sparse matrix) and \texttt{.parquet} (tabular) formats. Each year's folder includes:
    \begin{itemize}
        \item \texttt{X\_train.parquet}
        \item \texttt{X\_test.parquet}
    \end{itemize}
    Metadata columns include: \texttt{hash}, \texttt{label}, \texttt{family}, \texttt{vt\_count}, \texttt{year\_month}, followed by binary features.
    
    \item \textbf{Scalability Support:} We have released global vocabulary, selected features, and preprocessing objects (e.g., \texttt{VarianceThreshold}) to enable integration with ML pipelines, including Hugging Face.
\end{itemize}

\subsection{Accessibility and Reproducibility}

The dataset has been made publicly available on Hugging Face at \url{https://huggingface.co/datasets/IQSeC-Lab/LAMDA} and has been assigned a permanent Digital Object Identifier (DOI): \url{https://doi.org/10.57967/hf/5563}. Furthermore, a dedicated GitHub project page has been created at \url{https://iqsec-lab.github.io/LAMDA/}, which includes detailed instructions and code to reproduce the reported results.

We are committed to the long-term preservation of our dataset through regular checks aimed at identifying and rectifying any data anomalies. Moreover, we are dedicated to the continuous maintenance of this resource by promptly addressing user inquiries and issues, and by releasing updates and enhancements informed by user feedback.

\end{document}